\documentclass[12pt]{revtex4-2}
\usepackage[normalem]{ulem}
\usepackage{natbib}
\usepackage{graphicx}
\usepackage{amsfonts}
\usepackage{amsmath,amssymb} 
\usepackage{bm}
\usepackage{xcolor}
\usepackage{hyperref}
\usepackage{caption}
\usepackage{subcaption}
\usepackage{tabularx}
\usepackage{mathtools}
\usepackage{comment}
\input{CQGformat.input}

\newcommand{\En}{\mathcal{E}}
\newcommand{\Lz}{\mathcal{L}}
\newcommand{\Q}{\mathcal{Q}}

\renewcommand{\sp}{\epsilon} 
\newcommand{\mr}{\epsilon} 

\newcommand{\avg}[1]{\left\langle #1 \right\rangle}

\newcommand{\osc}[1]{\Breve{#1}}
\newcommand{\nit}[1]{\Tilde{#1}}
\newcommand{\partialnit}[1]{\hat{#1}}
\newcommand{\PD}[2]{\frac{\partial #1}{\partial #2}}
\newcommand{\HOT}[1]{\mathcal{O}( \mr ^{ #1 } )}

\begin{document}


\title{Fast inspirals and the treatment of orbital resonances}


\author{Philip Lynch}
\ead{philip.lynch@aei.mpg.de}
\address{Max Planck Institute for Gravitational Physics (Albert Einstein Institute), Potsdam-Golm, Berlin}
\address{School of Mathematics and Statistics, University College Dublin, Belfield, Dublin 4, Ireland}

\author{Vojt{\v e}ch Witzany}
\address{Institute of Theoretical Physics, Faculty of Mathematics and Physics, Charles University, CZ-180 00 Prague, Czech Republic}

\author{Maarten van de Meent}
\address{Max Planck Institute for Gravitational Physics (Albert Einstein Institute), Potsdam-Golm, Berlin}
\address{Niels Bohr International Academy, Niels Bohr Institute, Blegdamsvej 17, 2100 Copenhagen, Denmark}

\author{Niels Warburton}
\address{School of Mathematics and Statistics, University College Dublin, Belfield, Dublin 4, Ireland}

\date{\today}

\begin{abstract}
	Extreme mass ratio inspirals (EMRIs), where a compact object orbits a massive black hole, are a key source of gravitational waves for the future Laser Interferometer Space Antenna (LISA). 
	Due to their small mass ratio, ($\mr \sim 10^{-4}$--$10^{-7}$), the binary evolves slowly and EMRI signals will be in-band for years. 
	Additionally, astrophysical EMRIs are expected to have complex dynamics featuring both spin-precession and eccentricity.
	A standard approach to modelling these inspirals is via the method of osculating geodesics (OG) which we employ along with a toy model for the gravitational self-force. 
	Using this method requires resolving tens of thousands radial and polar orbital librations over the long duration of the signal which makes the inspiral trajectory expensive to compute.
	In this work we accelerate these calculations by employing Near-Identity (averaging) Transformations.
	However, this averaging technique breaks down at orbital resonances where the radial and polar frequencies are an integer ratio of each other.
	Thus, we switch to a partial averaging transformation in the vicinity of the resonance where the dynamics are characterised by the slow evolution of the so-called ``resonant phase". 
	Additionally, we develop an optimal switching criterion to minimise the computation time while maximising accuracy.
	We find the error in the waveform phase is improved from $\mathcal{O}(\mr^{-1/2})$ in the fully averaged scheme to $\mathcal{O}(\mr^{4/7})$ in the switching scheme.
	At the same time, this scheme improves the scaling of the computation time from being inversely proportional to $\mr$ using OG, to a very weak scaling with $\mr$.
	This results in a speed-up of at least two orders of magnitude for LISA EMRIs with room for further optimisation.
\end{abstract}


\maketitle

\section{Introduction} \label{section:Introduction}

Spaced based gravitational wave (GW) detectors, such as the Laser interferometer space antenna (LISA)  \cite{Baker:2019nia,Colpi:2024xhw}, will be sensitive to much lower frequency GWs than the current ground based detectors of LIGO-Vigro-Kagra collaboration. 
This will enable it to detect entirely new sources of GWs involving heavier masses and sources at larger separation than those detected by ground-based observatories \cite{Amaro-Seoane:2012vvq}. 

One particularly intriguing class of such sources are extreme mass ratio inspirals (EMRIs) \cite{Berry2019}. 
These consist of a massive black hole (MBH) primary with a mass $M \sim 10^5 - 10^7 M_\odot$ and a stellar mass compact object (CO) secondary (either a black hole or neutron star) with a mass $\mu \sim 1- 10^2 M_\odot$, resulting in a binary with (small) mass ratio $\mr \coloneqq \mu / M \sim 10^{-4} - 10^{-7}$. 
The secondary loses energy and angular momentum due to the emission of GWs, which leads to the gradual decay of the orbit and its final plunge into the central MBH. LISA will be sensitive to the outgoing GWs for months or years during the slow inspiral \cite{Babak2017a}, resulting in a precise mapping of the spacetime of the MBH \cite{Gair:2004iv}. 
Detection and analysis of these signals will result in unrivalled precision in MBH parameter estimation and our most rigorous strong-field tests of general relativity to date \cite{Gair:2012nm,Barausse:2020rsu}.

To achieve these aims EMRI waveform models need to meet three important criteria~\cite{LISAConsortiumWaveformWorkingGroup:2023arg}. 
They must be fast to compute, ideally in a fraction of a second, so that they can be used with Markov chain Monte Carlo Bayesian inference methods~\cite{Burke:2023lno}. 
They must extend throughout the entire EMRI parameter space, including the spins of the primary and the secondary, eccentricity and orbital inclination with respect to the plane of the primary \cite{Hopman:2005vr,Babak2017a, Burke:2023lno}.
Finally, they must be accurate enough not to bias parameter estimation, which means maintaining phase accuracy to within a fraction of a radian throughout the entire inspiral. 
Using a two-timescale analysis \cite{Hinderer2008} one can show that the orbital (and thus GW) phase elapsed during the inspiral between a referential orbital state and the plunge can be written as a post-adiabatic (PA) expansion:
\begin{equation}
	\varphi = \mr^{-1} \varphi^{0\text{PA}} + \mr^{-1/2} \varphi^{\text{res}}  +   \varphi^{1 \text{PA}}  + \HOT{1}.
\end{equation}
The leading order adiabatic (0PA) contribution can be determined by balancing the fluxes of energy, angular momentum and Carter constant lost by the binary with the flux radiated to infinity and down the horizon of the primary \cite{Mino2003,Drasco:2005is,Hughes2005, Sago:2005fn, Isoyama:2018sib}.
Currently, there exist fast and extensive kludge models \cite{Kennefick2002,Barack2004, Babak2007,Sopuerta2011,Chua2015} which have found use in mock LISA data challenges  \cite{Arnaud2006, Babak2009,Chua:2021aah}. 
However, they use non-relativistic assumptions that limit their utility for LISA data analysis \cite{Chua2017}. 
Fully-relativistic, adiabatic models that cover the entire parameter space have been computed \cite{Poisson:1993vp, Cutler:1994pb, Detweiler1978,Finn:2000sy, Kennefick2002, Hughes2000,Hughes2001, Hughes2021} but fast and practical implementations of these models currently only cover eccentric, Schwarzschild (non-spinning primary) inspirals \cite{Katz2021, Chua2021a}, or quasi-circular Kerr inspirals (spinning primary) \cite{Nasipak:2023kuf} thus far. 
A fast, analytic model that is extensive in the parameter space also exists, but the model relies on a slow velocity (post-Newtonian) expansion that is ill-suited to modelling the late-inspiral \cite{Isoyama:2021jjd}. 

To obtain the sub-radian accuracy we require, one must go beyond adiabatic models and develop post-adiabatic (1PA) waveforms. This necessitates knowledge of the local force on the secondary induced by its own gravitational field. 
This back-reaction force known as the gravitational self-force (GSF) \cite{Barack:2018yvs,Pound2021}. 
It is calculated via a perturbative expansion in the small mass ratio of the system ($\epsilon$) and, to obtain the 1PA contribution to the phases, one requires the complete first-order GSF along with an orbit averaged contribution from the second-order GSF \cite{Hinderer2008}.

Calculations of the GSF require knowledge of the entire past history of the inspiral, which makes these computations highly non-trivial. 
One approach is to assume the secondary is on a fixed geodesic and calculate the first order gravitational self-force for that geodesic in the frequency domain. 
This has yielded results for quasi-circular \cite{Barack2007} and eccentric \cite{Barack2010b,Akcay2013,Osburn2014} Schwarzschild orbits and eccentric \cite{VandeMeent2016}, inclined \cite{Lynch:2023gpu} and generic Kerr orbits \cite{VanDeMeent2018}. 
To compute an inspiral, these numerical results can be interpolated over the parameter space of geodesic orbits so that they can be rapidly evaluated when solving for the inspiral dynamics \cite{Warburton2012,Osburn2016}.

This approach must be modified at second order in mass ratio, and current calculations make use of a two-timescale approximation which can account for the slow inspiral of the source \cite{Miller2021}. Currently, the complete list of contributions necessary for 1PA-accurate phasing is available only for the simplest case of quasi-circular, Schwarzschild inspirals \cite{Pound2019,Warburton2021,Durkan:2022fvm,Wardell:2021fyy,Miller:2023ers,Spiers:2023mor}, but recent work has also included the effects of an (anti-)aligned spinning secondary \cite{Burke:2023lno}. Methods are under development to extend these calculations to Kerr space-time \cite{Spiers:2023cip} and to include eccentricity and generic secondary spins in the inspirals \cite{Leather:2023dzj,Witzany:2019nml,Skoupy:2021asz,Skoupy:2023lih}.

With a fast to evaluate model for the force, inspiral trajectories can be computed using the method of osculating geodesics (OG) \cite{Pound2008,Gair2011}. 
This method models the inspiral trajectory as a smooth evolution through geodesic orbits which are instantaneously tangent to the inspiralling motion.  
As a result the equations of motion are recast as a series of coupled first order differential equations for the evolving orbital elements $P_j$ and orbital phases $q_i$. 
This technique has been utilised for modelling inspirals with orbital eccentricity \cite{Warburton2012, Osburn2016,Lynch:2021ogr} and/or inclination \cite{Lynch:2023gpu, Drummond:2023loz}, but the resulting models are very slow to evaluate.
Since the solutions to the equations of motion oscillate with the orbital phases and one has to model $\sim \epsilon^{-1}$ orbital cycles, these equations of motion can take minutes to hours to solve numerically for a single EMRI. 

To overcome this problem, the long-standing technique of near-identity (averaging) transformations has been applied to EMRI systems to great effect \cite{NITs,kevorkian12, Lynch:2022zov}. 
One applies a small transformation to the variables to be solved such that the resulting equations for the new variables, $\nit{P}_j$ and $\nit{q}_i$, are independent of the orbital phases while accurately capturing the long-term secular behaviour of the system. 
The result is 1PA accurate EMRI models which can incorporate eccentricity and/ or inclination that can be numerically evaluated in a fraction of a second. 
NITs have been successfully applied to Schwarzschild inspirals with low eccentricity ($e\le 0.2$) \cite{NITs} and high eccentricity ($e\le 0.75$)\cite{McCart2021}, as well as Kerr inspirals with eccentricity ($e\le0.5$) \cite{Lynch:2021ogr}, inclination \cite{Lynch:2023gpu}, and both (generic) with GW fluxes and the Mathisson-Papapetrou-Dixon (MPD) force of a spinning secondary \cite{Drummond:2023wqc}. The use of NITs for more general perturbations of EMRIs was discussed in Refs.~\cite{Lukes-Gerakopoulos2021,Pan:2023wau}.

One of the leading formation channels for EMRIs predicts binaries which are both highly eccentric and inclined with respect to the orbital plane while in the LISA band \cite{Gair:2004iv,Babak2017a}, and so the focus of this work is to accurately and efficiently incorporate both of these effects.
For eccentric and inclined (generic) orbits, one finds that there are subspaces of the parameter space where the radial frequency $\Upsilon_r^{(0)} $ becomes an integer ratio of the polar frequency $\Upsilon_\theta^{(0)} $ , i.e., $ \Upsilon_\perp = \vec{\kappa} \cdot \vec{\Upsilon} =  \kappa_r \Upsilon_r^{(0)} + \kappa_\theta \Upsilon_\theta^{(0)}  = 0$, where $\kappa_r, \kappa_\theta \in \mathbb{Z}$. 
As a result, in this region the so-called resonant phase given as $q_\perp = \kappa_r q_r + \kappa_\theta q_\theta$, where  $q_r,q_\theta$ are the radial and orbital phases respectively, stops evolving, and this state is resolved only by the slow drift of frequencies due to radiation reaction. 
Since quantities such as energy and angular-momentum flux depend also on the value of $q_\perp$ \cite{Flanagan:2012kg}, the inspiral generally leaves the resonance with an $\mathcal{O}(\mr^{1/2})$ spread of possible energies and angular momenta depending on the precise of value of $q_\perp$ at which it crossed the resonance \cite{Flanagan:2010cd}. 
Even though this is a smooth evolution that transpires over $\mathcal{O}(\mr^{-1/2})$ orbital periods, this is sometimes viewed as an $\mathcal{O}(\mr^{1/2})$ ``jump'' in energies and angular momenta when resolving the inspiral over $\mathcal{O}(\mr^{-1})$ orbital cycles.
Finally, when evolved over the $ \mathcal{O}(\mr^{-1})$ inspiral timescale, the phase contribution of the resonant ``jump'' generally accumulates with a scaling of $\mathcal{O}(\mr^{-1/2})$ \cite{Flanagan:2010cd}. 
Failing to accurately model these passages through resonances will bias parameter estimation and lead to a loss of detection of EMRI signals by LISA \cite{Berry2016}.  
Therefore, understanding and modelling these effects accurately is a top priority \cite{Flanagan:2012kg,Ruangsri:2013hra,Nasipak:2021qfu,Nasipak:2022xjh}. 

Note that the resonances we are dealing with here should not be confused with tidal transient resonances due to the presence of a third body perturber when the radial, polar, and azimuthal frequencies are a small integer ratio of each other \cite{Bonga:2019ycj,Gupta:2021cno}. Tidal resonances have been efficiently modelled along with 0PA effects radiation reaction effects \cite{Gupta:2022fbe}. Similarly, the scalings we assume also preclude the modelling of inspirals through resonances under other perturbations such as non-Kerr multipoles of the massive primary or other external or internal perturbations (see, e.g., \cite{Apostolatos:2009vu,Polcar:2022bwv,Destounis:2023khj}). In this work, we focus purely on resonance effects that arise out of the orbital dynamics and the GSF, and aim to model these in a way that is both computationally efficient and maintains sub-radian phase accuracy.

We do this by introducing four separate models. The first uses the OG equations with a model for GSF to drive the inspiral. Unfortunately, our generic Kerr first order GSF code \cite{VanDeMeent2018} is too computationally expensive to tile even a small subsection of the generic Kerr parameter space, and so we use a toy model that combines information from interpolated eccentric \cite{Lynch:2021ogr} and quasi-circular inclined \cite{Lynch:2023gpu} orbits. We also emulate the second order GSF by rescaling our first order toy model and multiplying by an additional factor of the mass ratio.  The OG equations driven with a toy model can be used to simulate resonant effects, but the evolution is very slow to evaluate when directly integrated by standard numerical integration methods. Though the resulting inspiral trajectories are not to be taken as physically valid, for the purposes of this work we treat these as the ``true" inspirals against which we test faster models.

The second model averages away all dependence on the orbital phases from the OG equations of motion, which we denote the ``Full NIT".  While this model can rapidly produce inspiral trajectories in less than a second for any mass ratio, it has terms in both the averaged equations of motion and the transformation terms that become singular when a low-order orbital resonance is encountered. Thus, formally, this model cannot evolve through a resonance, though due to our use of interpolation, in practice our numerical integrator can cross resonant surfaces but with a severe loss of accuracy.

This necessitates the production of  a third model that removes all phase dependence apart from combinations of the resonant phase $q_\perp$ (and multiples there of) which we denote the ``Partial NIT". This accurately captures the resonant effects but is not as fast as the Full NIT. 

Finally we combine these two models into a fourth model which we denote the ``Switch NIT", where the Full NIT is used away from resonances and the Partial NIT is used to evolve through the resonance. A critical component of this model is our novel criterion for switching, which follows the general arguments laid out in Ref.~\cite{Lukes-Gerakopoulos2021} and which is designed to maximize the $\mr$-scaling in accuracy while minimizing the integration spent in the relatively expensive Partial NIT.
This final model accurately captures the effects of the orbital resonance while dramatically decreasing the computation time of the trajectory calculation.

We start by restating the form of the OG equations for generic Kerr inspirals in Sec.~\ref{section:OG}. 
We then give a brief overview of the phenomenon of orbital transient resonances in Kerr spacetime in Sec.~\ref{section:Resonances}.
In Sec.~\ref{section:FullNIT}, we summarize the details of the Full NIT applied the case of generic Kerr inspirals in the absence of any low order orbital resonances.
We then outline the Partial NIT procedure in Sec.~\ref{section:PartialNIT} before describing our Switch NIT procedure in Sec.~\ref{section:SwitchNIT}.
In Sec.~\ref{section:Implementation} describe our practical implementation  of the online and offline steps required for these three NIT variants and discuss how we generate and evaluate waveforms. 
We then present the numerical results of our implementation by examining the convergence of the error induced by the Full, Partial, and Switch NITs as a function of mass ratio. We then discuss how the time for the trajectory calculation varies with mass ratio for each procedure. 
Once, we are satisfied with the accuracy and speed of our Switch NIT procedure, we test it on a pair of year-long EMRIs, one which evolves through a single low order resonance in Sec.~\ref{section:single_res} and one which evolves through two low order resonances in Sec.~\ref{section:double_res}. While this implementation must be optimised further before it could be recommenced for data analysis applications, these tests confirm that the Switch NIT can accurately capture resonance crossings while drastically speeding up EMRI trajectory calculations. 

Throughout this work, we use a toy  force model which is informed by eccentric and spherical GSF. Details of its construction can be found  in \ref{section:gen_Toy_Model}. A full derivation of the partial NIT can be found in \ref{section:near-resonant-NIT}. Finally, the derivation of the switching condition and the associated error scalings can be found in \ref{section:transition_condition}. This work uses geometrized units where $ c = G=  1$.

\section{Inspirals in Kerr Spacetime} \label{section:OG}

We wish to describe the motion of a secondary of mass $\mu$ into a rotating black hole of mass $M$ and spin parameter $a = |J|/M$, where  $J$ is its spin angular momentum. For this, we make use of the method of osculating geodesics (OG) which has been very successful in describing both Schwarzchild \cite{Pound2008} and Kerr \cite{Gair2011} inspirals. One assumes the inspiral is smoothly evolving from one geodesic orbit to the next, which allows for the recasting of the forced geodesic equation into a system of first order ordinary differential equations for the ``orbital elements" which uniquely identify the geodesic orbit that is instantaneously tangent to the inspiral. 

There are many possible choices of orbital elements and in this work we use the quantities $\vec{P} = \{p,e,x\}$, where $p$ is the semilatus rectum, $e$ is eccentricity and $x$ is a measure of orbital inclination. These can be defined in terms of the minimum and maximum values of the radial ($r$) and polar ($\theta$) Boyer-Lindquist coordinates via:
	\begin{subequations}\label{eq:RootDefintions}
		\begin{gather}
			p =  \frac{2 r_{\text{max}} r_{\text{min}}}{(r_{\text{max}} + r_{\text{min}}) M}  \text{ ,}  \quad e =  \frac{ r_{\text{max}}- r_{\text{min}}}{r_{\text{max}} + r_{\text{min}} } \text{ , \quad and}  \quad  x = \pm \sqrt{1 - \cos^2 \theta_{\text{min}}}, \tag{\theequation a-c}
		\end{gather}
	\end{subequations}
	where $x$ is positive for prograde orbits and negative for retrograde orbits.
We also use (Carter-)Mino time, $\lambda$, as our time parameter as this decouples the radial and polar geodesic motion \cite{Carter:1968rr,Mino2003}. This is related to proper time, $\tau$, via:
\begin{equation}\label{eq:Mino}
	d\tau = (r^2 + a^2 \cos^2\theta) d \lambda.
\end{equation}

With this in hand, we parametrize the radial and polar motion using the Mino time action angles for the geodesic motion $\vec{q} = (q_r,q_\theta)$. For geodesic motion these are simply described by:
\begin{subequations}\label{eq:PhaseDefinitions}
	\begin{gather}
		q_r = \Upsilon_r ^{(0)}\lambda + q_{r,0}, \quad q_\theta = \Upsilon_\theta^{(0)} \lambda + q_{\theta,0},\tag{\theequation a-b}
	\end{gather}
\end{subequations}
where $q_{r,0}$ and $q_{\theta,0}$ are the initial values of the phases at $\lambda = 0$. We also denote $\Upsilon_r ^{(0)}$ and $\Upsilon_\theta ^{(0)}$ as the Mino time fundamental radial and polar frequencies respectively, which have known analytic expressions in terms of $p,e$ and $x$ \cite{Fujita2009a}. 
This allows us to make use of the analytic solutions for the radial, $r(a,p,e,x,q_r)$, and polar, $\theta(a,p,e,x,q_\theta)$, coordinates which are given in Refs.~\cite{Fujita2009a,VandeMeent2020} and implemented in the \texttt{KerrGeodesics} package \cite{2023zndo...8108265W} as part of the Black Hole Perturbation Toolkit \cite{BHPToolkit}. 
Finally, we also require evolution equations for ``extrinsic quantities" that don't show up on the right hand side of the equations of motion due to the symmetry of Kerr spacetime, but are still necessary to compute the waveform. 
In this case, these are the time and azimuthal coordinates of the secondary which, as a set, we denote by $\vec{S} = \{t,\phi\}$.

For this work we assume our secondary is under the influence of a force that resembles the GSF and experiences an acceleration away from geodesic motion with the form $a_\mu = \mr a^{(1)}_{\mu} + \mr^2 a^{(2)}_{\mu} + \HOT{3}$. As such the OG equations of motion accurate to 1PA order can be expressed as:
\begin{subequations}\label{eq:Generic_EMRI_EoM}
	\begin{align}
		\begin{split}
			\dot{P_j} &= \mr F_j^{(1)} (\vec{P}, \vec{q}) + \mr^2  F_j^{(2)} (\vec{P}, \vec{q}) + \HOT{3},
		\end{split}\\
		\begin{split}
			\dot{q_i} &=  \Upsilon_i^{(0)} (\vec{P}) + \mr  f_i ^{(1)}(\vec{P}, \vec{q}) + \HOT{2},
		\end{split}\\
		\begin{split}
			\dot{S_k} &= s_k(\vec{P}, \vec{q}) + \HOT{2} .
		\end{split}
	\end{align}
\end{subequations}
For the full form and derivation of these equations see Ref.~\cite{Lynch:2021ogr}. Note that an alternative form of these equations exist that are parametrized in terms of quasi-Keplerian angles \cite{Gair2011}. As that form is more computationally efficient, for our numerical comparisons we solve those equations instead and then convert to the Mino-time action angles after the fact.  

Before we can calculate inspirals, we first need a model for the secondary's four-acceleration. 
Creating an interpolated GSF model for generic Kerr inspirals is computationally unfeasible at this time, due to the cost of computing the generic Kerr self-force for a single point in the parameter space and the need to tile in three dimensions instead of two dimensions required for the equatorial and spherical cases (after fixing the spin of the primary to a single value).
Instead we construct a self-force inspired toy model for generic orbits by combining our interpolated self-force models for eccentric \cite{Lynch:2021ogr} and spherical \cite{Lynch:2023gpu} orbits in such a way that we have radial and polar cross terms in the Fourier expansion of the force components which will give rise to resonant effects. The resulting model is an analytic expression in terms of $p,e,x,q_r,$ and $q_\theta$. Since one will need to compute derivatives of the equations of motion with respect to these variables, this model allows for analytic calculations for these derivatives which are computationally cheaper and more accurate than taking numerical derivatives. Further details on the model's construction can be found in~\ref{section:gen_Toy_Model}.

	We also compare the the size of resonant effects on the integrals of motion induced by the purely dissipative parts of the toy model at different orbital resonances and found that toy model produces qualitatively comparable behaviour to that observed in in Ref.~\cite{Flanagan:2012kg} using GW flux calculations. However, our model has a tendency to overestimate the effect of the lowest order resonance and underestimate higher order resonances.

\section{Transient Orbital Resonances in Kerr spacetime} \label{section:Resonances}
	\begin{figure}
	\begin{subfigure}[b]{0.49\textwidth}
		\centering
		\includegraphics[trim={2cm 1.5cm 2cm 1.5cm},clip,width=0.9\textwidth]{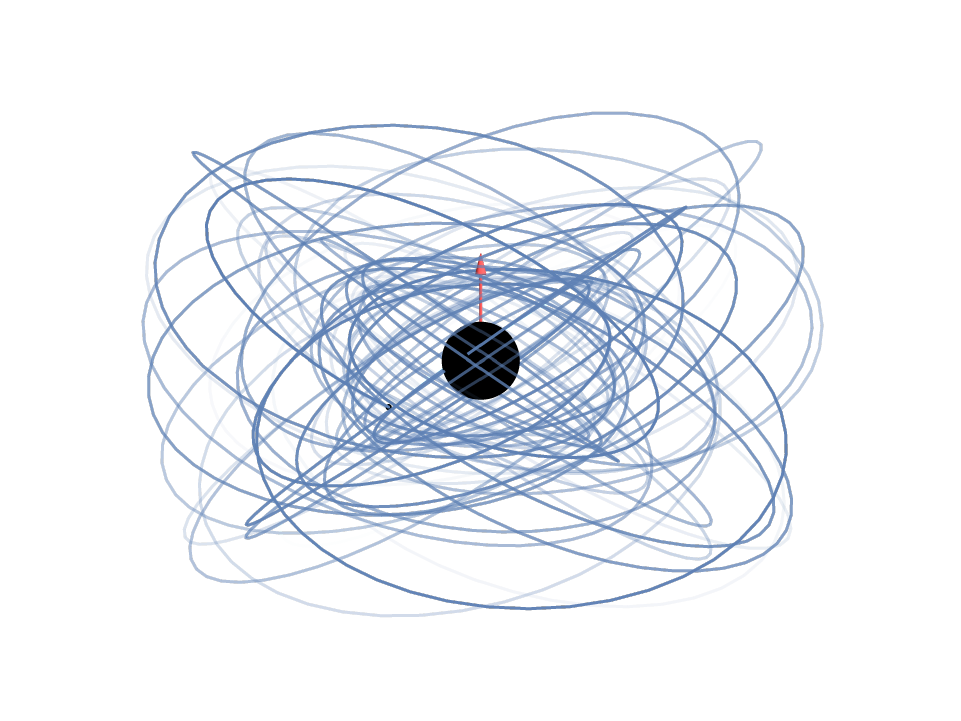}
		\caption{Generic orbit}
		\label{fig:GeodesicGeneric2}
	\end{subfigure}
	\hfill
	\begin{subfigure}[b]{0.49\textwidth}
		\centering
		\includegraphics[width=0.9\textwidth]{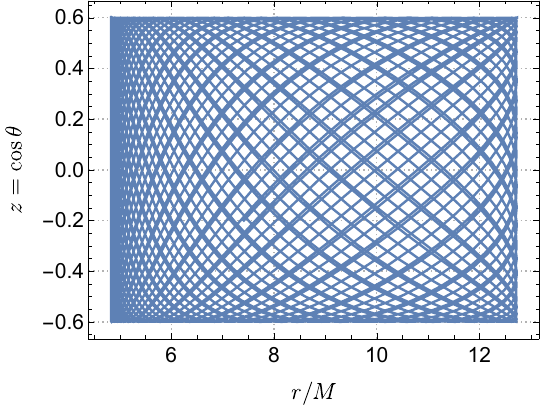}
		\caption{Generic phase space}
		\label{fig:PhaseSpaceGeneric}
	\end{subfigure}
	\hfill
	\begin{subfigure}[b]{0.49\textwidth}
		\centering
		\includegraphics[clip,width=0.9\textwidth]{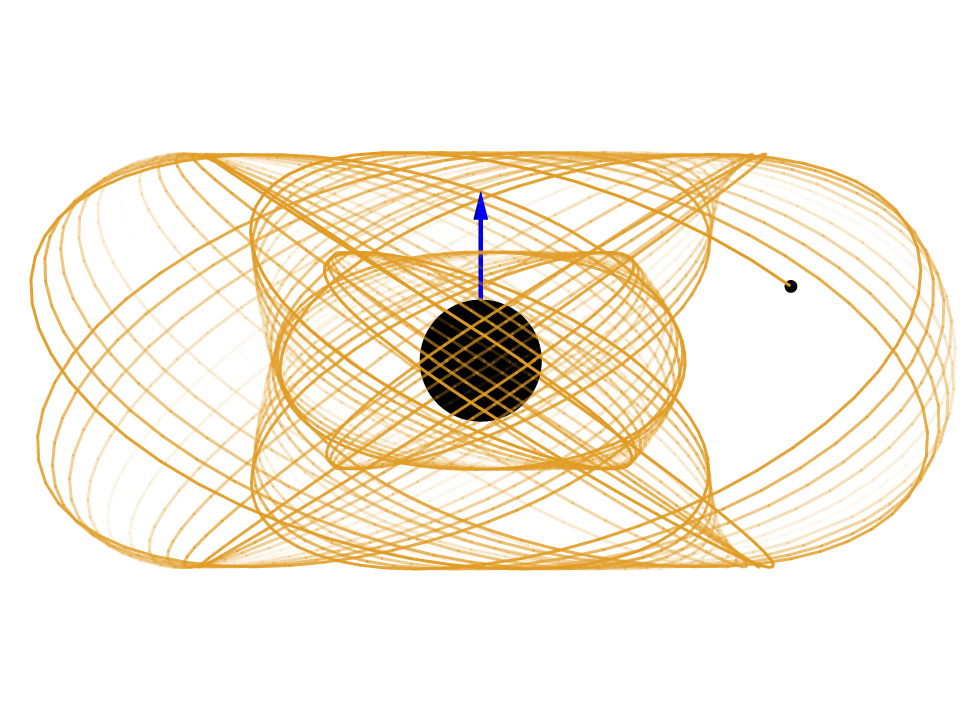}
		\caption{Resonant orbit: $q_{r,0} = 0, q_{\theta,0} = 0$}
		\label{fig:GeodesicResonant}
	\end{subfigure}
	\hfill
	\begin{subfigure}[b]{0.49\textwidth}
		\centering
		\includegraphics[width=0.9\textwidth]{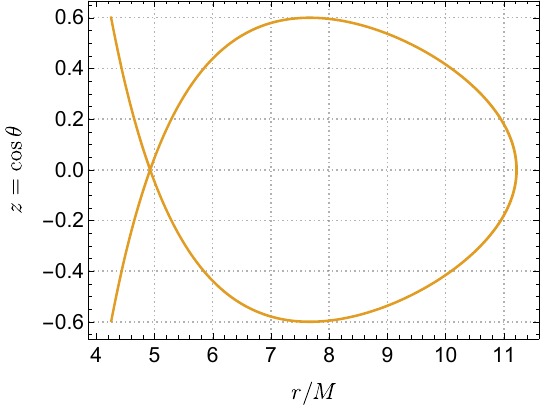}
		\caption{Resonant phase space: $q_{r,0} = 0, q_{\theta,0} = 0$}
		\label{fig:PhaseSpaceRes}
	\end{subfigure}
	\hfill
	\hfill
	\begin{subfigure}[b]{0.49\textwidth}
		\centering
		\includegraphics[clip,width=0.9\textwidth]{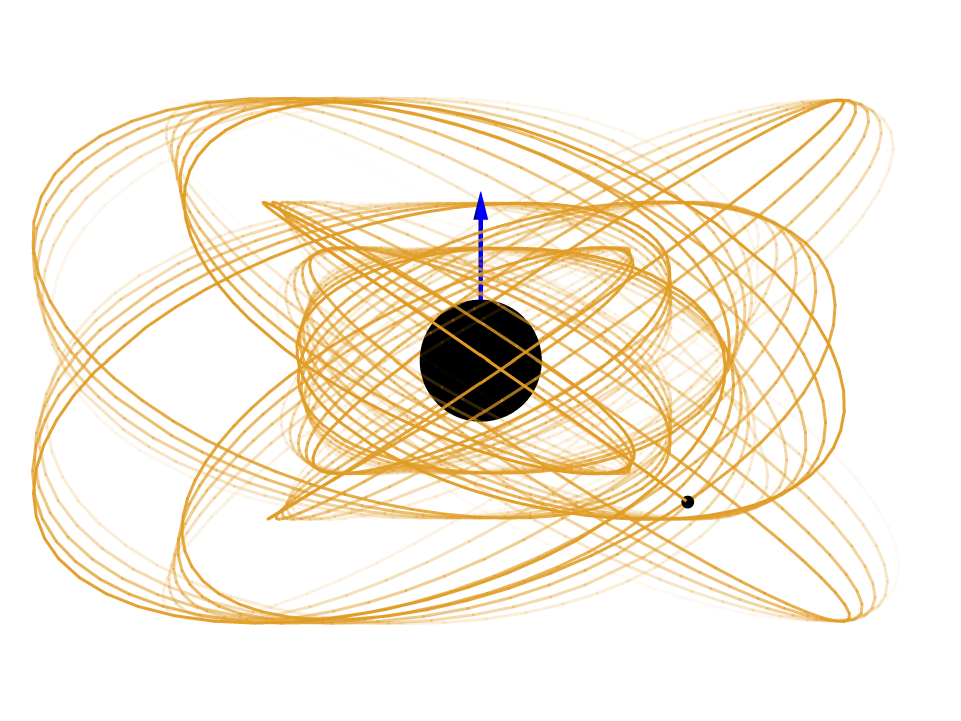}
		\caption{Resonant orbit: $q_{r,0} = 0, q_{\theta,0} = \pi/4$}
		\label{fig:GeodesicResonant2}
	\end{subfigure}
	\hfill
	\begin{subfigure}[b]{0.49\textwidth}
		\centering
		\includegraphics[width=0.9\textwidth]{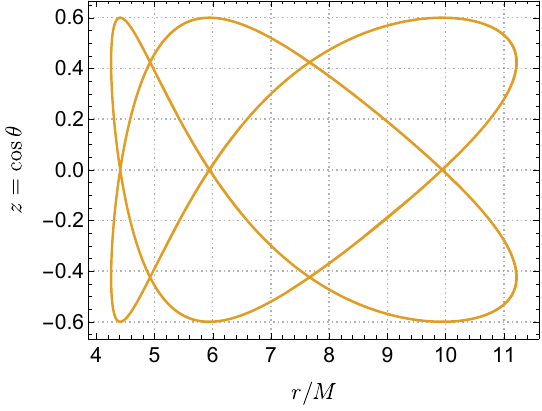}
		\caption{Resonant phase space: $q_{r,0} = 0, q_{\theta,0} = \pi/4$}
		\label{fig:PhaseSpaceRes2}
	\end{subfigure}
	\caption{Parametric plots of $r$ vs $z$ for a generic orbit and resonant orbits which all have $a = 0.9M$, $e = 0.45$ and $x = 0.8$, with initial phases $q_{r,0} = q_{\theta,0} = 0$. The non-resonant orbit has $p = 7$ and the resonant orbits has $p = 6.171$ which is the location of the $2/3$ orbital resonance. The two resonant orbits differ only by their initial phases.}
	\label{fig:GeodesicPhaseSpace}
\end{figure}

A resonant orbit occurs whenever the radial phase is related to the polar phase by a small number integer ratio, i.e., $\vec{\kappa}_\text{res} \cdot \vec{\Upsilon}^{(0)} = \kappa_r \Upsilon_r^{(0)} + \kappa_\theta \Upsilon_\theta ^{(0)} = 0$ for $\kappa_r,\kappa_\theta  \in \mathbb{Z}$. 
We denote a specific orbital resonance using the fraction $ \Upsilon^{(0)}_\theta /\Upsilon^{(0)}_r = |\kappa_\theta |/|\kappa_r|$.

As illustrated in Figs.~\ref{fig:GeodesicGeneric2} and \ref{fig:PhaseSpaceGeneric}, if a generic orbit is allowed to evolve for infinitely many orbits, it will eventually fill the entirety of the $(r,z = \cos \theta)$ space bounded between $r_\text{max}$ and $r_\text{min}$, and $z_\text{max}$ and $-z_\text{max}$.
Such an orbit is said to be ergodic in the phase space.
This allows us to equate the infinite Mino-time average for a geodesic with an integral over the 2-torus of the action angles $q_r$ and $q_\theta$ \cite{Drasco:2003ky}, i.e.,
\begin{equation}
	\avg{A} = \lim_{\lambda \rightarrow \infty} \frac{1}{2 \lambda} \int_{-\lambda}^\lambda A(\lambda') d\lambda' =    \frac{1}{(2 \pi)^2} \int_0^{2\pi} \int_0^{2\pi} A(q_r,q_\theta) dq_r dq_\theta = A_{0,0},
\end{equation}
where $A_{0,0}$ is the zeroth Fourier coefficient. We also define the purely oscillatory piece of a function to be $\osc{A} \coloneqq A - \avg{A}$.

However, as seen in Figs.~\ref{fig:PhaseSpaceRes} and \ref{fig:PhaseSpaceRes2}, a resonant orbit does not fill the $r,z$ space and instead repeatedly traces out the same trajectory in this space. 
Moreover, the dissimilarity between these figures demonstrates that the phase space trajectory is affected by the initial conditions for the phases, i.e., $q_{r,0}$ and $q_{z,0}$.
Thus, one cannot equate the infinite Mino-time average for a resonant geodesic with the 2-torus average of the action angles and instead one gets \cite{Grossman:2011im,Flanagan:2012kg}
\begin{equation}
	\avg{A}_{\text{res}} = \lim_{\lambda \rightarrow \infty} \frac{1}{2 \lambda} \int_{-\lambda}^\lambda A(\lambda') d\lambda' = \sum_{N \in \mathbb{Z}} A_{N\kappa_r, N \kappa_\theta} e^{i N \left(\kappa_r q_{r,0}+ \kappa_\theta q_{\theta,0} \right)}.
\end{equation}
As such, any averaging procedure done in the presence of an orbital resonance will have to account for this new definition of orbit average.

However, notice that since the Fourier coefficients of a $C^{\infty}$-function fall-off exponentially, the difference between $\langle A\rangle_{\rm res}$ and $\langle A\rangle$ is exponentially suppressed for smooth functions $A$ and growing resonant order, in other words 
\begin{align}
\langle A \rangle_{\rm res} - \langle A \rangle \lesssim C \exp\left[-B(|\kappa_r| + |\kappa_\theta|)\right]\,,\;{\rm for } \; |\kappa_r| + |\kappa_\theta| \gg 1\,,
\end{align}
where $B,C$ are some constants.
Thus, even though rational numbers are dense in the real numbers, and hence there are an infinite number of potential orbital resonances, one only has to worry about a finite number of resonances with a low order. In practice we take this cut off to be $\max \left( |\kappa_r| , |\kappa_\theta|\right) \lesssim 10$. This will be justified a posteriori by our results.

One should also note that there is a subtlety when evaluating the order of the resonance due to the symmetries of the problem \cite{THprivate}. 
The abstract space of Kerr geodesics has the reflection symmetry about the plane $\theta=\pi/2$ in the sense that when we take any initial polar phase $q_{\theta,0}$ and shift it by $\pi$, we get the same orbit, just reflected by the plane $\theta = \pi/2$. 
Indeed, for generic orbits, this new orbit is essentially the same geodesic as before the shift. 
On the other hand, for resonant orbits the $\theta = \pi/2$ reflected orbit is generally a topologically disparate orbit. 
Since to leading order the GSF is a functional computed along such geodesics, it inherits this symmetry, and it will be $\pi$-periodic in the polar $q_\theta$ angle, whereas it will generically be only $2 \pi$-periodic in radial angle $q_r$. 
This results in the odd Fourier modes of the GSF with respect to $q_\theta$ (or even modes in the case of $F_\theta$) being zero, as demonstrated in Figs.~9-12 of Ref.~\cite{VanDeMeent2018}. 
This means that any resonance with a ratio with an odd polar number $\kappa_\theta$ would be more accurately described as having twice that ratio, as there are no odd powered polar modes contributing to the strength of the resonance. 
For example: in the $1/2$ resonance, the leading order contribution comes from the $\| \kappa_\theta \| = 2$ and $\| \kappa_r \| = 4$ modes, making it effectively a  $2/4$ resonance. 
As such, throughout this paper we will adopt this convention for naming the resonances as it more accurately conveys their actual strength, e.g., $1/2 = 2/4, 3/4 = 6/8,$ etc.

\begin{figure}
	\centering
	\includegraphics[width = 0.9 \textwidth]{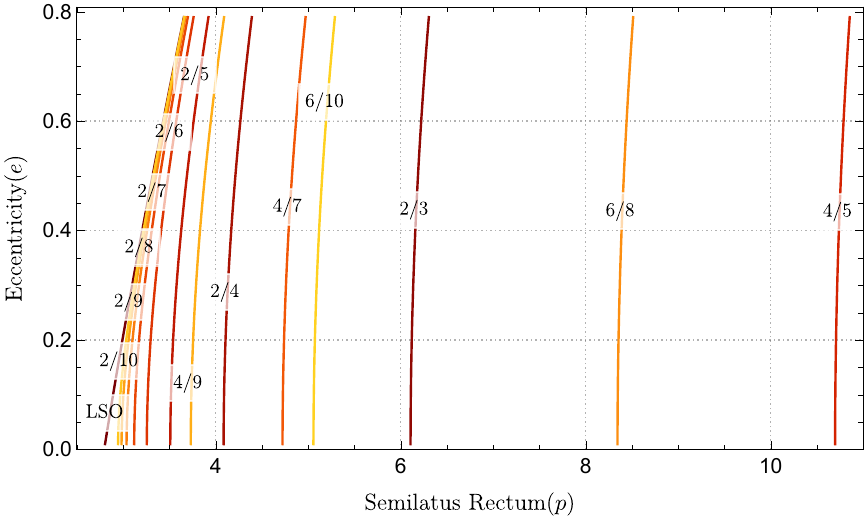}
	\caption{The locations of the resonant surfaces through $(p,e)$ space for $a = 0.9M$ and $x = 0.8$ along with the location of the last stable orbit (LSO). The curves are coloured such that the lower order the resonance, the darker the colour.}
	\label{fig:ResonanceLocations}
\end{figure}

 Generally, inspirals in Kerr space-time are very likely to pass through at least some low-order resonances, which can be seen as follows \cite{Ruangsri:2013hra}. 
The ratio $\kappa_\theta/\kappa_r$ corresponds to the frequency ratio as $|\kappa_\theta/\kappa_r| = \Omega_r/\Omega_\theta$ at resonance. 
In the Newtonian limit obtained, e.g., by taking $p\to \infty$, the geodesics become closed Keplerian ellipses which all have $\Omega_r/\Omega_\theta =1$. 
In the last stage of the inspiral the orbit reaches the surface of the last stable orbits (LSOs) characterized by $\Omega_r =0$ while $\Omega_\theta$ stays finite, or $\Omega_r/\Omega_\theta =0$. 
As a result, in an ideal inspiral reaching from a $p\to \infty$ orbit to its LSO, the frequencies are \textit{guaranteed} to pass through \textit{every} resonance with $\kappa_r>\kappa_\theta$. In particular, \textit{every} inspiral has to pass through the $2/3$ resonance on its way to the last stable orbit. 
Practically, however, one needs to ask whether such resonant passages will happen while the inspiral is in LISA band, and the answer seems to be affirmative for most LISA EMRIs \cite{Ruangsri:2013hra,Berry2016}.
 
The precise characterization of resonances is non-trivial, since they form a 3D hyper-surface in the 4D generic Kerr orbit parameter space \cite{Brink:2015roa}.
To get an idea of where these resonances occur in the parameter space, Fig.~\ref{fig:ResonanceLocations} illustrates the location in $(p,e)$ space in a 2D slice of the parameter space where we fixed $a = 0.9M$ and $x = 0.8$.
As we can see, most low order resonances occur near the location of the LSOs, which can be understood from the convergence of $\Omega_r$ to zero at the LSOs as mentioned above.
Since our inspiral models start to break down in this region anyway, these resonances are not the biggest concern.
What is more concerning are the resonances that occur at a significant distance from the last stable orbit, as the effect of inaccurately modelling these resonances can accumulate over a large number of orbits.
As such, the lowest order resonance of concern that we expect most EMRIs to pass through is the $2/3$ resonance.
Any averaging procedure employed to efficiently model EMRI trajectories will have to carefully account for the presence of these resonances in order to maintain subradian accuracy in the orbital phases.

	The size of the resonant effects at a given orbital resonance will also vary throughout the parameter space since they scale with the magnitude of the Fourier modes of the forcing terms in the equations of motion. Radial modes scale with eccentricity as $e^{\kappa_r}$, while polar modes scale with $(a  \cos( \theta_\text{min}))^{\kappa_\theta}$. As such, we see that the scaling of the resonant terms in the self-force will be 
	\begin{align}
		\frac{\langle A \rangle_{\rm res} - \langle A \rangle}{\langle A \rangle} \lesssim \tilde{C} e^{\kappa_r} (a  \cos( \theta_\text{min}))^{\kappa_\theta} \,,\;{\rm for } \; e \to 0\,,\;{\rm and }\; \theta_\text{min} \to \frac{\pi}{2} \; {\rm or } \; a\to 0 \,. \label{eq:epresoscaling}
	\end{align}
	Since the $1:1$ resonance is not present in the epicyclic oscillations of near-circular and near-equatorial Kerr geodesics apart from $p\to \infty$ \cite{Bardeen:1973tla}, resonant effects will smoothly vanish as one approaches near-equatorial and near-circular motion, and harmonics corresponding to higher-order resonances will vanish faster.\footnote{Note that these properties were not taken into account in Ref. \cite{Speri21} when parameterizing the size of the resonant terms, which probably led to overestimates of the importance of resonances for orbits at low eccentricity and inclination.}
		
	We can also use Eq.~\eqref{eq:epresoscaling} to deduce that resonances are suppressed in the weak field as follows. The first power of $a$ appears at the $1.5$PN order in the spin-orbital term, the $a^2$ in the 2PN spin-spin terms and the pattern is such that any appearance of $a^n$ is in a term of at least $n$-PN order (see, e.g., Ref. \cite{Levi:2016ofk}). Even more, the equations of motion of spinning binaries have recently been shown to be integrable to 2PN order \cite{Tanay:2020gfb}, which implies that all resonant terms in the equations of motion vanish at 2PN. Nevertheless, the scaling is complicated by the fact that radiation-reaction itself is suppressed in the weak-field and appears only at $2.5$PN in the equations of motion. In other words, if the resonance takes place at a larger $p$, the orbit spends more cycles evolving through the resonance \cite{Ruangsri:2013hra,Speri21}, which also contributes to the overall size of the resonance effects.
	
	Another factor to consider is that EMRIs will pass through multiple low order resonances before plunge. Since the resonance crossing  is not perfectly resolved, the resulting phase error will propagate to the next resonance, compounding the phase error with each resonance crossing. Specifically, if the inspiral accumulates a phase error of order $\mathcal{O}(1)$ or higher, we lose all predictive power about the resonant effects when evolving through the next resonance since we are randomly picking a trajectory on the resonant torus for the evolution. While we can estimate and control the error through a single resonance crossing, the presence of multiple resonance crossings make these estimates much more difficult, reinforcing the need to model each resonance crossing as accurately as possible.

\section{Near identity averaging transformations for generic Kerr inspirals} \label{section:NITs}

\subsection{Review of non-resonant averaging transformations (Full NIT)} \label{section:FullNIT}
Averaging transformations for a generic EMRI system in the absence of transient resonances was first given in Ref.~\cite{NITs} where a full derivation can be found. 
We now summarize the main findings of that work.

The NIT variables, $\nit{P}_j$, $\nit{q}_i$ and $\nit{S}_k$, are related to the OG variables  $P_j$, $q_i$ and $S_k$ via
\begin{subequations}\label{eq:transformation}
	\begin{align}
		\begin{split}\label{eq:transformationP}
			\nit{P}_j &= P_j + \sp Y_j^{(1)}(\vec{P},\vec{q}) + \sp^2 Y_j^{(2)}(\vec{P},\vec{q}) + \HOT{3},
		\end{split}\\
		\begin{split}\label{eq:transformationq}
			\nit{q}_i &= q_i + \sp X_i^{(1)}(\vec{P},\vec{q}) +\sp^2 X_i^{(2)}(\vec{P},\vec{q}) + \HOT{3},
		\end{split}\\
		\begin{split}\label{eq:extrinsic_transformation}
			\nit{S}_k &= S_k + Z_k^{(0)}(\vec{P},\vec{q}) +\sp Z_k^{(1)}(\vec{P},\vec{q}) + \HOT{2}.
		\end{split}
	\end{align}
\end{subequations}
Here, the transformation functions $Y_j^{(n)}$, $X_i^{(n)}$, and $Z_k^{(n)}$ are required to be smooth, periodic functions of the orbital phases $\vec{q}$. We also make the choice that the orbit averaged pieces of these functions:  $\avg{Y_j^{(n)}} = \avg{X_i^{(n)}} = \avg{Z_k^{(n)}} = 0$. Other choices for these pieces can be made, resulting in different equations of motion, as explored in Ref.~\cite{NITs}. 
At leading order, Eqs.~\eqref{eq:transformation} are identity transformations for $P_k$ and $q_i$ but not for $S_k$ due to the presence of a zeroth order transformation term $Z_k^{(0)}$. 

The inverse transformations can be found for $P_k$ and $q_i$ by requiring that their composition with the transformations in Eqs.~\eqref{eq:transformation} must give the identity transformation. Expanding order by order in $\epsilon$, this gives us
\begin{subequations} \label{eq:invNITdef}
	\begin{align}
		& P_j = \tilde{P}_j + \epsilon \tilde{Y}_j^{(1)}(\vec{\tilde{P}},\vec{\tilde{q}}) + \epsilon^2 \tilde{Y}_j^{(2)}(\vec{\tilde{P}},\vec{\tilde{q}})+ \mathcal{O}(\epsilon^3) \,,
		\\ 
		& q_i = \tilde{q}_i + \epsilon \tilde{X}_j^{(1)}(\vec{\tilde{P}},\vec{\tilde{q}}) + \epsilon^2 \tilde{X}_j^{(2)}(\vec{\tilde{P}},\vec{\tilde{q}})+ \mathcal{O}(\epsilon^3)\,,
	\end{align}
\end{subequations}
where the inverse transformation vectors are 
\begin{subequations}\label{eq:invNITvecs}
	\begin{align}
		& \tilde{Y}_j^{(1)} = - Y_j^{(1)}(\vec{\nit{P}},\vec{\nit{q}}) \,,
		\\ 
		& \tilde{Y}_j^{(2)} =- Y_j^{(2)}(\vec{\nit{P}},\vec{\nit{q}}) + \PD{Y_j^{(1)}(\vec{\nit{P}},\vec{\nit{q}})}{\nit{P_k}} Y_k^{(1)}(\vec{\nit{P}},\vec{\nit{q}}) + \PD{Y_j^{(1)}(\vec{\nit{P}},\vec{\nit{q}})}{\nit{q_k}} X_k^{(1)}(\vec{\nit{P}},\vec{\nit{q}}) \, ,
		\\
		& \tilde{X}_i^{(1)} = - X_i^{(1)}(\vec{\nit{P}},\vec{\nit{q}}) \,, \\ 
		& \tilde{X}_i^{(2)} = - X_i^{(2)}(\vec{\nit{P}},\vec{\nit{q}}) + \PD{X_i^{(1)}(\vec{\nit{P}},\vec{\nit{q}})}{\nit{P_j}} Y_j^{(1)}(\vec{\nit{P}},\vec{\nit{q}}) + \PD{X_i^{(1)}(\vec{\nit{P}},\vec{\nit{q}})}{\nit{q_k}} X_k^{(1)}(\vec{\nit{P}},\vec{\nit{q}})  \,.
	\end{align}
\end{subequations}
	
To find the equations of motion for the NIT variables $\nit{P}_j, \nit{q}_i$ and $\nit{S}_k$, one takes the time derivative of Eqs.~\eqref{eq:transformation}, substitutes in the equations of motion Eqs.~\eqref{eq:Generic_EMRI_EoM}, then uses the inverse transformations Eq.~\eqref{eq:invNITdef} to make sure the right hand side is in terms of only the transformed variables. Then one uses the oscillatory parts of  $Y_j^{(n)}$, $X_i^{(n)}$, and $Z_k^{(n)}$  to cancel out all of the oscillatory terms at each order of $\mr$. The result is equations of motion which take the following form:
\begin{subequations}\label{eq:transformed_EoM}
	\begin{align}
		\begin{split}
			\frac{d \nit{P}_j}{d \lambda} &=\epsilon \nit{F}_j^{(1)}(\vec{\nit{P}}) + \epsilon^2 \nit{F}_j^{(2)}(\vec{\nit{P}}) +  \mathcal{O}(\epsilon^3),
		\end{split}\\
		\begin{split}
			\frac{d \nit{q}_i}{d\lambda} &= \Upsilon_i^{(0)}(\vec{\nit{P}}) +\epsilon \nit{f}_i^{(1)}(\vec{\nit{P}}) + \mathcal{O}(\epsilon^2),
		\end{split}\\
		\begin{split}
			\frac{d \nit{S}_k}{d \lambda} &= \Upsilon_k^{(0)}(\vec{\nit{P}}) + \epsilon \nit{s}_k^{(1)}(\vec{\nit{P}}) +  \mathcal{O}(\epsilon^2).
		\end{split}
	\end{align}
\end{subequations}
Crucially, these equations of motion are now independent of the orbital phases $\vec{q}$. 
The terms in the averaged equations of motion are related to the terms in the OG equations of motion via
\begin{subequations} \label{eq:NIT_EoM}
	\begin{gather}
		\nit{F}_j^{(1)} = \left<F_{j}^{(1)}\right>, \quad \nit{f}_{i}^{(1)}= \left<f_{i}^{(1)}\right>, \quad \Upsilon_{k}^{(0)} = \left<s_{k} ^{(0)}\right>, \tag{\theequation a-c}
	\end{gather}
	\begin{equation}
		\nit{F}_j^{(2)} = \left<F_j^{(2)} \right> + \left<\frac{\partial \osc{Y}_j^{(1)}}{\partial \nit{q}_i} \osc{f}_i^{(1)} \right> + \left<\frac{\partial \osc{Y}_j^{(1)}}{\partial \nit{P}_k} \osc{F}_k^{(1)} \right>, \tag{\theequation d}
	\end{equation}
	\begin{equation}
		\nit{s}_k^{(1)} = - \left<\frac{\partial \osc{s}_k^{(0)}}{\partial \nit{P}_j} \osc{Y}_j^{(1)} \right> - \left<\frac{\partial \osc{s}_k^{(0)}}{\partial \nit{q}_i} \osc{X}_i^{(1)} \right>. \tag{\theequation e}
	\end{equation}
\end{subequations}
In deriving these equations of motion, we have constrained the oscillating pieces of the first order NIT transformation functions to be
\begin{equation}\label{eq:NIT_Y}
	Y_j^{(1)} =\sum_{\vec{\kappa} \neq \vec{0}} \frac{i}{\vec{\kappa} \cdot \vec{\Upsilon}^{(0)}} F_{j,\vec{\kappa}}^{(1)} e^{i \vec{\kappa} \cdot \vec{q}},
\end{equation}

\begin{equation}\label{eq:NIT_X}
	X_i^{(1)} =\sum_{\vec{\kappa} \neq \vec{0}}\left( \frac{i}{\vec{\kappa} \cdot \vec{\Upsilon}^{(0)}} f_{i,\vec{\kappa}}^{(1)} + \frac{1}{(\vec{\kappa} \cdot \vec{\Upsilon}^{(0)})^2} \frac{\partial \Upsilon_i}{\partial P_j}F_{j,\vec{\kappa}}^{(1)} \right) e^{i \vec{\kappa} \cdot \vec{q}}.
\end{equation}
For our purposes, we only need the second order transformation of the orbital elements which we constrain to be
	\begin{align} \label{eq:NIT_Y2}
	\begin{split}
		Y^{(2)}_{j} = & \sum_{\vec{\kappa} \neq \vec{0}} \frac{i e^{i \vec{\kappa} \cdot \vec{q}} }{\vec{\kappa} \cdot \vec{\Upsilon}^{(0)}} \Biggl( 
		F_{j,\vec{\kappa}}^{(2)} + \PD{\avg{Y_j^{(1)}}}{\nit{P}_k} F^{(1)}_{k,\vec{\kappa}} - i \PD{\avg{F_j^{(1)}}}{\nit{P}_k} \frac{F^{(1)}_{k,\vec{\kappa}}}{\vec{\kappa} \cdot \vec{\Upsilon}^{(0)}}
		\\ & +   \sum_{\vec{\kappa}' \neq \vec{0}} \biggl(   i \frac{F^{(1)}_{k,\vec{\kappa} - \vec{\kappa}^\prime }}{\vec{\kappa}' \cdot \vec{\Upsilon}^{(0)}} \left( \PD{F^{(1)}_{j,\vec{\kappa}'}}{\nit{P}_k} - \frac{F^{(1)}_{j,\vec{\kappa}^\prime}}{\vec{\kappa}' \cdot \vec{\Upsilon}^{(0)}} \PD{(\vec{\kappa}' \cdot \vec{\Upsilon}^{(0)}) } {\nit{P}_k}  \right) - \frac{\vec{\kappa}' \cdot \vec{f}^{(1)}_{\vec{\kappa} - \vec{\kappa}'}}{\vec{\kappa}' \cdot \vec{\Upsilon}^{(0)}} F^{(1)}_{j,\vec{\kappa}'}
		\biggr) \Biggr).
	\end{split}
\end{align}
The average of $Y^{(2)}$ above is chosen to be zero here so the corresponding term does not need to be included.
Substituting these expressions for the transformation terms into the expressions for the sub-leading terms in the averaged equations of motion allows us to express them in the simplified form:
\begin{subequations}
	\begin{gather}
		\nit{F}_j^{(2)} = \avg{F_j^{(2)}} + \mathcal{N}_\text{Full}( F_j^{(1)}), \quad 	\nit{s}_k^{(1)} =  \mathcal{N}_\text{Full}( s_k^{(0)}).
	\end{gather}
\end{subequations}
The contribution from the $\vec{\kappa}$ Fourier modes to the $\mathcal{N}_\text{Full}$ operator are given by
\begin{align} \label{eq:gen_N_Operator}
	\begin{split}
		\mathcal{N}_{\vec{\kappa}}(A) & \coloneqq  \frac{i}{\vec{\kappa} \cdot \vec{\Upsilon}^{(0)} } \Biggl[
		i A_{\vec{\kappa}}\left( \vec{\kappa} \cdot \vec{f}_{-\vec{\kappa}}^{(1)}\right) 
		+ \PD{A_{\vec{\kappa}} }{\nit{P}_j} F_{j,-\vec{\kappa}}^{(1)} 
		-  \frac{A_{\kappa}}{\vec{\kappa} \cdot \vec{\Upsilon}^{(0)} } \Biggl( \left( \PD{(\vec{\kappa} \cdot \vec{\Upsilon}^{(0)})  }{\nit{P}_j}  \right) F_{j,-\vec{\kappa}}^{(1)} \Biggr) \Biggr],
	\end{split}
\end{align}
 and thus the $\mathcal{N}_\text{Full}$ operator is given by the sum over these contributions
 \begin{align} \label{eq:N_Operator_Full}
 	\begin{split}
 	\mathcal{N}_\text{Full}  & \coloneqq \sum_{\vec{\kappa} \neq \vec{0}} 	\mathcal{N}_{\vec{\kappa}}(A) .
 	\end{split}
 \end{align}
After numerically solving the equations of motion, computing a waveform only requires knowledge of the transformations in Eq.~\eqref{eq:transformation} to zeroth order in the mass ratio so that the error is $\mathcal{O}(\mr)$, i.e.,
\begin{subequations}
	\begin{align}
		\begin{split}
			P_j &= \nit{P_j} + \mathcal{O}(\epsilon),
		\end{split}\\
		\begin{split}
			q_i &= \nit{q_i} + \mathcal{O}(\epsilon),
		\end{split}\\
		\begin{split}
			S_k &= \nit{S}_k -Z_k^{(0)}(\vec{\nit{P}},\vec{\nit{q}} ) +  \mathcal{O}(\epsilon).
		\end{split}
	\end{align}
\end{subequations}
where the zeroth order transformation term for the extrinsic quantities $\osc{Z}_k^{(0)}$ is known analytically as it is related to the analytic solutions for the geodesic equations for $t$ and $\phi$ derived in Ref.~\cite{Fujita2009a} by
\begin{equation}\label{eq:Z_solution}
	\osc{Z}_k^{(0)} = - \osc{S}_{k,r} (q_r) - \osc{S}_{k,\theta} (q_\theta).
\end{equation}
Furthermore, to be able to directly compare between OG and NIT inspirals, we will need to match their initial conditions to sufficient accuracy. 
In Refs.~\cite{Lynch:2021ogr,Lynch:2023gpu,Drummond:2023wqc} it was stated that, since we only require the result to be accurate to 1PA order, we only need to calculate the initial conditions of the phases and extrinsic quantities to within an $\mathcal{O}(\epsilon)$ error and the initial conditions of the orbital elements to within $\HOT{2}$ error. 
While this is still true, our calculation for the near-resonant switching criteria assumes that we  carry out the the transformation to the orbital elements and phases through one order higher in the mass ratio. Thus, to make our initial condition calculation consistent with this, we always calculate the initial conditions via:
\begin{subequations}  \label{eq:NIT_ICs}
	\begin{align}
		\begin{split}
			\nit{P}_j(0) &= P_j(0) +\mr  Y^{(1)}_j(\vec{P}(0), \vec{q}(0)) +  \mr^2  Y^{(2)}_j(\vec{P}(0), \vec{q}(0)) +  \HOT{3},
		\end{split}\\
		\begin{split}
				\nit{q}_i(0) &= q_i(0) +\mr  X^{(1)}_i(\vec{P}(0), \vec{q}(0)) +   \HOT{2},
		\end{split}\\
		\begin{split}
			\nit{S}_k(0) &= \nit{S}_k(0) -Z_k^{(0)}(\vec{\nit{P}}(0),\vec{\nit{q}} (0)) +   \mathcal{O}(\mr).
		\end{split}
	\end{align}
\end{subequations}

\subsection{Partial NIT when near orbital resonances} \label{section:PartialNIT}
As discussed in Sec.~\ref{section:Introduction}, the Full NIT can only be applied for generic orbits away from low-order resonances.
In the presence of one of the low order resonances, $\nit{F}_{j}^{(2)}$ (Eq.~(\ref{eq:NIT_EoM}d)), $Y_{j}^{(1)}$ (Eq.~\eqref{eq:NIT_Y}) and  $X_{i}^{(1)}$ (Eq.~\eqref{eq:NIT_X}) all exhibit singular behaviour.
As such, we adopt a ``Partial NIT" formulation when in the vicinity of an orbital resonance.  The concept of partially averaging for a resonant system is a long standing approach in classical mechanics \cite{arnold06, kevorkian12}, but was first introduced in the EMRI context in Ref.~\cite{VanDeMeent2014a}.  While the resulting inspirals will not be as quick to compute as the Full NIT (but still much faster than the OG equations), the resulting inspiral quantities should still be accurate to the OG inspiral to linear order in mass ratio. We present the full derivation of Partial NIT in \ref{section:near-resonant-NIT} and summarize the main results below.

The Partial NIT variables, $\partialnit{P}_j$, $\partialnit{q}_i$ and $\partialnit{S}_k$, are related to the OG variables  $P_j$, $q_i$ and $S_k$ via
\begin{subequations}\label{eq:partial_transformation}
	\begin{align}
		\begin{split}\label{eq:transformation1}
			\partialnit{P}_j &= P_j + \sp \partialnit{Y}_j^{(1)}(\vec{P},\vec{q} ) + \sp^2 \partialnit{Y}_j^{(2)}(\vec{P}, \vec{q} ) + \HOT{3},
		\end{split}\\
		\begin{split}
			\partialnit{q}_i &= q_i + \sp \partialnit{X}_i^{(1)}(\vec{P},\vec{q}) +\sp^2 \partialnit{X}_i^{(2)}(\vec{P},\vec{q} ) + \HOT{3},
		\end{split}\\
		\begin{split}\label{eq:extrinsic_transformation}
			\partialnit{S}_k &= S_k + \partialnit{Z}_k^{(0)}(\vec{P},\vec{q} ) +\sp \partialnit{Z}_k^{(1)}(\vec{P},\vec{q} ) + \HOT{2},
		\end{split}
	\end{align}
\end{subequations}
In summary, the equations of motion for the partial NIT variables now take the form
\begin{subequations}\label{eq:res_transformed_EoM}
	\begin{align}
		\begin{split}
			\frac{d \partialnit{P}_j}{d \lambda} &=\epsilon \partialnit{F}_j^{(1)}(\vec{\partialnit{P}},\partialnit{q}_\perp) + \epsilon^2 \partialnit{F}_j^{(2)}(\vec{\partialnit{P}},\partialnit{q}_\perp) +  \mathcal{O}(\epsilon^3),
		\end{split}\\
		\begin{split}
			\frac{d \partialnit{q}_i}{d\lambda} &= \Upsilon_i^{(0)}(\vec{\partialnit{P}}) +\epsilon \partialnit{f}_i^{(1)}(\vec{\partialnit{P}},\partialnit{q}_\perp) + \mathcal{O}(\epsilon^2),
		\end{split}\\
		\begin{split}
			\frac{d \partialnit{S}_k}{d \lambda} &= \Upsilon_k^{(0)}(\vec{\partialnit{P}}) + \epsilon \partialnit{s}_k^{(1)}(\vec{\partialnit{P}}) +  \mathcal{O}(\epsilon^2).
		\end{split}
	\end{align}
\end{subequations}
Crucially, these equations of motion only depend on the slowly evolving orbital elements~$\vec{\partialnit{P}}$ and the resonant phase~$\partialnit{q}_\perp$ but not on any of the other rapidly oscillating orbital phases~$\vec{q}$.

We still choose the average pieces of the transformation terms to be $\avg{\partialnit{Y}^{(1)}_j} = \avg{\partialnit{Y}^{(2)}_j}= \avg{\partialnit{X}^{(1)}_i}=\avg{\partialnit{Z}^{(0)}_k} = \avg{\partialnit{Z}^{(1)}_k} =0$ and so the transformed forcing functions are related to the original functions by 
\begin{subequations} \label{eq:Partial_NIT_EoM}
	\begin{gather}
		\partialnit{F}_j^{(1)} = \sum_{N} F^{(1)}_{j,N \vec{\kappa}_{\text{res}}} e^{i N q_\perp}, \quad \partialnit{f}_{i}^{(1)}=  \sum_{N} f^{(1)}_{i,N \vec{\kappa}_{\text{res}}} e^{i N q_\perp}, \quad \Upsilon_{k}^{(0)} = \left<s_{k} ^{(0)}\right>, \tag{\theequation a-c}
	\end{gather}
	\begin{equation} \label{eq:Paartial_NIT_Second_Order_Dependence_On_Phaases}
		\partialnit{F}_j^{(2)} = \sum_{N} F^{(2)}_{j,N \vec{\kappa}_{\text{res}}} e^{i N q_\perp} + \left<\frac{\partial \partialnit{Y}_j^{(1)}}{\partial \nit{q}_i} \osc{f}_i^{(1)} \right> + \left<\frac{\partial \partialnit{Y}_j^{(1)}}{\partial \nit{P}_k} \osc{F}_k^{(1)} \right>, \tag{\theequation d}
	\end{equation}
	\begin{equation}
		\partialnit{s}_k^{(1)} = - \left<\frac{\partial \osc{f}_k^{(0)}}{\partial \nit{P}_j} \partialnit{Y}_j^{(1)} \right> - \left<\frac{\partial \osc{f}_k^{(0)}}{\partial \nit{q}_i} \partialnit{X}_i^{(1)} \right>. \tag{\theequation e}
	\end{equation}
\end{subequations}
In deriving these equations of motion, we have constrained the oscillating pieces of the partial NIT transformation functions to be
\begin{equation}\label{eq:PartialNIT_Y}
	\partialnit{Y}_j^{(1)} \coloneqq \sum_{\vec{\kappa} \in R} \frac{i}{\vec{\kappa} \cdot \vec{\Upsilon}} F_{j,\vec{\kappa}}^{(1)} e^{i \vec{\kappa} \cdot \vec{q}},
\end{equation}

\begin{equation}\label{eq:PartialNIT_X}
	\partialnit{X}_i^{(1)} \coloneqq \sum_{\vec{\kappa} \in R}\left( \frac{i}{\vec{\kappa} \cdot \vec{\Upsilon}} f_{i,\vec{\kappa}}^{(1)} + \frac{1}{(\vec{\kappa} \cdot \vec{\Upsilon})^2} \frac{\partial \Upsilon_i}{\partial P_j}F_{j,\vec{\kappa}}^{(1)} \right) e^{i \vec{\kappa} \cdot \vec{q}},
\end{equation}
where $R$ is the set $\{ \vec{\kappa} \in \mathbb{Z}^2 | \vec{\kappa} \neq N \vec{\kappa}_{\text{res}}, \forall N \in \mathbb{Z} \}$ of all non-resonant 2-tuples and $\kappa_\text{res} = (\kappa_r,\kappa_\theta)$ is such that $\kappa_\text{res} \cdot \vec{\Upsilon}^{(0)} = 0$. 

The second order transformations terms for the orbital elements is given by:
	\begin{align} \label{eq:PartialNIT_Y2}
	\begin{split}
		\partialnit{Y}^{(2)}_{j} = & \sum_{\vec{\kappa} \in R} \frac{i e^{i \vec{\kappa} \cdot \vec{q}} }{\vec{\kappa} \cdot \vec{\Upsilon}^{(0)}} \Biggl( 
		F_{j,\vec{\kappa}}^{(2)} + \PD{\avg{Y_j^{(1)}}}{\nit{P}_k} F^{(1)}_{k,\vec{\kappa}} - i \PD{\avg{F_j^{(1)}}}{\nit{P}_k} \frac{F^{(1)}_{k,\vec{\kappa}}}{\vec{\kappa} \cdot \vec{\Upsilon}^{(0)}}
		\\ & +   \sum_{\vec{\kappa}' \in R} \biggl(   i \frac{F^{(1)}_{k,\vec{\kappa} - \vec{\kappa} '}}{\vec{\kappa}' \cdot \vec{\Upsilon}^{(0)}} \left( \PD{F^{(1)}_{j,\vec{\kappa}'}}{\nit{P}_k} - \frac{F^{(1)}_{j,\vec{\kappa} '}}{\vec{\kappa}' \cdot \vec{\Upsilon}^{(0)}} \PD{(\vec{\kappa}' \cdot \vec{\Upsilon}^{(0)}) } {\nit{P}_k}  \right) - \frac{\vec{\kappa}' \cdot \vec{f}^{(1)}_{\vec{\kappa} - \vec{\kappa}'}}{\vec{\kappa}' \cdot \vec{\Upsilon}^{(0)}} F^{(1)}_{j,\vec{\kappa}'}
		\biggr) \Biggr).
	\end{split}
\end{align}
Substituting these expressions for the transformation terms into the expressions for the sub-leading terms in the averaged equations of motion allows us to write them in the simplified form 

\begin{equation}
	\partialnit{F}_j^{(2)} = \sum_{N} F^{(2)}_{j,N \vec{\kappa}_{\text{res}}} e^{i N q_\perp} + \mathcal{N}_{\text{Partial}}( F_j^{(1)}),
\end{equation}
where $\mathcal{N}_\text{Partial}$ is similar to the $\mathcal{N}_\text{Full}$ operator but one only sums over the contributions form the non-resonant modes, i.e.,
 \begin{align} \label{eq:partialNIT_N_Operator}
	\begin{split}
		\mathcal{N}_\text{Partial}  & \coloneqq \sum_{\vec{\kappa} \in R} 	\mathcal{N}_{\vec{\kappa}}(A) .
	\end{split}
\end{align}
Since the rates of change of the extrinsic quantities ($s_t^{(0)}$ and $s_\phi^{(0)}$)  decouple into a purely radial piece and purely polar piece, there are no cross terms  that would be effected by evolving through the resonance. 
As a result, the leading-order transformation terms and the terms in averaged equations of motion  remain unchanged from the Full NIT case, i.e., $\partialnit{Z}_k^{(0)} = Z_k^{(0)}$  and  $\partialnit{s}_k^{(1)} = \nit{s}_k^{(1)}$. 

Furthermore, to be able to directly compare between OG and NIT inspirals, we will need to match their initial conditions to sufficient accuracy. 
For the same reasons as with the Full NIT, we use the following prescription for the initial conditions:
\begin{subequations}  \label{eq:partialNIT_ICs}
	\begin{align}
		\begin{split}
			\partialnit{P}_j(0) &= P_j(0) +\mr  \partialnit{Y}^{(1)}_j(\vec{P}(0), \vec{q}(0)) +  \mr  \partialnit{Y}^{(2)}_j(\vec{P}(0), \vec{q}(0)) +  \HOT{3},
		\end{split}\\
		\begin{split}
			\partialnit{q}_i(0) &= q_i(0) +\mr  \partialnit{X}^{(1)}_i(\vec{P}(0), \vec{q}(0)) +   \HOT{2},
		\end{split}\\
		\begin{split}
			\partialnit{S}_k(0) &= S_k(0) -Z_k^{(0)}(\vec{\nit{P}}(0),\vec{\nit{q}} (0)) +   \mathcal{O}(\mr).
		\end{split}
	\end{align}
\end{subequations}

\subsection{Switching from Full NIT to Partial NIT (Switch NIT)} \label{section:SwitchNIT}
The Partial NIT allows us to evolve inspirals at the same accuracy as the OG evolution, including through the resonance, while greatly speeding up the calculation in the vicinity of the resonance where $q_\perp$ varies slowly. Far away, from the resonance $q_\perp$ resumes varying rapidly on the orbital time scale producing $\mathcal{O}(\mr^{-1})$ oscillations, significantly slowing the calculation of a full inspiral.

One way this can be sped-up further is by using the Full NIT when far away from a resonance and then switching to the Partial NIT equations of motion when in the vicinity of the resonance. We call this approach the ``Switch NIT".
While this observation may seem obvious, deciding exactly when to transition between the two equations of motion is highly non-trivial, and is explored in full in \ref{section:transition_condition}. 

Summarizing the results of that analysis, we conclude that one should transition from the Full NIT to the Partial NIT when
\begin{equation}\label{eq:res_condition}
		\| \Upsilon_\perp \| \leq C \|  \mr^\beta \PD{\Upsilon_\perp }{P_j } \nit{F}^{(1)}_j T \| = C \| \mr^\beta \Upsilon_\perp' T \|,
\end{equation}
where  $C$ is a dimensionless constant and we have introduced the prime for any function $A$ as
\begin{align}
	A' \equiv  \frac{\partial A}{\partial P_j} \avg{F_j^{(1)}} + \mathcal{O}(\mr)\,.
\end{align}
In \ref{section:transition_condition}, we derive that  $\beta = 2/7$ and the optimal switching timescale $T$ is given by
\begin{equation}\label{eq:res_timescale}
	T = \left[ \frac{ \delta^2} {\left(\Upsilon'_\perp \right)^5} \frac{\left(\Upsilon_i^{(0)}\right)^2 \tilde{F}^{(1)}_j}{ \tilde{F}'^{(2)}_j  + \tilde{F}^{(1)}_j \nit{f}^{(1)}_i } \right]^{1/7}.
\end{equation}
Here $\delta$ is the ratio of the resonant harmonic modes to the orbit average of the forcing terms. 
We calculate this using the oscillatory $L^2$ norm divided by the orbit average which is given by:
\begin{equation}
	\delta = \sum_j \frac{1}{ F^{(1)}_{j, \vec{0}}} \left( \sqrt{ \int_{0}^{2\pi} (\osc{F}^{(1)}_j )^2 d q_\perp } \right) = \sum_j \frac{1}{ F^{(1)}_{j, \vec{0}}} \left( \sqrt{ \sum_{N = - \infty}^{\infty} F^{(1)}_{j, N \vec{\kappa}_\text{res}} F^{(1)}_{j, - N \vec{\kappa}_\text{res} }}   \right) .
\end{equation}
With this definition for the transition region, we find empirically that the choice of $C = 1$ works very well for the region of parameter space explored in this work. However, other choices for the definition of $\delta$ can be made which require fitting $C$ to recover the same results.

When transitioning from the Full NIT to the Partial NIT we must apply a near-identity transformation to the orbital elements and orbital phases in order to capture the change in the averaging transformation used in each region. When entering the near-resonance region, we employ the inverse transformation Eqs.~\eqref{eq:invNITdef} to go from Full NIT variables to OG variables, and then use the Partial NIT Eqs.~\eqref{eq:partial_transformation} to go from OG variables to Partial NIT variables. Combining these two steps gives the following transformation: 
\begin{subequations} \label{eq:Switch_Jump1}
	\begin{align}
		\begin{split}
			\nit{P}_j \rightarrow \nit{P}_j -\mr \left( Y_j^{(1)} - \partialnit{Y}_j^{(1)}  \right) - \mr^2 \left(Y_j^{(2)} - \partialnit{Y}_j^{(2)}  - \PD{Y_j^{(1)} }{\nit{P}_k } Y^{(1)}_k - \PD{Y_j^{(1)} }{\nit{q}_i} X^{(1)}_i  \right)  +  \HOT{3}
		\end{split}\\
		\begin{split}
			\nit{q}_i \rightarrow \nit{q}_i - \mr \left( X_i^{(1)} - \partialnit{X}_i^{(1)}  \right)  + \HOT{2}.
		\end{split}
	\end{align}
\end{subequations}
Similarly, when exiting the near-resonance region, we  apply the inverse partial transformation Eqs.~\eqref{eq:res_inverse_trasformaiton} to go from Partial NIT variables to OG variables, and then use the Full NIT Eqs.~\eqref{eq:transformation} to go from OG variables to Full NIT variables:
\begin{subequations} \label{eq:Switch_Jump2}
	\begin{align}
		\begin{split}
			\nit{P}_j \rightarrow \nit{P}_j + \mr \left( Y_j^{(1)} - \partialnit{Y}_j^{(1)}  \right) +  \mr^2 \left(Y_j^{(2)} - \partialnit{Y}_j^{(2)}  + \PD{\partialnit{Y}_j^{(1)} }{\partialnit{P}_k } \partialnit{Y}^{(1)}_k + \PD{\partialnit{Y}_j^{(1)} }{\partialnit{q}_i} \partialnit{X}^{(1)}_i  \right)  +\HOT{3}	\end{split}\\
		\begin{split}
			\nit{q}_i \rightarrow \nit{q}_i + \mr \left( X_i^{(1)} - \partialnit{X}_i^{(1)}  \right)  + \HOT{2}.
		\end{split}
	\end{align}
\end{subequations}
Note that since there is no change to either the averaging transformation or the averaged equations of motion of the extrinsic quantities $t$ and $\phi$ when entering or leaving the resonance region, we do not need to apply a transformation to these variables.

In \ref{section:transition_condition}, we derive that the error induced in the orbital elements by using the Switch NIT with this switching condition  that should scale as $\sim \mr^{11/7}$. Thus when evolved for a timescale $\sim \epsilon^{-1} $, the resulting phase error scales as $ \sim \mr^{4/7}$. For a typical EMRI detectable by LISA, this error scaling in the phase should be acceptable for achieving the sub-radian accuracy goal required of LISA data science.

We will demonstrate in Sec.~\ref{section:Results} that using the prescription for the Switch NIT can significantly reduce the runtime for an EMRI trajectory while introducing no significant error in the waveform, and that we can repeat this procedure to account for multiple resonance crossings. 
Since the evolution through the resonance is sensitive to the orbital phases when crossing the resonance, the effect of the phase error resulting from each resonance crossing will amplify with the next one. However, since $\delta$ is expected to be small in practice, we expect the Switch NIT procedure to keep the phase error small enough we are able to maintain an accuracy that is sufficient for LISA data analysis.

\section{Implementation} \label{section:Implementation}
Combining the GSF inspired toy model along with our action angle formulation of the OG equations gives us all the information required to calculate the Full, Partial, and Switch NIT equations of motion. 
We first evaluate and interpolate the various terms in the averaged equations of motion across the parameter space.
While this offline process can be expensive, it only needs to be completed once and sets up all three NIT formulations. The main differences between these formulations are most evident in their online steps which we will outline separately.

\subsection{Offline Steps}
\begin{enumerate}
	
	\item  We begin by selecting a grid which covers the section of parameter space we are interested in. 
	For this work we fix the spin of the primary to be $a = 0.9 M$ and choose an equally spaced grid of $ \vec{P} = (p,e,x)$ values on which to evaluate the terms in the NIT equations of motion.
	We pick $p$ to range from $3.75$ to $7.25$ in steps of $0.05$, $e$ to range from $0.25$ to $0.41$ in steps of $0.01$, and $x$ range from $0.79$ to $0.81$ in steps of $0.005$ for a total of $1020$ grid points.
	
	\item For each of these grid points we evaluate the functions $F_{j}^{(1)}$,$F_{j}^{(2)}$, $f_{i}^{(1)}$, and $s_k^{(0)}$, along with their derivatives with respect to $P_j$, for 21 equally spaced values of both $q_r$ and $q_\theta$ ranging from $0$ to $2\pi$ each for a total of 441 evaluations for each function and partial derivative.
	
	\item We then perform a fast Fourier transform on the output data to obtain the Fourier coefficients of the forcing functions and their derivatives. With 21 equally spaced points in both phases, one obtains Fourier coefficients up to and including order $\pm 10$. 
	
	\item We repeat this across the parameter space and store the values of each of these Fourier coefficients.
	
	\item We then interpolate the Fourier coefficients of $F_{j}^{(1\backslash2)}$ and $f_{i}^{(1)}$ using Hermite polynomials. This allows us to construct not only the transformation terms $Y_{j}^{(1)}$ and $X_{i}^{(1)}$, but will also allow us to quickly evaluate the semi-oscillating terms in the Partial NIT $\sum_{N} F^{(1\backslash 2)}_{j ,N k_r, N \kappa_\theta} e^{i N q_\perp} $ and $\sum_{N} f^{(1)}_{i ,N \kappa_r, N \kappa_\theta} e^{i N q_\perp}$.
	
	\item Using the rest of the stored Fourier coefficients, one can then use the definition of the $\mathcal{N}_\text{Full}$ operator given by Eq.~\eqref{eq:N_Operator_Full} to construct values of $\nit{F}_{j}^{(2)}$, $\nit{f}_{i}^{(1)}$, and $\nit{s}_{k}^{(1)}$ and  and at each grid point, which are then interpolated using Hermite polynomials and stored to use with the Full NIT.

	\item Similarly, using the definition of the near-resonance $\mathcal{N_{\text{partial}}}$ operator given by Eq.~\eqref{eq:partialNIT_N_Operator}, we construct the remaining averaged parts of $\partialnit{F}_{j}^{(2)}$ at each grid point, which are then interpolated using Hermite polynomials and stored for use with the Partial NIT.
	
	\item The above step can be repeated if we wish to interpolate the Partial NIT functions for a different orbital resonance.
	
\end{enumerate}
We implemented the above algorithm in \textit{Mathematica} 13.1 and find that when parallelized across 40   Intel Xeon E5-2698V4s @ 2.20GHz, the calculation takes about 7 hours to calculate the Fourier coefficients, about an 30 minutes to interpolate the Fourier coefficients of $F_{j}^{(1\backslash2)}$ and $f_{i}^{(1)}$ and then about an hour to interpolate the Partial NIT terms for a single resonance.

\subsubsection{Full NIT Online Steps}

\begin{enumerate}
	
	\item We load in the interpolants for $\nit{F}_{j}^{(2)}$,$\nit{f}_{i}^{(1)}$, and $\nit{s}_{k}^{(1)}$, and define the Full NIT equations of motion given in Eqs.~\eqref{eq:NIT_EoM}.

	\item In order to make comparisons between OG and Full NIT inspirals we also load interpolants of the Fourier coefficients of  $F_{j}^{(1\backslash2)}$ and $f_{i}^{(1)}$ and their derivatives with respect to $P_j$ in order to construct $Y_{j}^{(1\backslash2)}$ and $X_{i}^{(1)}$.
	
	\item We then state the initial conditions of the OG inspiral $(p(0),e(0), x(0), q_r(0), q_\theta(0))$ and use  Eq.~\eqref{eq:NIT_ICs} to set initial conditions of the Full NIT inspiral.
	
	\item We then evolve the full NIT equations of motion using an ODE solver (in this work we always use Mathematica's \texttt{NDSolve} function). 
	
\end{enumerate}

As with the offline steps we implement the online steps in \textit{Mathematica} 13.1. Note that steps (ii) and (iii) are only necessary because we want to make direct comparisons between Full NIT and OG inspirals with the same initial conditions. Away from resonance, the difference between the NIT and OG variables will always be $\mathcal{O}(\mr)$, and so performing the NIT transformation or inverse transformation to greater than zeroth order in mass ratio will not be necessary when producing waveforms to 1PA accuracy.

\subsubsection{Partial NIT Online Steps}

\begin{enumerate}
	
	\item  For the partial NIT we import the interpolants for the orbit average pieces of $\partialnit{F}_{j}^{(1\backslash2)}$, $\partialnit{f}_{i}^{(1)}$ and $\nit{s}_k^{(1)}$.
	
	\item We also import interpolants of the Fourier coefficients of  $F_{j}^{(1\backslash2)}$ and $f_{i}^{(1)}$ and their derivatives with respect to $p,e$ and $x$. This allows us to both define the transformation terms, $\partialnit{Y}_{j}^{(1 \backslash2)}$ and $\partialnit{X}_{i}^{(1)}$, and the oscillatory pieces of the Partial NIT equations of motion (Eqs.~\eqref{eq:Partial_NIT_EoM}).
	
	\item We then state the initial conditions of the OG inspiral $(p(0),e(0), x(0), q_r(0), q_\theta(0))$ and use Eq.~\eqref{eq:partialNIT_ICs} to set initial conditions of the Partial NIT inspiral.
	
	\item We then evolve the Partial NIT equations of motion using an ODE solver. 
		
\end{enumerate}

 As before, step (iii) is only necessary because we want to make direct comparisons between Partial NIT and switch NIT or OG inspirals with the same initial conditions.

\subsection{Switch NIT}

\begin{enumerate}
	
	\item For the Switch NIT, we first load in all of the interpolants that we have calculated. 
	
	\item We define the right-hand side (RHS) of the Switch NIT equations of motion as the $b$ times RHS of the Full NIT equations of motion $+ (1 - b)$ times the RHS of the Partial NIT equations of motion where $b$ is a binary parameter that is set to $1$ if outside the resonance region and set to $0$ if inside the resonance region.
	
	 \item We then state the initial conditions of the inspiral $(p(0),e(0), x(0), q_r(0), q_\theta(0))$ and check using Eq.~\eqref{eq:res_condition} to see if the inspiral is starting inside the near-resonance region or not and use either Eq.~\eqref{eq:NIT_ICs} or Eq.~\eqref{eq:partialNIT_ICs} respectively to set the initial conditions and our starting value of $b$.
	 
	 \item We again solve the equations of motion numerically using \texttt{NDSolve}, but make use of the method \texttt{WhenEvent} to switch the equations of motion at two events for each resonance we wish to model:
	 
	 \begin{enumerate}
	 	\item When $\| \Upsilon_\perp \| \leq \| \mr^\beta \Upsilon_\perp' T \|$, we have entered the resonance region and so we apply the transformations Eqs.~\eqref{eq:Switch_Jump1} and set $b= 0$.
	 	
	 	\item  When $\| \Upsilon_\perp \| \geq \| \mr^\beta \Upsilon_\perp' T \|$, we have exited the resonance region and so we apply the transformations Eqs.~\eqref{eq:Switch_Jump2} and set $b = 1$.
	 \end{enumerate}
	
\end{enumerate}

\subsection{Waveform Generation}

In order to generate waveforms, one must first parametrize the orbit in terms of the coordinate time variable $t$. 
This can be done by performing an additional transformation to the equations of motion and solving these new equations as was first prescribed in Ref.~\cite{Pound2021} and implemented in \cite{Lynch:2023gpu} and \cite{Drummond:2023wqc}. 
However, since we make use of the Partial NIT where one still has to include the orbital phases in the equations of motion, one would have to find an invertible transformation between the Mino-time action angles and the Boyer-Lindquist coordinate time action angles which is currently unknown (see \cite{Witzany:2022vck} for a perturbative construction in Schwarzschild).
For this reason, we opt to instead take our solution for $t(\lambda)$ and resample it in steps of $\delta \lambda = 0.01$ in order to produce an interpolant for $\lambda(t)$. 
With this, one can reparametrize the remaining solutions for the orbital elements as functions of $t$. 

We currently do not have access to interpolated Teukolsky amplitudes for generic Kerr inspirals to produce waveforms from our calculated the trajectories. Thus we use the same method as the numerical kludge EMRI model \cite{Babak2007} and use the quadrupole formula where one makes the approximation of equating flat space spherical polar coordinates with Boyer-Lindquist coordinates. While this process does not perfectly capture the waveform from a source that is deep in the strong field, it has still been shown to fare surprisingly well against Teukolsky waveforms \cite{Babak2007, McCart2021}. We sample our waveforms every $\delta t = 2 M$ which for the $10^6 M_\odot$ MBH primary corresponds to once every~$\sim10s$.  We use this same waveform generation scheme for all of our inspirals so that any difference between the resulting waveforms is purely a result of the difference in the trajectories.  

We calculate the waveform mismatch $\mathcal{M}$ between our two waveforms which varies from 0 (perfectly overlapping signals) to 1 (completely orthogonal signals). We make use of the \texttt{SimulationTools} Mathematica package to calculate our waveform mismatches~\cite{SimulationTools} and use a flat noise curve.

\section{Results} \label{section:Results}
We now present the results of our inspiral evolution schemes. We start by demonstrating the convergence with mass ratio of the Full NIT compared to OG both away from and in the presence of an orbital resonance. We then demonstrate the convergence of the Partial NIT compared to OG near the 2/3 resonance. We then look at the convergence of the Switch NIT compared with the Partial NIT and show that the errors are consistent with the error scalings we derived for our choice of transition region. Finally, we look at two examples of year long EMRI signals, one which evolves through just a 2/3 resonance and one that evolves through the 2/3 and 2/4 resonance which tests the effectiveness of our Switch NIT compared to using either the Partial NIT or OG.

\subsection{Full NIT convergence}
In order to test that we have implemented the NIT equations of motion correctly, we examine how the differences between the OG and NIT inspiral quantities vary with the mass ratio.
If implemented correctly, the difference in the phases and extrinsic quantities should scale linearly while the difference in the orbital elements (after inverting the transformation to leading order) should scale quadratically. Any deviation from this would indicate either a bug in our code, a large interpolation error for the terms in our NIT equations of motion, or error in the numerical solver.

As such, we start an inspiral at $\left( p(0),e(0),x(0) \right) = \left(9.5,0.2,0.8 \right)$ and evolve the inspiral until $p = 9$ using both the OG and Full NIT equations of motion and varying the mass ratio from $10^{-1}$ to $10^{-4}$. We examine these mass ratios not because we believe our models to be accurate in this entire range. Instead, we explore this range because it is sufficient for demonstrating the scaling of the residuals between the OG and NIT models while avoiding the computational cost of the OG model at extreme mass ratios.   The initial conditions were specifically chosen to avoid encountering any low order resonances during the inspiral.

\begin{figure}
	\begin{subfigure}[b]{0.49\textwidth}
		\includegraphics[width=1\textwidth]{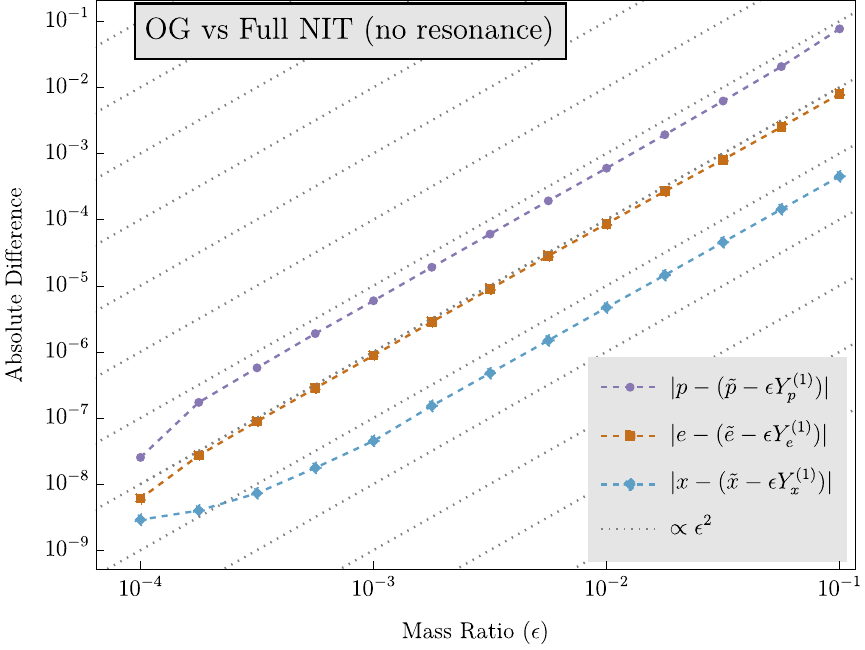}
		\caption{Orbital elements.}
		\label{fig:OffResonanceConvergenceOrbitalElements}
	\end{subfigure}
	\hfill
	\begin{subfigure}[b]{0.49\textwidth}
		\includegraphics[width=1\textwidth]{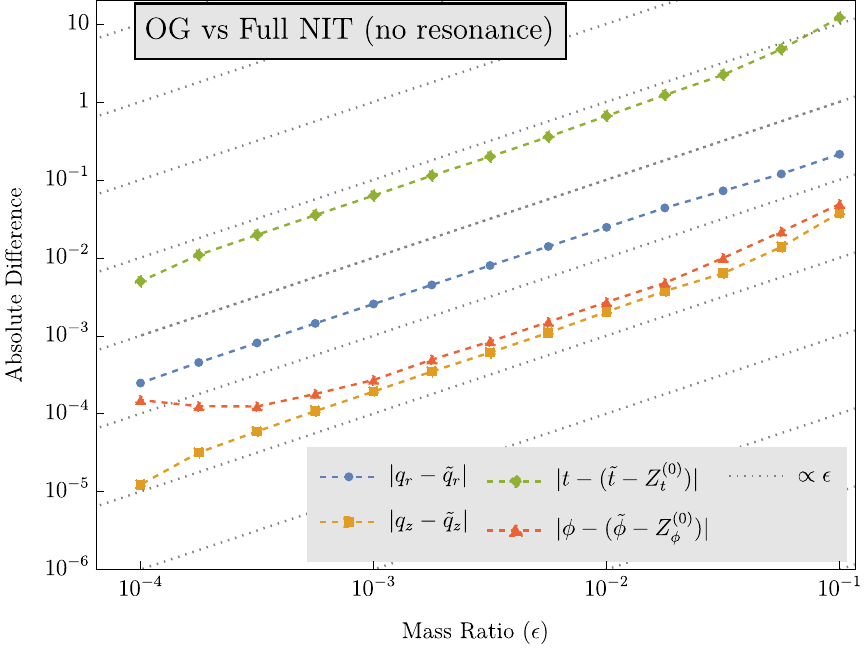}
		\caption{Orbital phases and extrinsic quantities.}
		\label{fig:OffResonanceConvergencePhases}
	\end{subfigure}
	\hfill
	\caption{The absolute difference in the quantities of an inspiral with $a = 0.9 M$ and initial conditions $(e_0,  x_0) = (0.2,0.8)$ and evolved from $p = 9.5$ to $p = 9$ for different values of the mass ratio when calculated using either the OG or NIT equations of motion. 
	As expected, the differences in the phases and extrinsic quantities scale linearly with the mass ratio while the orbital elements scale quadratically. \label{fig:OffResonanceConvergence}}
\end{figure}

As shown in Fig.~\ref{fig:OffResonanceConvergence}, the differences between the OG and NIT orbital elements generally\footnote{The smallest values of mass ratio tested here do not align with this trend as the error due to the NIT becomes subdominant to the errors in the interpolating functions used for the terms in the equations of motion and/ or the numerical error in the ODE solver.} scale quadratically while the phases and the extrinsic values generally scale linearly with the mass ratio. 
This demonstrates that in the absence of low order orbital resonances, the full NIT formulation is valid and the equation implemented correctly in the code for generic Kerr inspirals.

\begin{figure}
	\begin{subfigure}[b]{0.49\textwidth}
		\includegraphics[width=1\textwidth]{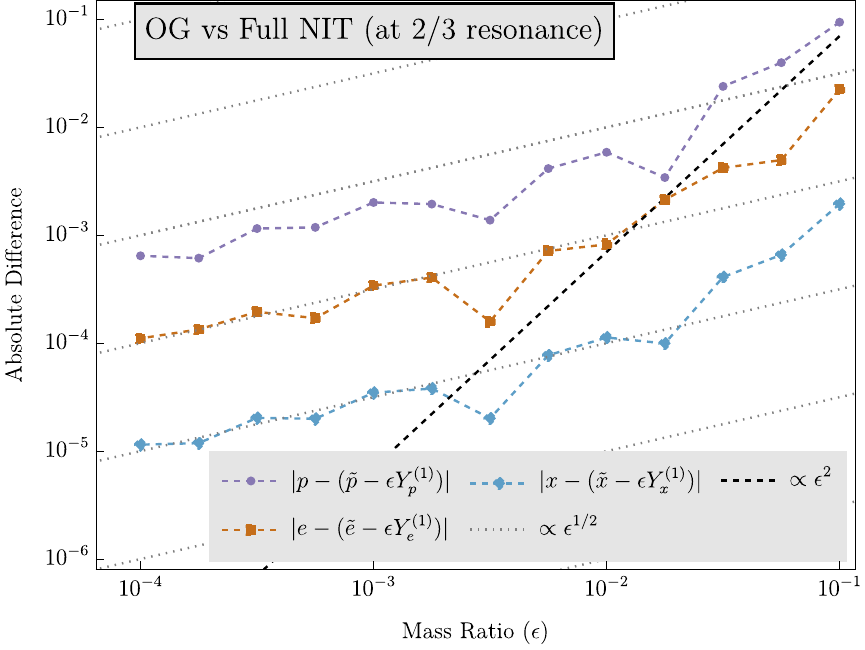}
		\caption{Orbital elements.}
		\label{fig:NITvsOGONResonanceConvergenceOrbitalElements}
	\end{subfigure}
	\hfill
	\begin{subfigure}[b]{0.49\textwidth}
		\includegraphics[width=1\textwidth]{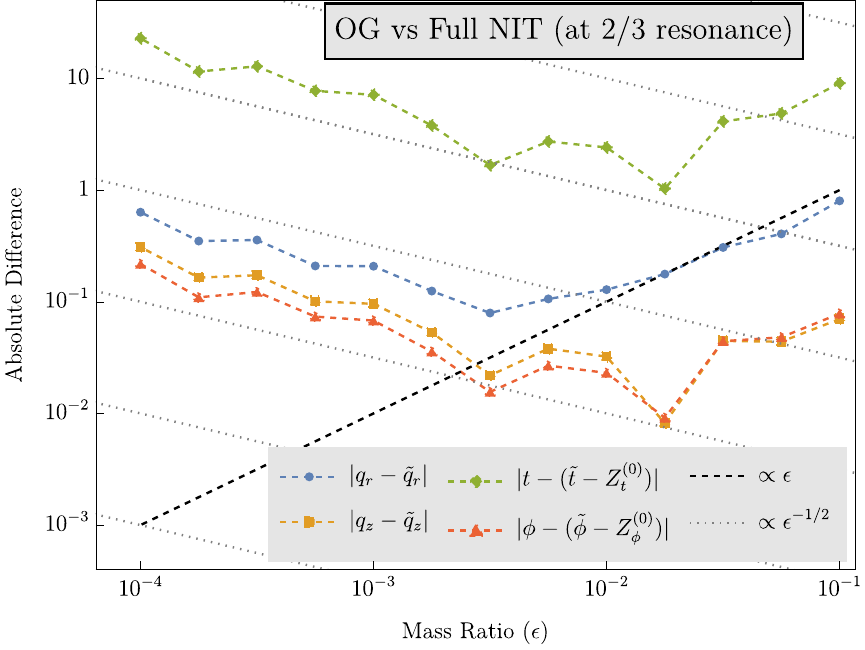}
		\caption{Orbital phases and extrinsic quantities.}
		\label{fig:NITvsOGONRResonanceConvergencePhases}
	\end{subfigure}
	\hfill
	\caption{The absolute difference in the quantities of an inspiral with $a = 0.9 M$ and initial conditions $(e_0,  x_0) = (0.2,0.8)$ and evolved from $p = 6.5$ to $p = 5.8$ for difference values of the mass ratio when calculated using either the OG or NIT equations of motion. The orbit now evolves through the 2/3 orbital resonance and so now the difference in the orbital elements scales as $\mr^{1/2}$ while the difference in the orbital phases scales as $\mr^{-1/2}$. \label{fig:NITvsOGONResonanceConvergence}}
\end{figure}

 When we repeat this same analysis in a part of the parameter space near a low-order resonance the above scaling is not observed, as expected due to presence of the resonance crossing. 
In particular, keeping the initial values of $e$ and $x$ the same, we now evolve $p$ from $6.5$ to $5.8$ so that the inspirals now pass through the $2/3$ orbital resonance.
As seen in Fig.~\ref{fig:NITvsOGONResonanceConvergence}, when the NIT inspiral crosses a resonance we encounter different scalings with the mass ratio. 
This results in an error in the orbital elements that scales as $\mr^{1/2}$ and an error in the orbital phases and extrinsic quantities that scales as $\mr^{-1/2}$, as predicted in Ref.~\cite{Flanagan:2010cd}. 
Interestingly, one does not see this scaling for all values of the mass ratio. For mass ratios larger than $10^{-2}$, the inherent $\mr$ or $\mr^2$ error of performing the NIT is the dominant source of error. This suggests that resonant effects need to be included in order to obtain accurate models for both IMRI and EMRI systems.

\subsection{Partial NIT convergence}

As before, we now test the implementation of the Partial NIT by investigating how the differences between the OG and NIT orbital elements should scale quadratically with mass ratio and the phases and extrinsic quantities should scale linearly with the mass ratio.
We start each inspiral at $\left( p(0),e(0),x(0) \right) = \left(6.5,0.2,0.8 \right)$ and evolve the system until $p = 5.8$ while varying the mass ratio from $10^{-1}$ to $10^{-4}$, which ensures that the inspirals always pass through the $2/3$ orbital resonance.

\begin{figure}
	\begin{subfigure}[b]{0.49\textwidth}
		\includegraphics[width=1\textwidth]{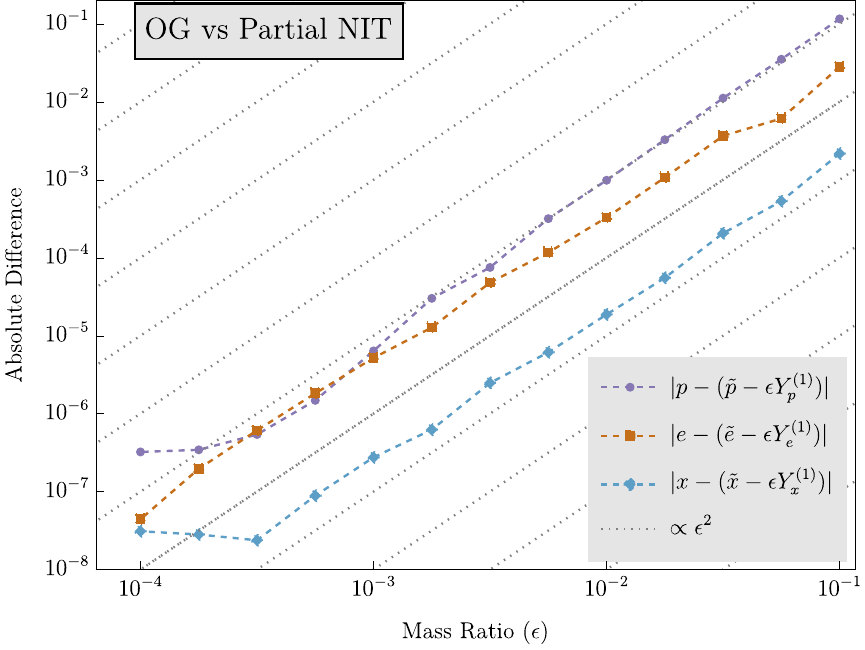}
		\caption{Orbital elements.}
		\label{fig:ResonanceConvergenceOrbitalElements}
	\end{subfigure}
	\hfill
	\begin{subfigure}[b]{0.49\textwidth}
		\includegraphics[width=1\textwidth]{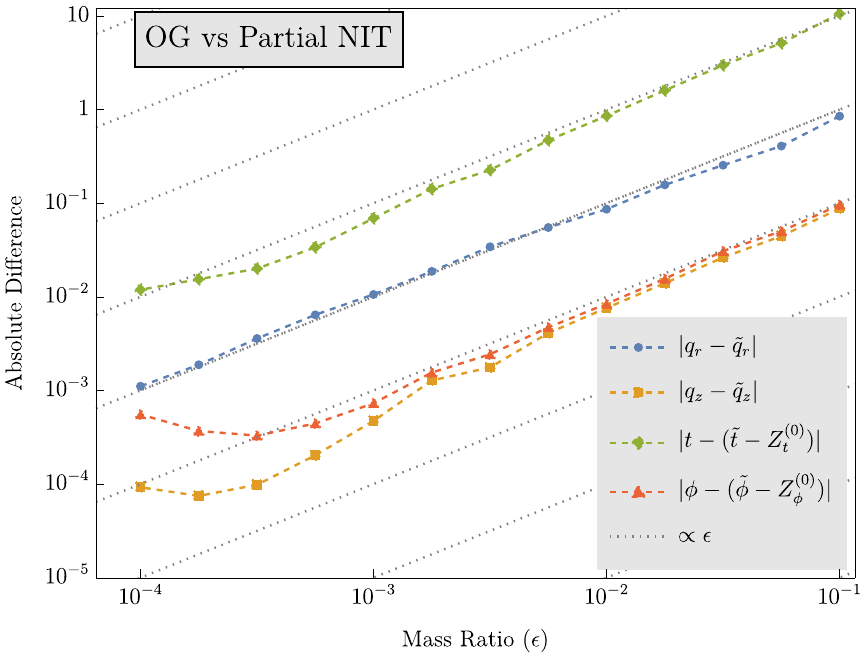}
		\caption{Orbital phases and extrinsic quantities.}
		\label{fig:ResonanceConvergencePhases}
	\end{subfigure}
	\hfill
	\caption{The absolute difference in the quantities of an inspiral with $a = 0.9 M$ and initial conditions $(e_0,  x_0) = (0.2,0.8)$ and evolved from $p = 6.5$ to $p = 5.8$ for difference values of the mass ratio when calculated using either the OG or Partial NIT equations of motion. As expected, the differences scale in the phases and extrinsic quantities scale linearly with the mass ratio while the orbital elements scale quadratically.\label{fig:ResonanceConvergence}}
\end{figure}
The differences between using the OG and Partial NIT equations of motion are displayed in Fig.~\ref{fig:ResonanceConvergence}.
In Fig.~\ref{fig:ResonanceConvergenceOrbitalElements}, we see that the difference in the orbital elements (after performing the first order inverse transformation) generally scales quadratically with the mass ratio, which is accordance with the error scaling seen in Eqs.~\eqref{eq:partial_transformation}.
The differences in the orbital phases and extrinsic quantities still generally scale linearly with the mass ratio.
This is due to these quantities being more sensitive than the orbital elements to the dependence on the Partial NIT equations of motion on the resonant phase $q_\perp$. Overall, this test demonstrates that the differences scale as expected and assures us that we have implemented the Partial NIT correctly. 

\subsection{Switch NIT convergence}
 We look to verify the convergence of the Switch NIT with our choice of transition condition.
  Note that we have opted not to compare against the OG solutions. While this reduces the computational cost of the comparison, the primary reason is that for larger mass ratios, the differences are dominated by the $\mr^2$ and $\mr$ scalings we saw in the previous subsections from applying any sort of averaging scheme. 
  This makes it difficult to discern the scaling of the differences which arise solely from the switching procedure. As such, we compare the inspiral solutions from the Switch NIT to those from the Partial NIT.
  
   We set our initial eccentricity and inclination to be $(e_0, x_0) = (0.2,0.8)$ and evolve inspirals from an initial orbital separation of $p = 7$ to $p= 4.5$ while varying the mass ratio from $\mr = 10^{-1} - 10^{-4}$ such that we should encounter the $2/3$ resonance about half way through the inspiral. 
   Note that this is wider than before as it is important to start and end the inspirals outside the near-resonance region. 
 
 We found the differences to be very oscillatory as the difference is dependent on the value of the resonant phase $q_\perp$ when crossing the resonance. 
 While we cannot directly control this value, we can change its initial value $q_{\perp,0}$ and have found that the resulting error behaves like a sinusoidal function of $q_{\perp,0}$. 
 Thus, for each value of mass ratio, we take $15$ equally spaced values of $q_{\perp,0}$ from $0$ to $2\pi$. and calculate $L^2$ norm of the differences for that value of mass ratio and along with the minimum and maximum value obtained, to give the reader an idea of the variance in the differences.

\begin{figure} \
	\begin{subfigure}[b]{0.49\textwidth}
		\includegraphics[width=1\textwidth]{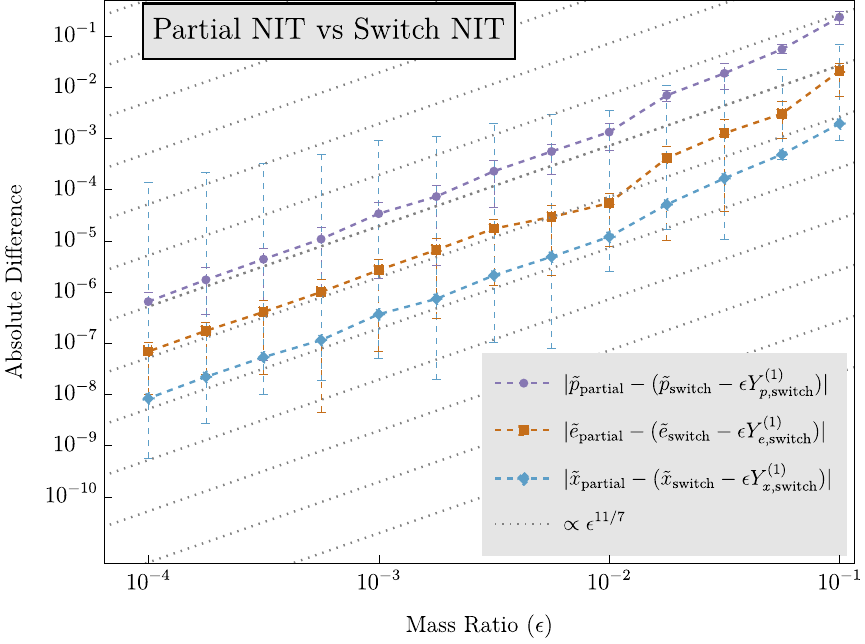}
		\caption{Orbital elements.}
		\label{fig:TransitionVsResNITConvergenceOrbitalElements}
	\end{subfigure}
	\hfill
	\begin{subfigure}[b]{0.49\textwidth}
		\includegraphics[width=1\textwidth]{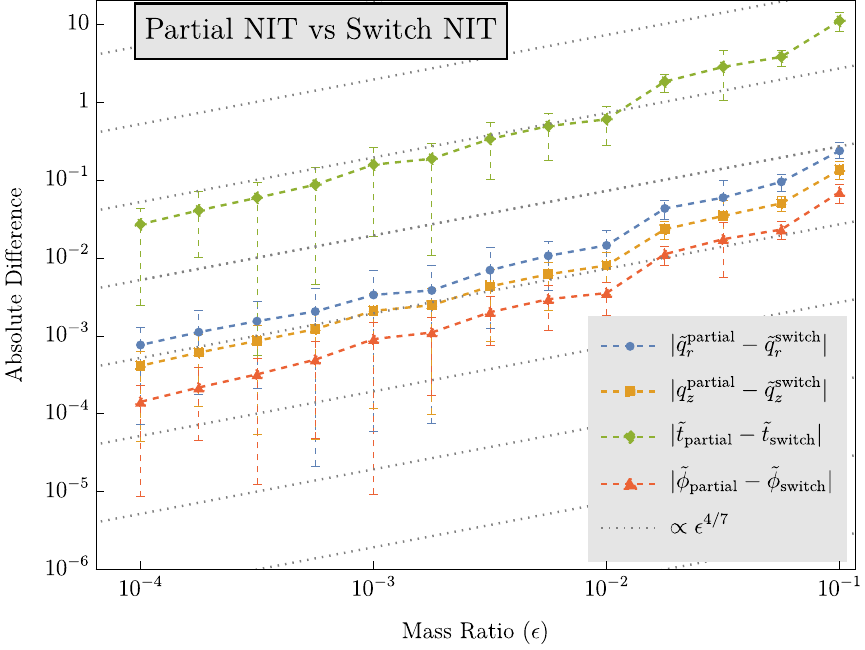}
		\caption{Orbital phases and extrinsic quantities.}
		\label{fig:TransitionVsResNITConvergencePhases}
	\end{subfigure}
	\hfill
	\caption{The absolute difference in the quantities of an inspiral with $a = 0.9 M$ and initial conditions $(e_0,  x_0) = (0.2,0.8)$ and evolved from $p = 7$ to $p = 4.5$ for difference values of the mass ratio when calculated using either the switch NIT or the partial NIT. At each value of $\mr$ we equally sample the initial value of $q_\perp$ 15 times from 0 to $2\pi$ and display the $L_2$ of the data as the central dot while the error bars display the minimum and maximum value obtained for the value obtained. \label{fig:TransitionVsResNITConvergence}}
\end{figure}

We display the results in Fig.~\ref{fig:TransitionVsResNITConvergence}, with the central point representing the $L^2$ norm and the error bars showing the minimum and maximum error obtained for each mass ratio. These results are consistent with the theoretical scalings of $\mr^{11/4}$ for the orbital elements and $\mr^{4/7}$ for the phases and extrinsic quantities. 

\subsection{Runtime}

\begin{figure}
	\includegraphics[width=0.9\textwidth]{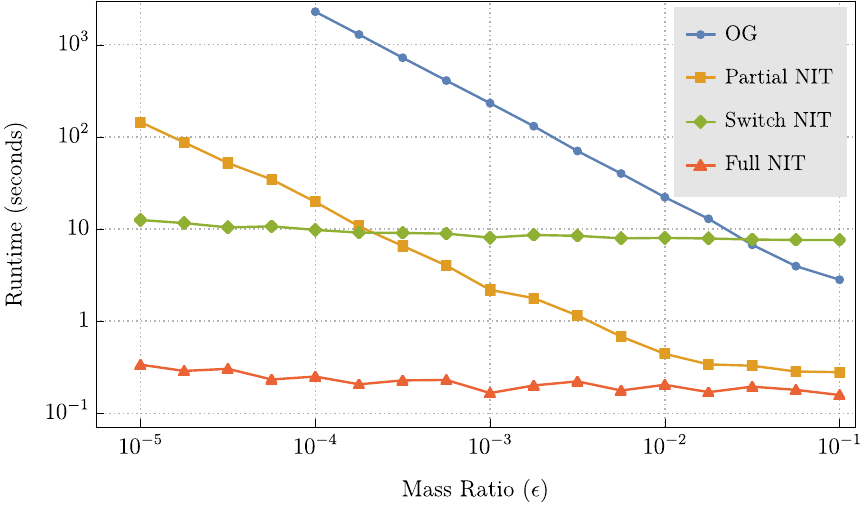}
	\caption{Runtime as recorded on an Apple M1 Max chip as a function of mass ratio for the OG, Partial NIT, Switch NIT and Full NIT while evolving from $p = 7$ to $p = 4.5$ and passing through the 2/3 resonance with $(e_0, x_0) = (0.2, 0.8)$. }
	\label{fig:Timings}
\end{figure}

Having demonstrated that our various NIT schemes obtain the accuracy we would expect, we now investigate the effect that each of them have on the time required to calculate a single inspiral for different values of the mass ratio. 
We set the initial conditions to be  $(p_0,e_0, x_0) = (7,0.2, 0.8)$ and evolve the inspiral until $p = 4.5$, passing through the 2/3 resonance roughly halfway through the inspiral. 
We repeat this for several values of the mass ratio in the range $10^{-5} \le \mr \le 10^{-1}$ (with the exception of the the OG inspirals as they take too long to finish beyond $\mr \leq 10^{-4}$). 
In each case we use \textit{Mathematica}'s \texttt{NDSolve} with \texttt{PrecisionGoal} (relative accuracy) set to $\infty$ and \texttt{AccuracyGoal} (absolute accuracy) set to 7 running on an Apple M1 Max @ 3.22 GHz.

The results are displayed in Fig.~\ref{fig:Timings}. 
First we note that the OG is by far the slowest. 
Its timing is inversely proportional to the mass ratio as the number of cycles the solver has to resolve scales inversely with the mass ratio over a fixed frequency window.
We also note that the fastest timing comes from the Full NIT, whose timing is independent of the mass ratio as it does not resolve any orbital cycles. However, as we've seen this scheme is not accurate when crossing an orbital resonance.

As such, we must rely on the Partial NIT which provides a consistent order of magnitude speed up over the OG calculations. 
However, it suffers from the same scaling with mass ratio as OG. 
Even though resonant oscillations are be easier to resolve, the number of oscillations still increases as the mass ratio gets smaller.

Finally, we see that the timing of the Switch NIT still increases mildly as mass ratio decreases. Nevertheless, it has a much more favourable scaling. Within the resonance region, the resonant phase evolves on a semi-fast timescale as opposed to a fast timescale, resulting in fewer integration steps needed to resolve the oscillations. However, this contribution to the computation time is much smaller than the constant $\mathcal{O}(10s)$ of overhead caused by the event locator used to find the location of the resonance region during the evolution.
This means that for larger mass ratios the Switch NIT is significantly slower than the Partial NIT and only becomes faster for mass ratios $\lesssim 10^{-3.75}$ in our implementation. 
This current iteration of the Switch NIT provides at least two orders of magnitude of speed up over using the OG equations for EMRIs with $\epsilon<10^{-4}$. However, this is still too slow for LISA data analysis, but it is the most viable procedure out of the ones explored in this work, especially if the computational overhead of the switching procedure can be reduced. 

\subsection{Evolving through a single low-order resonance}\label{section:single_res}

We now examine the case of a canonical EMRI consisting of a $10^6 M_\odot$ primary and a $10 M_\odot$ secondary for a mass ratio of $\mr = 10^{-5}$. 
The inspiral has initial conditions $(p_0,e_0,x_0) = (7.045,0.45,0.8)$ and evolved until the semilatus rectum reaches the value $p = 4.74$. 
These values were chosen so that the inspiral would last a little over one year, and so that the inspiral passes through the low-order $|\kappa_\theta|/|\kappa_r  |= 2/3$ resonance.
It is worth noting that the inspiral also crosses through the $6/10$ and $4/7$ resonances.
Technically, there are infinitely many resonance crossings since natural numbers are dense in the reals.
However, we only list resonances for which $\max \left( \kappa_r, \kappa_\theta \right) \leq 10$ since we truncate our Fourier expansions after the 10th coefficient and higher order resonances are exponentially suppressed, {as discussed in Sec.~\ref{section:Resonances}.
Thus, in our numerical implementation, the terms in the NIT and averaged equations of motion are not directly effected by any resonance with a value larger than 10. 
Moreover, the lowest order resonance is the $2/3$ resonance and so we expect that to have the largest effect on the inspiral.
As such, we only use the Partial NIT to account for the $2/3$ resonance and neglect all others in order to understand the effect this will have on the accuracy of our inspiral calculations.  

We first compute the year long inspiral using the OG equations using \texttt{NDSolve} with the \texttt{AccuracyGoal} and \texttt{PrecisionGoal} settings set to $13.5$. This took just over two days to compute on a single core of an Intel Xeon E5-2698V4 @ 2.20GHz.
Using this as our point of comparison, we evolved inspirals with equivalent initial conditions and accuracy and precision goals utilizing the Partial NIT and the Switch NIT. 
We also evolved an adiabatic inspiral in order to subtract this contribution (i.e., $p_{\text{Ad}}, e_{\text{Ad}},$ and $x_{\text{Ad}}$) away from the post-adiabatic inspirals to  highlight the effect of the resonance crossing on $\Delta p = p -p_{\text{Ad}}$, $\Delta e = e - e_{\text{Ad}}$, and $\Delta x = x-x_{\text{Ad}}$ \cite{Flanagan:2010cd}. 
Our results are presented in Fig.~\ref{fig:SingleRes}.
While the inspiral also crosses through other resonances, Fig.~\ref{fig:SingleRes} makes it clear that the $2/3$ resonance has by far the largest effect on the orbital elements, with the effects of the other resonance crossings being far too small to resolve.
Furthermore, this figure demonstrates how the Partial NIT and the Switch NIT capture the ``resonance jump" experienced by the OG inspiral while including far fewer orbital oscillations.

\begin{figure}
	\begin{subfigure}[b]{0.95\textwidth}
		\includegraphics[width=1\textwidth]{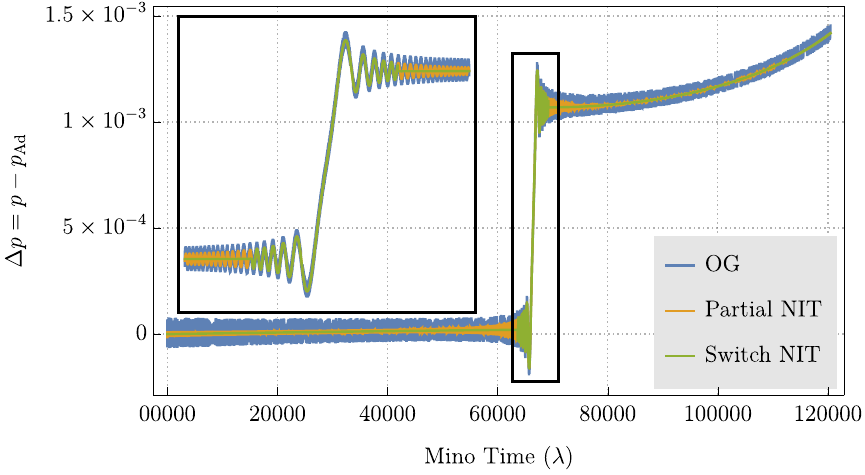}
		\caption{The difference in $p$.}
		\label{fig:SingleResdp}
	\end{subfigure}
	\hfill
	\begin{subfigure}[b]{0.495\textwidth}
		\includegraphics[width=1\textwidth]{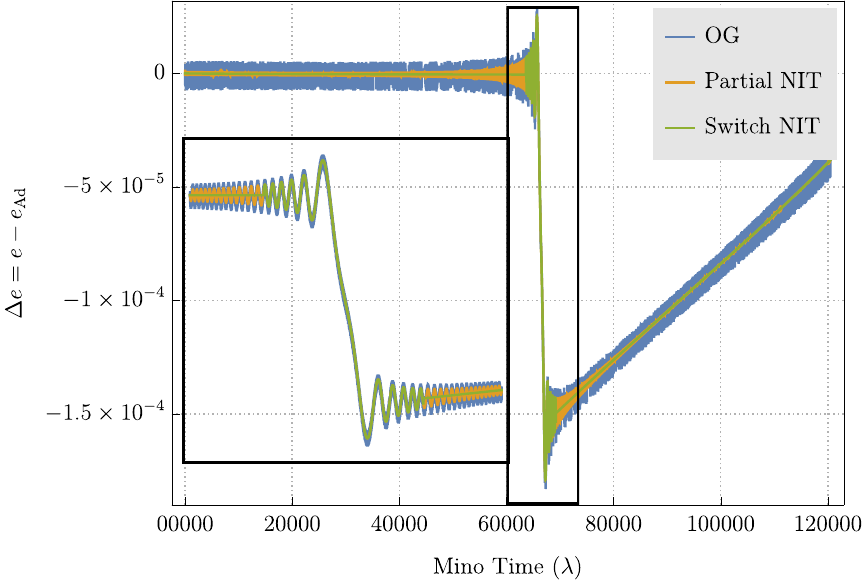}
		\caption{The difference in $e$.}
		\label{fig:SingleResde}
	\end{subfigure}
	\hfill
	\centering
	\begin{subfigure}[b]{0.495\textwidth}
		\centering
		\includegraphics[width=1\textwidth]{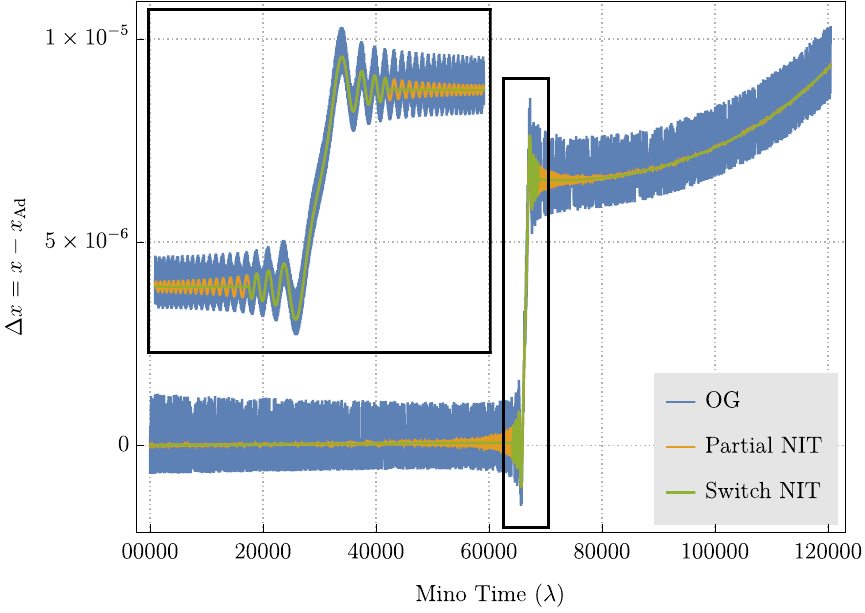}
		\caption{The difference in $x$.}
		\label{fig:SingleResdx}
	\end{subfigure}
	\hfill
	\caption{The difference between the evolution of orbital elements $\Delta \vec{P} = \vec{P} -\vec{P}_{\text{Ad}}$, where $\vec{P} = (p,e,x)$, as a function of Mino time $\lambda$ for a year long inspiral with $a = 0.9 M$, $\mr = 10^{-5}$ and initial conditions $(p_0,e_0,x_0) = (7.045,0.45,0.8)$. One can see that all three methods for evolving the inspiral accurately capture the effect of the $2/3$ orbital resonance, but the switch NIT does so without resolving as many oscillations.\label{fig:SingleRes}}
\end{figure}

\begin{figure}
	\begin{subfigure}[b]{0.49\textwidth}
		\includegraphics[width=1\textwidth]{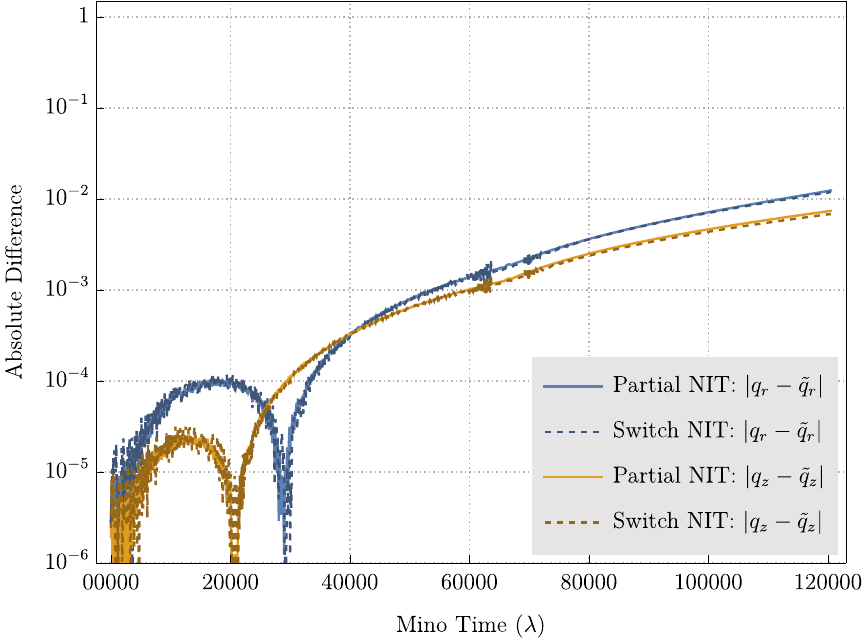}
		\caption{Orbital phases.}
		\label{fig:SingleResPhases}
	\end{subfigure}
	\hfill
	\begin{subfigure}[b]{0.49\textwidth}
		\includegraphics[width=1\textwidth]{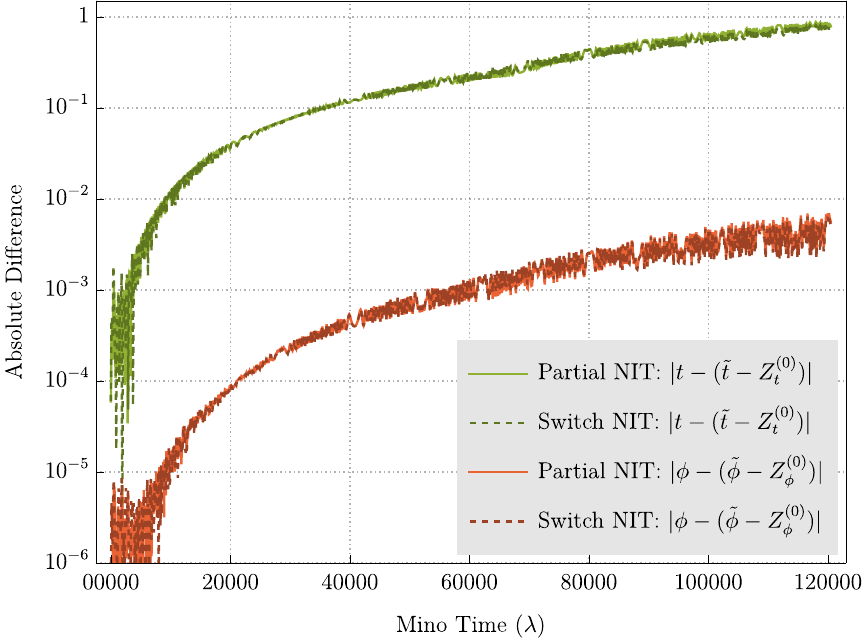}
		\caption{Extrinsic quantities.}
		\label{fig:SingleResExtrinsic}
	\end{subfigure}
	\hfill
	\caption{The absolute difference in the phases and extrinsic quantities between the OG and NIT equations of motion for a year long inspiral with $a = 0.9 M$, $\mr = 10^{-5}$ and initial conditions $(p_0,e_0,x_0) = (7.045,0.45,0.8)$. We observe that there is no difference in accuracy between using the partial NIT and the switch NIT.\label{fig:SingleResPhasesandExtrinsic}}
\end{figure}

We also examine the effect that including a resonance transition has on the accuracy of the orbital phases and extrinsic quantities which is displayed in Fig.~\ref{fig:SingleResPhasesandExtrinsic}.
We see a natural growth in the phase error over time. This may be due to accumulating numerical error from the numerical integrator, but may also be due to neglecting the effects of the other resonance crossings besides the $2/3$ resonance.
Importantly, we see no significant difference in accuracy when using either the Partial NIT or the Switch NIT.
Moreover, the end of a year-long inspiral the difference in the phases is $<2 \times 10^{-2}$ and the difference in $t/M$ is less than $1$, which should be accurate enough produce 1PA waveforms fit for LISA data science.

\begin{table}
	\begin{center}
		\begin{tabular}{c | c| c} 
			\hline
			Inspiral & Runtime & Mismatch  \\ [0.5ex] 
			\hline
			Full NIT & 0.544s & $0.343$ \\ 
			Partial NIT & 257s & $6.78 \times 10^{-5}$\\
			Switch NIT & 21.1s & $2.99 \times 10^{-4}$\\
			\hline 
		\end{tabular}
	\end{center}
	\caption{A table of the time taken to compute a year long inspiral with mass ratio $10^{-5}$ and initial conditions $(p_0,e_0,x_0) = (7.045,0.45,0.8)$ using different equations of motion as calculated numerically using NDSolve with AccuracyGoal of 7 as implemented in Mathematica 13 on an Apple M1 Max chip.
		The mismatches between the semi-relativistic quadrupole waveforms generated from these inspirals and the OG inspiral are also listed. \label{table:single_res_runtime_accuracy}}
\end{table}

This is confirmed by Table~\ref{table:single_res_runtime_accuracy}, which displays the time required to calculate inspirals using either the Full NIT, the Partial NIT or the Switch NIT and the associated mismatches between the semi-relativistic quadrupole waveforms generated from these inspirals when compared to the waveform produced from the OG inspiral.
We find the Full NIT to be the most computationally efficient but the resulting inspirals would not be sufficient for accurate parameter estimation for the LISA mission.
The Partial NIT produces inspirals and waveforms that are accurate enough for LISA data science even when neglecting all other resonance crossings. 
Unfortunately, with a single inspiral taking 257s (or 4 minutes 17s) to compute, the Partial NIT is substantially faster than the OG inspiral but still much too slow for practical waveform generation for data analysis.
Finally, we note the Switch NIT combines the best of both approaches, producing inspirals and waveforms that are almost as accurate as the Partial NIT while only taking $21.1s$ to compute an inspiral.
This is still slower than the sub-second computation time that one would need for data analysis, but this can likely be reduced further by optimising the resonance condition and using more efficient numerical methods.

In conclusion, this test case has confirmed two important insights.
First, one does not need to account for every orbital resonance to produce waveforms that are sufficiently accurate for LISA science. 
Modelling the lowest order ones will suffice.
Second, this demonstrates that the Switch NIT is the best strategy so far for accurately capturing resonant behaviour while reducing the computation time for calculating 1PA inspiral trajectories.

\subsection{Evolving through multiple low-order resonances} \label{section:double_res}
Armed with these two insights, we now look to a case where there is more than one low-order resonance crossing.
We wish to see if one can produce sufficiently accurate waveforms if one only employs a resonance transition for the $2/3$ resonance or if one needs to account for both the $2/3$ and $2/4$ resonances. 
We pick a canonical EMRI mass ratio of $\mr = 10^{-5}$, and chose initial conditions $(p_0,e_0,x_0) = (6.8,0.45,0.8)$ and evolve until $p = 3.75$ such that the resulting inspiral lasts for just over 1 year.
This time, however, the inspiral passes through the following of orbital resonances: $2/3, 6/10,4/7, 2/4,$ and $4/9$.
With the results of our last test in mind, we neglect all of the resonance crossings bar the $2/3$ and $2/4$ resonances.
Using this inspiral, we wish to investigate if we can accurately transition through more than one resonance, and how much accuracy is lost if one accounts for the $2/3$ resonance but neglects the $2/4$ resonance.

We first compute the year long inspiral using the quasi-Keplerian OG equations using NDSolve with accuracy and precision goals set to $13.5$ which took just over three days to compute on a single core of an Intel Xeon E5-2698V4 @ 2.20GHz.
Using this as our point of comparison, we evolved inspirals with equivalent initial conditions and accuracy and precision goals utilizing the Switch NIT with either a single switch for the $2/3$ resonance, or a switch for both the $2/3$ and $1/2$ resonances.

\begin{figure}
	\begin{subfigure}[b]{0.95\textwidth}
		\includegraphics[width=1\textwidth]{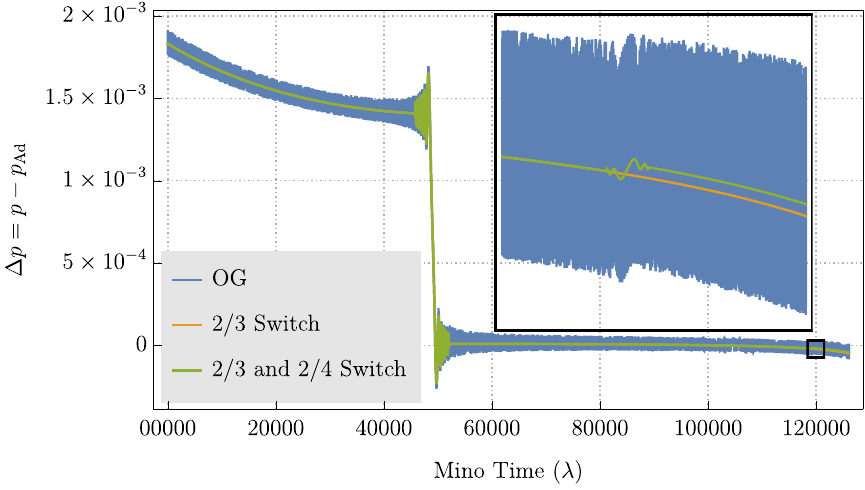}
		\caption{The difference in $p$.}
		\label{fig:DoubleResdp}
	\end{subfigure}
	\hfill
	\begin{subfigure}[b]{0.495\textwidth}
		\includegraphics[width=1\textwidth]{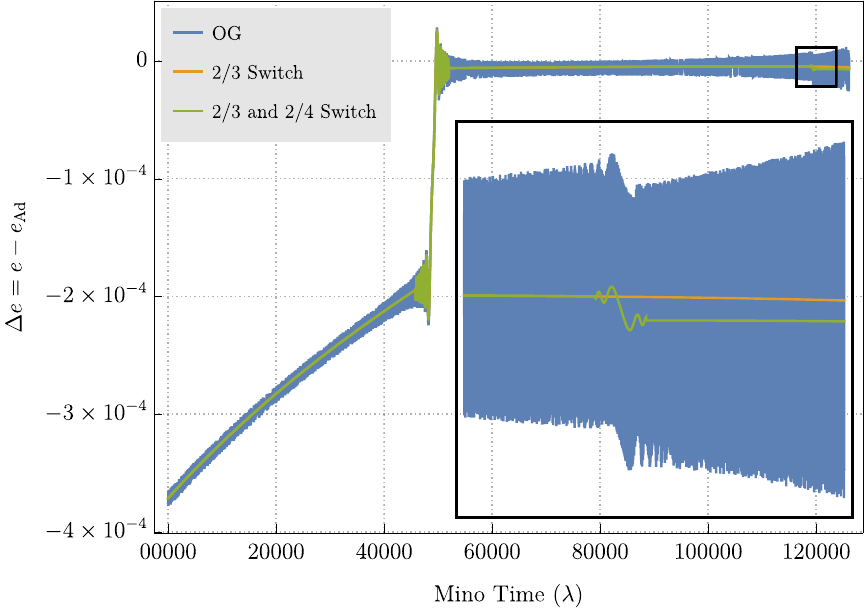}
		\caption{The difference in $e$.}
		\label{fig:DoubleResde}
	\end{subfigure}
	\hfill
	\centering
	\begin{subfigure}[b]{0.495\textwidth}
		\centering
		\includegraphics[width=1\textwidth]{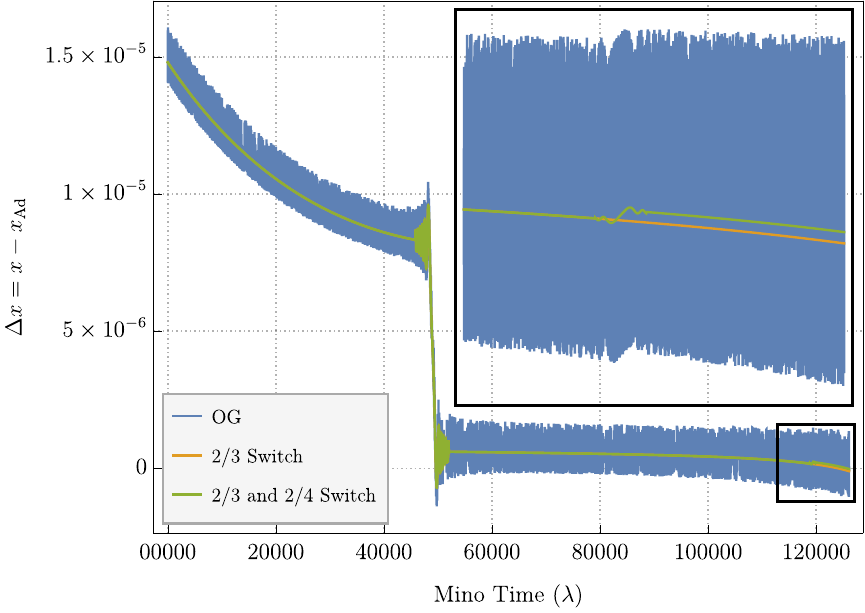}
		\caption{The difference in $x$.}
		\label{fig:DoubleResdx}
	\end{subfigure}
	\hfill
	\caption{The difference between the evolution of orbital elements $\Delta \vec{P} = \vec{P} -\vec{P}_{\text{Ad}}$ as a function of Mino time $\lambda$ for a year long inspiral with $a = 0.9 M$, $\mr = 10^{-5}$ and initial conditions $(p_0,e_0,x_0) = (6.8,0.45,0.8)$ using the OG equations of motion, the NIT equations of motion with a transition through the $2/3$ resonance and the NIT equations of motion with transitions for both the $2/3$ and $2/4$ resonances.\label{fig:DoubleRes}}
	
\end{figure}

To demonstrate the effect of the two resonances on the evolution of the semilatus rectum $p$, we match an adiabatic inspiral to the OG inspiral after the $2/3$ resonance but before the $2/4$ resonance and evolve it both forward and backwards in time. 
We then subtract the adiabatic solution for the semilatus rectum $p_{\text{Ad}}$ from the post-adiabatic inspirals solutions to illustrate the effects of both the $2/3$ and $2/4$ resonances as seen in Fig.~\ref{fig:DoubleRes}.
The figure shows how the $2/3$ has a significantly larger effect on the evolution of $p$ than the $2/4$ resonance, and that while there is an error induced by neglecting the $2/4$ resonance, it is comparably small.

\begin{figure}
	\begin{subfigure}[b]{0.49\textwidth}
		\includegraphics[width=1\textwidth]{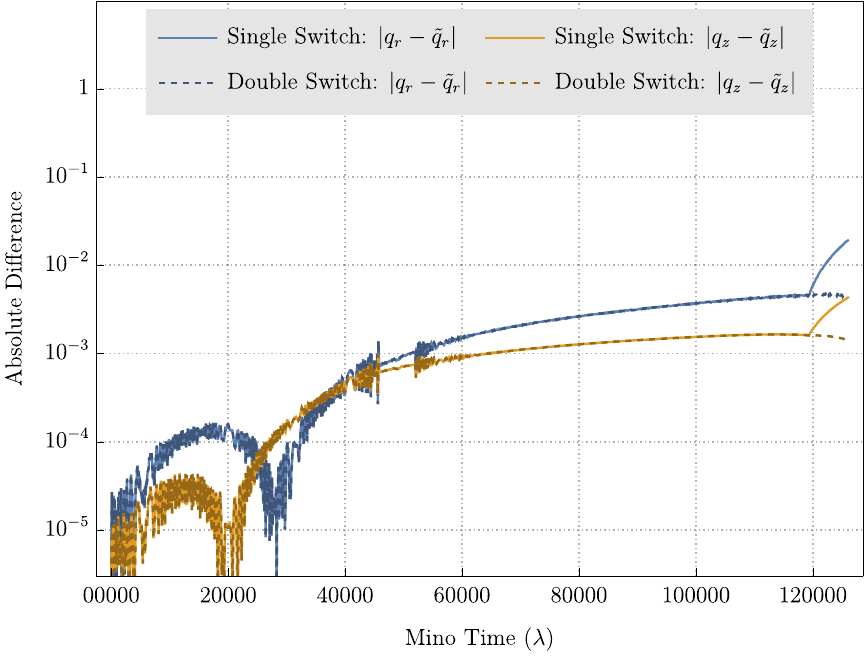}
		\caption{Orbital phases.}
		\label{fig:DoublePhases}
	\end{subfigure}
	\hfill
	\begin{subfigure}[b]{0.49\textwidth}
		\includegraphics[width=1\textwidth]{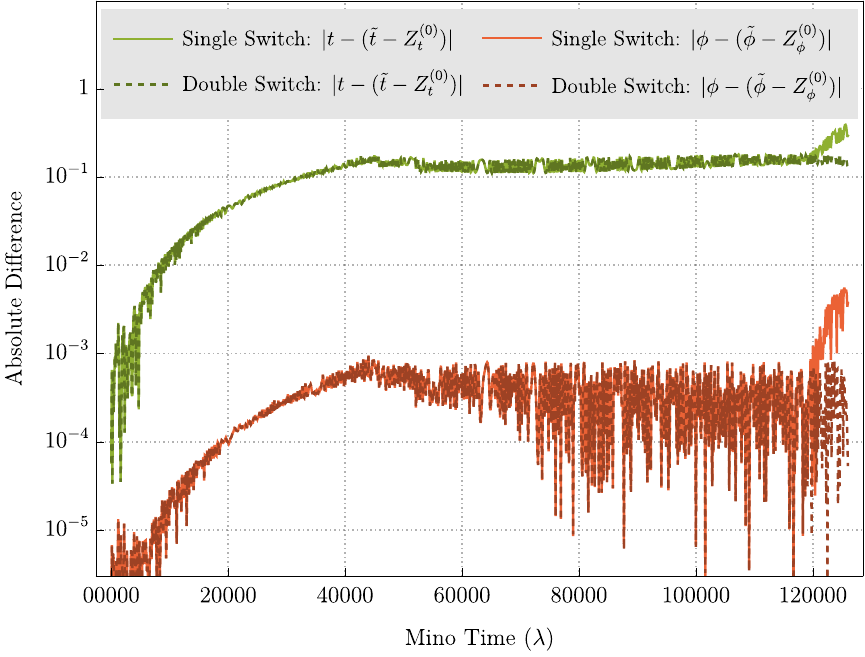}
		\caption{Extrinsic quantities.}
		\label{fig:DoubleExtrinsic}
	\end{subfigure}
	\hfill
	\caption{The absolute difference in the phases and extrinsic quantities between the OG and NIT equations of motion for a year long inspiral with $a = 0.9 M$, $\mr = 10^{-5}$ and initial conditions $(p_0,e_0,x_0) = (6.8,0.45,0.8)$. 
	Around $\lambda = 120,000$, one can clearly see the effect of neglecting the $2/4$ resonance.\label{fig:DoubleResPhasesandExtrinsic}}
\end{figure}

This is further supported when we look at the differences in the orbital phases and extrinsic quantities in Fig.~\ref{fig:DoubleResPhasesandExtrinsic}.
We see that neglecting the $2/4$ resonance induces a small but noticeable error in the orbital phases and extrinsic quantities towards the end of the inspiral. 
Since the inspiral terminates shortly after the $2/4$ resonance crossing, this error remains small. 
However, if the inspiral were to be evolved for longer, this error will accumulate and may become substantial.
This suggests that while incorporating the $2/4$ resonance may not provide a significant increase in accuracy for this particular inspiral, in general one may still need to account for it.

\begin{table}
	\begin{center}
		\begin{tabular}{c | c| c} 
			\hline
			Inspiral & Runtime & Mismatch  \\ [0.5ex] 
			\hline
			Full NIT & 4.0s & $0.569$ \\ 
			Single Switch NIT & 22.1s & $5.30 \times 10^{-5}$\\
			Double Switch NIT & 57.8s & $4.93\times 10^{-5}$\\
			\hline 
		\end{tabular}
	\end{center}
	\caption{A table of the time taken to compute a year long inspiral with mass ratio $10^{-5}$ and initial conditions $(p_0,e_0,x_0) = (6.8,0.45,0.8)$ using NIT equations of motion with different numbers of resonant transitions as calculated using NDSolve with an accuracy goal of 7 as implemented in Mathematica 13 on an Apple M1 Max Chip.
		The mismatches between the semi-relativistic quadrupole waveforms generated from these inspirals and the OG inspiral are also listed. \label{table:double_res_runtime_accuracy}}
\end{table}

Table~\ref{table:double_res_runtime_accuracy} shows the runtime and the waveform mismatch of each inspiral as compared with the waveform generated by the OG inspiral. 
We see that neglecting the $2/3$ resonance produces a waveform which agrees very poorly with the OG waveform. 
Including the transition through the $2/3$ resonance produces a waveform with a mismatch of only $5.3 \times 10^{-5}$ which is significantly smaller than the $3 \times 10^{-3}$ requirement to produce a waveform bank that can capture $90\%$ of signals \cite{Lindblom2008}. 
Including the transition through the $2/4$ resonance slightly decreases the mismatch to $4.93\times 10^{-5}$. From the differences in the phases and extrinsic quantities shown in Figs.~\ref{fig:DoubleResPhasesandExtrinsic}, if one were to run the the inspiral for longer, one would expect to see the error from neglecting the $2/4$ resonance have a larger effect. As it stands, it does not seem justifiable to more than double the runtime from $22.1s$ to $57.8s$ to include such a small resonance effect.

Both this and the previous test indicate that of the resonance crossings that we have examined, the most important to account for is the $2/3$ resonance. 
One may also need to account for other low order resonances such at the $2/4$ and $2/6$ resonances, but our preliminary results suggest that the $2/3$ resonance might be the only resonance that one must include to produce post-adiabatic waveforms accurate enough for LISA science. 
However, our toy force model has a tendency to overestimate the $2/3$ resonance effects while underestimating the rest and so a more robust study of inspirals throughout more of the parameter space with a more accurate model for the GSF is needed before such a strong conclusion can be drawn.

\section{Discussion and conclusions}  \label{section:Conclusion}

In this work, we present the first application of near-identity averaging transformations to generic Kerr inspirals in the vicinity of low-order orbital resonances. 
Generic Kerr GSF codes are too computational expensive to feasibly create an interpolated self-force model as was done in Refs.~\cite{Lynch:2021ogr,Lynch:2023gpu}. 
We circumvent this by combining the interpolated models for the eccentric equatorial GSF and the spherical GSF to create a toy model for the generic GSF which is quick to evaluate and has the qualitative behaviour one would expect of the generic GSF as one approaches the quasi-circular and equatorial limits. 
It is this toy model, along with the OG equations, that we use to drive our generic Kerr inspirals.  We use this toy model as a proof of concept of the methods we have developed for rapidly calculating Kerr inspirals. 
A follow on study with the GSF to post adiabatic order is needed to provide quantitative predictions for the transient resonances experienced by astrophysical EMRIs.

We then use near-identity averaging transformations to speed up these calculations.
We investigate how the accuracy and speed-up scales with the mass ratio and confirm that this technique works as expected, so long as one is not in the presence of a low order orbital resonance. 
Since the ``Full" NIT becomes singular in the event of an orbital resonance, we implement a ``Partial" NIT to be used in the vicinity of an orbital resonance for the first time and test the scaling of the accuracy and speed-up with the mass ratio in the presence of the $2/3$ resonance. 
From this, we find that while the phase difference between the Partial NIT and OG inspirals scales linearly with the mass ratio, as we would expect, and the Partial NIT provides an order of magnitude worth of speed-up, it would still be much too slow for data analysis applications. 

We note that this could be reduced further by utilizing the Full NIT when far from a resonance and then switching to the Partial NIT when in the vicinity of a resonance, which we refer to as the ``Switch" NIT. 
We test the convergence of this method against the partial NIT and find that our error is consistent with the predicted scalings of $\mr^{11/4}$ for the orbital elements and $\mr^{4/7}$ for both the phases and the $t$ and $\phi$ coordinates.
We test this procedure using two different, year-long EMRI trajectories; one which only evolves through the $2/3$ resonance, and one which also evolves through the $2/4$ resonance. 
From these tests, we confirm that one can use the Switch NIT with no significant loss of accuracy compared to only using the Partial NIT. 
Moreover,  our tests suggests that one can safely neglect higher order resonances without any significant loss of accuracy. 
Our results even suggest that one might be able to ignore all resonances bar the $2/3$ resonance since it has by far the largest impact on the inspiral, though further investigation is needed to ensure that this is really the case across the parameter space. Moreover, it is not clear how representative our toy model for the generic orbit GSF is for these results.

However, even with at least two orders of magnitude of speed-up from utilising the Switch NIT, our current implementation still takes $\mathcal{O}(10)$ seconds to compute a year long EMRI evolving through a single resonance, which, while a drastic improvement over the days required to solve the OG equations, is still not fast enough for LISA data analysis.
This can be reduced further by investigating and optimising evaluation of the transformation terms when the switch is made or by implementing the above procedure in a compiled language such as C/C++, though this still may not be fast enough.
It is possible that a NIT model with an empirically fitted resonant ``jump" derived from the Partial NIT could account for resonances while minimising runtime, but this will be left as future work.

\section*{Acknowledgements}
PL acknowledges support from the Irish Research Council under Grant GOIPG/2018/1978.
MvdM acknowledges financial support by the VILLUM Foundation (grant no. VIL37766) and the DNRF Chair program (grant no. DNRF162) by the Danish National Research Foundation.
NW acknowledges support from a Royal Society~-~Science Foundation Ireland University Research Fellowship.
This publication has emanated from research conducted with the financial support of Science Foundation Ireland under Grant numbers 16/RS-URF/3428, 17/RS-URF s-RG/3490 and 22/RS-URF-R/3825. We thank Ian Hinder and Barry Wardell for the SimulationTools analysis package. This work makes use of the Black Hole Perturbation Toolkit.
\vskip5mm

\bibliography{GenericKerrNIT}

\appendix 

\section{Gravitational self-force inspired toy model} \label{section:gen_Toy_Model}

The prohibitive computational cost of the generic GSF code presented in Ref.~\cite{VanDeMeent2018} necessitates that we produce a toy model for the force. The model we chose to construct is informed by GSF data in the equatorial and spherical limits and we impose that it recovers these limits exactly. Moreover, we impose that the model has a similar form to generic GSF data when Fourier decomposed as this is important for producing the effects from orbital resonances. However, the source data provides no direct information about the magnitude of the mixed radial-polar modes which are crucial for resonances. The mixed radial-polar modes are generated purely by the outer-product Ansatz described below. As a counter-example, the source data would also allow for fully separable Ansätze with no mixed modes, which would produce no resonances at all. Hence we refer to our choice as a toy model.

To construct our toy model, we must first recall that our first order eccentric orbit self-force model takes the form:
\begin{equation}
	a^{(1)}_\mu = A_\mu^0 (a,p,e) + \sum_{n = 1}^{15} A_\mu^n (a,p,e)  \cos(n q_r)  + B_\mu^n(a,p,e)  \sin(n q_r),
\end{equation}
where we have absorbed a rescaling factor described in Ref.~\cite{Lynch:2021ogr} into the coefficients $A^n_\mu$ and $B^n_\mu$. Note that these coefficients only depend on $(a,p,e)$ while for the generic orbit GSF, these would also depend on $x$.
We truncate the series at $n = 15$ as this provides sufficient accuracy.
Furthermore, our first order spherical orbit self-force model takes the form:
\begin{equation}
	a^{(1)}_\mu = C_\mu^0(a,p,x)  + \sum_{m = 1}^{24} C_\mu^m \cos(m q_\theta)(a,p,x)   + D_\mu^m(a,p,x)  \sin(m q_\theta)
\end{equation}
where again we have absorbed the rescaling factor described in Ref.~\cite{Lynch:2023gpu} into the coefficients $C^m_\mu$ and $D^m_\mu$.  Again, these coefficients only depend on $(a,p,x)$ while for the generic orbit GSF, these would also depend on $e$.

While it would be easier to combine these terms together to get cross terms if the Fourier series was expressed as a complex exponential series instead of a $\sin$ and $\cos$ series, it is possible to derive a Fourier series for a real valued 2D function my making use of trigonometric identities and simplifying:

\begin{subequations}
	\begin{align*}
		a_\mu^{(1)}(q_r, q_\theta) & = \sum_{n=0}^{\infty}\sum_{m=0}^{\infty}\mathcal{A}_\mu^{n m}\cos\left(n q_r\right)\cos\left(m q_\theta \right) \\
		& + \sum_{n=0}^{\infty}\sum_{m=0}^{\infty}\mathcal{B}_\mu^{n m}\cos\left(n q_r\right)\sin\left(m q_\theta \right) \\
		& + \sum_{n=0}^{\infty}\sum_{m=0}^{\infty}\mathcal{C}_\mu^{n m}\sin\left(n q_r\right)\cos\left(m q_\theta \right)  \\
		& + \sum_{n=0}^{\infty}\sum_{m=0}^{\infty}\mathcal{D}_\mu^{n m}\sin\left(n q_r\right)\sin\left(m q_\theta \right) \\
	\end{align*}
\end{subequations}
In our toy model, we make a simplification by defining the cross terms using an outer product, meaning that the coefficients $\mathcal{A}_\mu^{n m}$ , $\mathcal{B}_\mu^{n m}$ , $\mathcal{C}_\mu^{n m}$, and $\mathcal{D}_\mu^{n m}$ are given by:
\begin{subequations}
	\begin{align}
		\mathcal{A}_\mu^{n m} & \coloneqq A_\mu^n C_\mu^m \\ 
		\mathcal{B}_\mu^{n m} & \coloneqq  A_\mu^n D_\mu^m \\
		\mathcal{C}_\mu^{n m} & \coloneqq  B_\mu^n C_\mu^m \\
		\mathcal{D}_\mu^{n m} & \coloneqq  B_\mu^n D_\mu^m
	\end{align}
\end{subequations}
from our equatorial and spherical GSF models. Using what we know from the $n=0$ and $m = 0$ cases, we express our generic orbit force components as:
\begin{subequations}
	\begin{align*}
		a^{(1)}_\mu(q_r, q_\theta) & = A_\mu^0 + C_\mu^0 \\
		&+\sum_{n = 1}^{15} A_\mu^n \cos(n q_r)  + B_\mu^n \sin(n q_r) 
		+ \sum_{m = 1}^{24} C_\mu^m \cos(m q_\theta)  + D_\mu^m \sin(m q_\theta) \\
		&+\sum_{n=1}^{15}\sum_{m=1}^{24}A_\mu^n C_\mu^m\cos\left(n q_r\right)\cos\left(m q_\theta \right) 
		+ \sum_{n=1}^{15}\sum_{m=1}^{24}A_\mu^n D_\mu^m\cos\left(n q_r\right)\sin\left(m q_\theta \right) \\
		& + \sum_{n=1}^{15}\sum_{m=1}^{24}B_\mu^n C_\mu^m \sin\left(n q_r\right)\cos\left(m q_\theta \right) 
		+ \sum_{n=1}^{15}\sum_{m=1}^{24}B_\mu^n D_\mu^m\sin\left(n q_r\right)\sin\left(m q_\theta \right) \\
	\end{align*}
\end{subequations}
Note that all of the inclined orbit terms will vanish in the equatorial limit except for $C^0_\mu$ and all the eccentric orbit terms will vanish in the circular limit except for $A^0_\mu$. 
We wish to weight these orbit averaged terms so that we can recover the two limit cases accurately. 

To do this, we note that our equatorial model covers eccentricities ranging from $0<e<0.5$. 
Our spherical model is tiled in terms of a parameter $v$ where $v =\cos^2 \theta_\text{min} = 1-x^2$ such that our model covers $0< v <0.5$. 
This similarity will come in handy. 
We want the weighting factors to be smooth, recover the two limit factors, and we make the choice that when $e = v$ that the weighting factors are both $0.5$ so that the orbit averaged piece will be the mean of the equatorial and spherical contributions. 
As such, we chose following weighting functions:

\begin{equation}
	\alpha(e,v) = \Bigg\{  \begin{array}{lr}
		\frac{e}{e+v} & e\leq v \\
		1-\frac{v}{e+v} & e > v 
	\end{array} , \quad
	\beta(e,v) = \Bigg\{ \begin{array}{lr} 
	\frac{v}{e+v} & e\leq v \\
	1-\frac{e}{e+v} & e > v 
	\end{array}.
\end{equation}

Despite their piecewise definition, these functions are smooth and continuous everywhere except for the point $(e,v) = (0,0)$.
We can now write our generic force components as 
\begin{subequations}
	\begin{align*}
		a^{(1)}_\mu(q_r, q_\theta) = & \alpha A_\mu^0 + \beta C_\mu^0 \\
		&+\sum_{n = 1}^{15} A_\mu^n \cos(n q_r)  + B_\mu^n \sin(n q_r) 
		+ \sum_{m = 1}^{24} C_\mu^m \cos(m q_\theta)  + D_\mu^m \sin(m q_\theta) \\
		&+ \sum_{n=1}^{15}\sum_{m=1}^{24}A_\mu^n C_\mu^m\cos\left(n q_r\right)\cos\left(m q_\theta \right) 
		+ \sum_{n=1}^{15}\sum_{m=1}^{24}A_\mu^n D_\mu^m\cos\left(n q_r\right)\sin\left(m q_\theta \right) \\
		& + \sum_{n=1}^{15}\sum_{m=1}^{24}B_\mu^n C_\mu^m \sin\left(n q_r\right)\cos\left(m q_\theta \right) 
		+ \sum_{n=1}^{15}\sum_{m=1}^{24}B_\mu^n D_\mu^m\sin\left(n q_r\right)\sin\left(m q_\theta \right) .\\
	\end{align*}
\end{subequations}
Before now, we only ever needed the orbit averaged contribution from the second order self-force, which in the absence of any results for Kerr inspirals, we simply set to zero.
However, when evolving near an orbital resonance, Eq.~\eqref{eq:Paartial_NIT_Second_Order_Dependence_On_Phaases} involves the oscillatory part of the second order self-force.
Since we are already using a toy model for the first order self-force, we choose to also create a toy model for the second order self-force $a^{(2)}_\mu$.
However, we do not have any data for generic Kerr inspirals that can help inform such a model.
Thus, we have opted to use this first order toy model to inform the second order toy model, which we take to be:
\begin{equation}
	a^{(2)}_\mu = \frac{a^{(1)}_\mu}{r^2 \sqrt{1-\cos^2 \theta}},
\end{equation}
where the factor of $1/r^2$ is used to ensure that the second order self-force corresponds to the correct post-Newtonian order and the factor $1/\sqrt{1-\cos^2 \theta}$ prescribes the effect of the inclination, and implicitly spin of the primary, on the second order self-force.
With both of these terms, we express the self-force as $a_\mu = a_\mu^{(1)} + \mr a_\mu^{(2)}$.

Finally, in order for this toy model to work with the method of osculating geodesics, we require that it must satisfy the orthogonality condition with the geodesic four velocity, i.e., $a_\mu u^\mu = 0$. 
To enforce this relationship we project off any parts of the force that violate this condition using the following relationship:
\begin{equation}
	a_\mu^\perp = a_\mu + a_\nu u^\nu u_\mu.
\end{equation}

Using the projected force components with our osculating geodesic equations of motion we find inspirals that have qualitatively correct $p,e,$ and $x$ evolution for an EMRI under the effect of the gravitational self-force as well as strong resonant effects. 

To verify this, we derive the variation of the flux for the constants of motion $\vec{\mathcal{J}} = \{ \En,\Lz,\Q\}$ where $\En$ is the orbital energy per unit rest mass $\mu$, $\Lz$ is the z-component of the angular momentum per unit mass $\mu$ and K is the Carter constant divided by $\mu^2$. To obtain the rate of change of these quantities, we use the expressions for these constants as a function of $\vec{P}$ derived in Ref.~\cite{Schmidt2002} and use the chain rule, i.e 
	\begin{equation}
		\frac{d \mathcal{J}_i}{dt} = \dot{ \mathcal{J}}_i =  \frac{\partial \mathcal{J}_i}{\partial P_j} \frac{d P_j}{d \lambda} \frac{d \lambda}{d t} = \frac{\partial \mathcal{J}_i}{\partial P_j} \frac{F^{(1)}_j}{s_t}
	\end{equation}
 We are interested in how the flux of $\vec{\mathcal{J}}$ varies over an orbital resonance and so using our notion of a partial average, we obtain: 
	\begin{equation}
		 \avg{\dot{ \mathcal{J}}_i}_\text{res} =  \frac{\partial \mathcal{J}_i}{\partial P_j} \frac{\hat{F}^{(1)}_{j,\text{diss}}(q_\perp)}{\Upsilon^{(0)}_t}.
		\end{equation} 
 Note that we look to only include dissipative effects from the force, as so we make use of the dissipative-conservative split derived in Ref.~\cite{Hinderer2008}  i.e. $F_{j,\text{diss}}^{(1)} = \frac{1}{2}\left( F_{j}^{(1)} (q_r,q_\theta) + F_{j}^{(1)} (2\pi - q_r,2\pi - q_\theta) \right)$.
In line with the analysis performed in Ref.~\cite{Flanagan:2012kg}, we define the variation $\Delta$ of a function $X$ to be 
\begin{equation}
	\Delta X = \frac{| X_\text{max} | - | X_\text{min} |}{\left( | X_\text{max} | + | X_\text{min}|  \right) / 2},
\end{equation}
where the minimum and maximum values of the function are found via numerical root finding. 

\begin{table}
	\begin{center}
		\begin{tabular}{| c | c || c | c || c | c || c | c  |}  
			\hline
			  $p /M$ & $\Omega_r/\Omega_\theta$   &  $\Delta \langle  \dot{\En} \rangle _\text{res}$ &  $\Delta \dot{\En}$ Ref.~\cite{Flanagan:2012kg} &  $\Delta \langle  \dot{\Lz} \rangle _\text{res}$ &  $\Delta \dot{\Lz}$ Ref.~\cite{Flanagan:2012kg} &  $\Delta \langle  \dot{\Q} \rangle _\text{res}$Toy &  $\Delta \dot{\Q}$ Ref.~\cite{Flanagan:2012kg}   \\ [0.5ex] 
			\hline
			$2.9112$ & $1/3 (2/6)$ & $0.028 \% $ &$0.056 \% $ &   $0.035 \% $ & $0.070 \% $ &    $0.149 \% $ & $0.310 \% $  \\ 
			 $3.5560$& $1/2  (2/4)$ &  $0.084\% $ & $0.131 \% $ &  $0.063 \% $ &  $0.179 \% $ &  $0.203\% $ & $0.046 \% $   \\ 
			$5.3414$& $2/3 $ & $0.332 \% $ &  $0.102 \% $ &  $0.116\% $ &  $0.067\% $ & $0.687 \% $ & $0.208 \% $  \\ 
			 $7.4198$& $3/4 (6/8)$ &  $5 \times 10^{-8} \% $ & $0.001\% $ & $5 \times 10^{-9}  \% $ & $0.001 \% $  & $8 \times 10^{-6} \% $ & $0.006 \% $  \\ 
			\hline 
		\end{tabular}
	\end{center}
	\caption{Variation in flux for orbits with $e = 0.3$ and $\theta_{\text{min}} = 70^{\circ}$ about a black hole with spin $a = 0.9 M$ calculated from our toy GSF model and compared with the values obtained in Ref.~\cite{Flanagan:2012kg}. Our toy model tends to overestimate the variation for the $2/3$ resonance and significantly underestimate the variation for the higher order $2/6, 2/4,$  and $6/8$ resonances. \label{table:flux_variation_comparison}}
\end{table}

In Table~\ref{table:flux_variation_comparison}, we show the values of $\Delta \langle  \dot{\mathcal{J}} \rangle _\text{res} $ from our toy model against the values obtain in table IV of Ref.~\cite{Flanagan:2012kg} which was obtained from combining the Teukolsky fluxes down the horizon of the black hole and out to infinity. From this comparison, we see that our toy model produces variations in the fluxes that are at least qualitatively in line with what one would obtain from a realistic self-force model with the values all being within an order of magnitude of the values reported in Ref.~\cite{Flanagan:2012kg} (with the exception of the $6/8$ resonance). Our model has a tendency to overestimate the effect of the lowest order $2/3$ resonance while underestimating the effects of higher order resonances.

\section{Partial NIT Derivation}
\label{section:near-resonant-NIT}
The Full NIT will break down in the presence of orbital resonances where the radial and polar frequencies become commensurate i.e. $\kappa_r \Upsilon_r^{(0)} + \kappa_\theta \Upsilon_\theta^{(0)} = 0$ where $\kappa_r,\kappa_\theta \in \mathbb{Z}$.
As such, we will introduce a new averaging procedure which averages almost all dependence on the orbital phases, except for the resonant phase $q_\perp \coloneqq \kappa_r q_r + \kappa_\theta q_\theta$, which we call the Partial NIT.
This will mean that our equations will oscillate, and so will be slower to solve than the Full NIT equations of motion.
However, $q_\perp$ oscillates on a timescale between that of the slow evolution of the orbital elements and the rapidly oscillating orbital phases, and so can be thought of as a ``semi-fast" variable. This appendix serves to recast the appendix C of \cite{VanDeMeent2014a} in the notation of \cite{NITs}.

\subsection{Near Identity Transformation}
We will first focus on the evolution of the orbital elements and orbital phases and so we once again introduce the transformation 

\begin{subequations}\label{eq:res_transformation}
	\begin{align}
		\begin{split}
			\partialnit{P}_j &= P_j + \sp \partialnit{Y}_j^{(1)}(\vec{P},\vec{q}) + \sp^2 \partialnit{Y}_j^{(2)}(\vec{P},\vec{q}) + \HOT{3},
		\end{split}\\
		\begin{split}
			\partialnit{q}_i &= q_i + \sp \partialnit{X}_i^{(1)}(\vec{P},\vec{q}) +\sp^2 \partialnit{X}_i^{(2)}(\vec{P},\vec{q}) + \HOT{3},
		\end{split}\\
		\begin{split}
			\partialnit{q}_\perp &= q_\perp + \sp \partialnit{W}^{(1)}(\vec{P},\vec{q}) +\sp^2 \partialnit{W}^{(2)}(\vec{P},\vec{q}) + \HOT{3},
		\end{split}
	\end{align}
\end{subequations}
where we have implicitly imposed that none of the functions of the right hand side depend on the resonant phase $q_\perp$.
This transformation has an inverse that is given by

\begin{subequations}\label{eq:res_inverse_trasformaiton}
	\begin{align}
		\begin{split}
			P_j &= \partialnit{P_j} - \epsilon \partialnit{Y}_j^{(1)}(\vec{\partialnit{P}},\vec{\partialnit{q}}) \\ & - \epsilon^2 \left(  \partialnit{Y}_j^{(2)}(\vec{\partialnit{P}},\vec{\partialnit{q}}) - \PD{\partialnit{Y}_j^{(1)}(\vec{\partialnit{P}},\vec{\partialnit{q}})}{\partialnit{P_k}} \partialnit{Y}_k^{(1)}(\vec{\partialnit{P}},\vec{\partialnit{q}}) - \PD{\partialnit{Y}_j^{(1)}(\vec{\nit{P}},\vec{\nit{q}})}{\nit{q_k}} \partialnit{X}_k^{(1)}(\vec{\partialnit{P}}, \vec{\partialnit{q}})  \right) + \mathcal{O}(\epsilon^3),
		\end{split}\\
		\begin{split}
			q_i &= \partialnit{q_i} - \epsilon \partialnit{X}_i^{(1)}(\vec{\partialnit{P}},\vec{\partialnit{q}})  \\ &- \epsilon^2 \left(  \partialnit{X}_i^{(2)}(\vec{\partialnit{P}},\vec{\partialnit{q}}) - \PD{\partialnit{X}_i^{(1)}(\vec{\partialnit{P}},\vec{\partialnit{q}})}{\partialnit{P_j}} \partialnit{Y}_j^{(1)}(\vec{\partialnit{P}},\vec{\partialnit{q}}) - \PD{\partialnit{X}_i^{(1)}(\vec{\partialnit{P}},\vec{\partialnit{q}})}{\partialnit{q_k}} \partialnit{X}_k^{(1)}(\vec{\partialnit{P}},\vec{\partialnit{q}})  \right) + \mathcal{O}(\epsilon^3).
		\end{split}\\
		\begin{split}
			q_\perp &= \partialnit{q}_\perp - \epsilon \partialnit{W}_i^{(1)}(\vec{\partialnit{P}},\vec{\nit{q}})  \\ &- \epsilon^2 \left(  \partialnit{W}_i^{(2)}(\vec{\partialnit{P}},\vec{\partialnit{q}}) - \PD{\partialnit{W}_i^{(1)}(\vec{\partialnit{P}},\vec{\partialnit{q}})}{\partialnit{P_j}} \partialnit{Y}_j^{(1)}(\vec{\partialnit{P}},\vec{\partialnit{q}}) - \PD{\partialnit{W}_i^{(1)}(\vec{\partialnit{P}},\vec{\partialnit{q}})}{\partialnit{q_k}} \partialnit{X}_k^{(1)}(\vec{\partialnit{P}},\vec{\partialnit{q}})  \right) + \mathcal{O}(\epsilon^3).
		\end{split}
	\end{align}
\end{subequations}

\subsection{Transformed Equations of Motion}
By taking the time derivative of the NIT \eqref{eq:res_transformation}, substituting the EMRI equations of motion \eqref{eq:Generic_EMRI_EoM} and inverse NIT \eqref{eq:res_inverse_trasformaiton}, and expanding in powers of $\mr$ we obtain the NIT transformed equations of motions
\begin{subequations}\label{eq:res_transformed_EoM_oscillating}
	\begin{align}
		\begin{split}
			\frac{d \partialnit{P}_j}{d \lambda} &=\epsilon \partialnit{F}_j^{(1)}(\vec{\partialnit{P}},\vec{\partialnit{q}},\partialnit{q}_\perp) + \epsilon^2 \nit{F}_j^{(2)}(\vec{\partialnit{P}},\vec{\partialnit{q}},\partialnit{q}_\perp) +  \mathcal{O}(\epsilon^3),
		\end{split}\\
		\begin{split}
			\frac{d \partialnit{q}_i}{d\lambda} &= \Upsilon_i^{(0)}(\vec{\partialnit{P}}) +\epsilon \partialnit{f}_i^{(1)}(\vec{\partialnit{P}},\vec{\partialnit{q}},\partialnit{q}_\perp) + \mathcal{O}(\epsilon^2),
		\end{split}\\
		\begin{split}
			\frac{d \partialnit{q}_\perp}{d\lambda} &= \vec{\kappa}_{\text{res}} \cdot \vec{\Upsilon}^{(0)}(\vec{\partialnit{P}}) +\epsilon \vec{\kappa}_{\text{res}} \cdot \vec{\partialnit{f}}^{(1)}(\vec{\partialnit{P}},\vec{\partialnit{q}},\partialnit{q}_\perp) + \mathcal{O}(\epsilon^2),
		\end{split}
	\end{align}
\end{subequations}
where 
\begin{subequations}
	\begin{gather}
		\partialnit{F}_j^{(1)} = F^{(1)}_j + \PD{\partialnit{Y}_j^{(1)}}{\partialnit{q}_i} \Upsilon_i^{(0)}, \quad 
		\partialnit{f}_i^{(1)} = f^{(1)}_i + \PD{\partialnit{X}_i^{(1)}}{\partialnit{q}_k} \Upsilon^{(0)}_k - \PD{\Upsilon^{(0)}_i}{\partialnit{P}_j} \partialnit{Y}_j^{(1)}, \tag{\theequation a-b}
	\end{gather}
\end{subequations}
and 
\begin{align}
	\partialnit{F}_j^{(2)} & = F^{(2)}_j + \PD{\partialnit{Y}_j^{(2)}}{\partialnit{q}_i} \Upsilon_i^{(0)} + \PD{\partialnit{Y}_j^{(1)}}{\partialnit{q}_i} f_i^{(1)} + \PD{\partialnit{Y}_j^{(1)}}{\partialnit{P}_k} F_k^{(1)} -\PD{\partialnit{F}_j^{(1)}}{\partialnit{P}_k} \partialnit{Y}_k^{(1)} - \PD{\partialnit{F}_j^{(1)}}{\partialnit{q}_i} \partialnit{X}_i^{(1)} -\PD{\partialnit{F}_j^{(1)}}{\partialnit{q}_\perp} \partialnit{W}^{(1)}
\end{align}
Note that all functions on the right hand side are evaluated at $\vec{\partialnit{P}}$, $\vec{\partialnit{q}}$ and $\partialnit{q}_\perp$.

\subsection{Cancellation of oscillating terms at $\mathcal{O}(\mr)$}
We note that we can decompose any $2 \pi$ periodic function into its averaged, resonant oscillatory and non-resonant oscillatory pieces using a Fourier expansion:

\begin{equation}\label{eq:res_decomposition}
	A(\vec{P},\vec{q},q_\perp) = \avg{A}(\vec{P})  
	+ \sum_{N \neq 0} A_{N \vec{\kappa}_{\text{res}}}(\vec{P}) e^{i N q_\perp} 
	+  \sum_{\vec{\kappa} \in R} 
	A_{\vec{\kappa}}(\vec{P}) e^{i \vec{\kappa} \cdot \vec{q}}. 
\end{equation}
where R is the set $\{ \vec{\kappa} \in \mathbb{Z}^2 | \vec{\kappa} \neq N \vec{\kappa}_{\text{res}}, \forall N \in \mathbb{Z} \}$ of all non-resonant 2-tuples and $\kappa_\text{res} = (\kappa_r,\kappa_\theta)$ is such that $\kappa_\text{res} \cdot \vec{\Upsilon}^{(0)} = 0$. Applying this decomposition to $\partialnit{F}^{(1)}_j$, one obtains
\begin{align}
	\begin{split}
		\partialnit{F}^{(1)}_j &=  F^{(1)}_j + \PD{\partialnit{Y}_j^{(1)}}{\partialnit{q}_i} \Upsilon_i^{(0)} = F^{(1)}_j + \PD{\partialnit{Y}_j^{(1)}}{\partialnit{q}_i}\Upsilon_i^{(0)} \\
		&= \avg{F^{(1)}_j}  + \sum_{N \neq 0} F^{(1)}_{j,N \vec{\kappa}_{\text{res}}} e^{i N q_\perp} + 
		\sum_{\vec{\kappa} \in R} 
		\left( F^{(1)}_{j,\vec{\kappa}} 
		+ i \left(\vec{\kappa} \cdot \vec{\Upsilon}^{(0)}\right) \partialnit{Y}^{(1)}_{j,\vec{\kappa}} \right) 
		e^{i \vec{\kappa} \cdot \vec{q}}.
	\end{split}
\end{align}
As such, we can cancel the non-resonant oscillatory pieces of $\partialnit{F}_j^{(1)}$ by choosing the oscillatory part of $\partialnit{Y}_j^{(1)}$ to be
\begin{equation}
	\partialnit{Y}^{(1)}_{j,\vec{\kappa}} \coloneqq \frac{i}{\vec{\kappa} \cdot \vec{\Upsilon}^{(0)}} F^{(1)}_{j,\vec{\kappa}}(\vec{P}).
\end{equation}
for $\vec{\kappa} \neq N \vec{\kappa}_{\text{res}}$ and $0$ when $\vec{\kappa} = N \vec{\kappa}_{\text{res}}$.
Using the above choice for $\partialnit{Y}_j^{(1)}$, the equation for $\partialnit{f}_i^{(1)}$ becomes
\begin{align}
	\begin{split}
		\partialnit{f}_i^{(1)} =& f^{(1)}_i - \PD{\Upsilon^{(0)}_i}{\partialnit{P}_j} \partialnit{Y}_j^{(1)} + \PD{\partialnit{X}_i^{(1)}}{\partialnit{q}_k} \Upsilon^{(0)}_k  
		\\ =& \avg{f^{(1)}_i} - \PD{\Upsilon^{(0)}_i}{\nit{P}_j} \avg{\partialnit{Y}_j^{(1)}} 
		+ \sum_{N \neq 0}  f^{(1)}_{i,N \vec{\kappa}_{\text{res}}}  e^{i N \partialnit{q}_\perp}\\
		&+ \sum_{\vec{\kappa} \in R} 
		\left(f^{(1)}_{i,\vec{\kappa}} - \frac{i}{\vec{\kappa} \cdot \vec{\Upsilon}^{(0)}} \PD{\Upsilon^{(0)}_i}{\partialnit{P}_j} F_{j,\vec{\kappa}} +   i (\vec{\kappa} \cdot \vec{\Upsilon}^{(0)}) X^{(1)}_{i,\vec{\kappa}} \right) e^{i \vec{\kappa} \cdot \vec{q}}
	\end{split}
\end{align}
As a result, we can remove the oscillating pieces of $\partialnit{f}_i^{(1)}$ by choosing
\begin{equation}
	\partialnit{X}_{i, \vec{\kappa}}^{(1)} \coloneqq  \frac{i}{\vec{\kappa} \cdot \vec{\Upsilon}^{(0)}} f_{i,\vec{\kappa}}^{(1)} + \frac{1}{(\vec{\kappa} \cdot \vec{\Upsilon}^{(0)})^2} \frac{\partial \Upsilon_i^{(0)}}{\partial P_j}F_{j,\vec{\kappa}}^{(1)}.
\end{equation}
for $\vec{\kappa} \neq N \vec{\kappa}_{\text{res}}$.
Moreover, we can determine the transformation term $W$ by examining the equation  the equation for $\vec{\kappa}_\text{res} \cdot \vec{\partialnit{f}}^{(1)}$:
\begin{align}
	\begin{split}
		\vec{\kappa}_\text{res}  \cdot \vec{\partialnit{f}}^{(1)} = & \vec{\kappa}_\text{res}  \cdot \vec{f}^{(1)} - \PD{\vec{\kappa}_\text{res}  \cdot \vec{\Upsilon}^{(0)}}{\partialnit{P}_j} \partialnit{Y}_j^{(1)} + \PD{\partialnit{W}^{(1)}}{\partialnit{q}_k} \Upsilon^{(0)}_k  
		\\ =& \avg{\vec{\kappa}_\text{res}  \cdot \vec{f}^{(1)}} - \PD{(\vec{\kappa}_\text{res}  \cdot \vec{\Upsilon}^{(0)})}{\partialnit{P}_j} \avg{\partialnit{Y}_j^{(1)}}  
		+ \sum_{N \neq 0}  (\vec{\kappa}_\text{res}  \cdot \vec{f}^{(1)})_{N \vec{\kappa}_{\text{res}}}  e^{i N \nit{q}_\perp}\\
		&+ \sum_{\vec{\kappa} \in R} 
		\left((\vec{\kappa}_\text{res}  \cdot \vec{f}^{(1)})_{\vec{\kappa}} - \frac{i}{\vec{\kappa} \cdot \vec{\Upsilon}^{(0)}} \PD{(\vec{\kappa}_\text{res}  \cdot \vec{\Upsilon}^{(0)})}{\partialnit{P}_j} F_{j,\vec{\kappa}} +   i (\vec{\kappa} \cdot \vec{\Upsilon}^{(0)}) \partialnit{W}_{\vec{\kappa}} \right) e^{i \vec{\kappa} \cdot \vec{\partialnit{q}}}
	\end{split}
\end{align}
As such, to remove the oscillating pieces of $\vec{\kappa} \cdot \vec{\Upsilon}^{(1)}$, the oscillatory piece of $\partialnit{W}^{(1)}$ must take the form
\begin{equation}
	\partialnit{W}_{\vec{\kappa}}^{(1)} \coloneqq  \frac{i}{\vec{\kappa} \cdot \vec{\Upsilon}^{(0)}} (\vec{\kappa}_\text{res}  \cdot \vec{f}^{(1)})_{\vec{\kappa}}  + \frac{1}{(\vec{\kappa} \cdot \vec{\Upsilon}^{(0)})^2} \frac{\partial (\vec{\kappa}_\text{res}  \cdot \vec{\Upsilon}^{(0)})}{\partial P_j}F_{j,\vec{\kappa}}^{(1)} = \vec{\kappa}_\text{res} \cdot \vec{\partialnit{X}}_{\vec{\kappa}}^{(1)}.
\end{equation}
for $\vec{\kappa} \neq N \vec{\kappa}_{\text{res}}$.
Note that this derivation is consistent with the fact that since $\nit{q}_\perp = \vec{\kappa} \cdot \vec{\nit{q}}$ then by Eq.~\eqref{eq:res_transformation}, $\partialnit{W}_{\vec{\kappa}}^{(1)} = \vec{\kappa} \cdot \vec{\partialnit{X}}^{(1)}$. Note that in practice, this means we not need to include a separate equation to evolve $\partialnit{q}_\perp$ if we are already evolving $\partialnit{q}_i$ separately.

\subsection{Cancellation of oscillating terms at $\HOT{2}$}
Using the above choice for the oscillatory part of $\partialnit{Y}_j^{(1)}$, we can express the non-resonant oscillatory part of the expression for $\partialnit{F}_j^{(2)}$ as
\begin{align}
	\begin{split}
		\osc{\partialnit{F}}^{(2)}_j = & \osc{F}_j^{(2)} + \PD{\partialnit{Y}_j^{(2)}}{\partialnit{q}_i} \Upsilon_i^{(0)} + \left\{\PD{\partialnit{Y}_j^{(1)}}{\partialnit{q}_i} f_i^{(1)} \right\} + \left\{\PD{\partialnit{Y}_j^{(1)}}{\partialnit{P}_k} F_k^{(1)} \right\} - \PD{\avg{F_j^{(1)}}}{\partialnit{P}_k} \osc{\partialnit{Y}}_k^{(1)}
		\\ =& \sum_{\vec{\kappa} \in R}  \Biggl( 
		F_{j,\vec{\kappa}}^{(2)} + i (\vec{\kappa} \cdot \vec{\Upsilon}^{(0)}) \partialnit{Y}^{(2)}_{j,\vec{\kappa}}
		+ \PD{\avg{\partialnit{Y}_j^{(1)}}}{\partialnit{P}_k} F^{(1)}_{k,\kappa} - i \PD{\avg{F_j^{(1)}}}{\partialnit{P}_k} \frac{F^{(1)}_{k,\vec{\kappa}}}{\vec{\kappa} \cdot \vec{\Upsilon}^{(0)}}
		\\ & +   \sum_{\vec{\kappa}' \in R} \biggl(   i \frac{F^{(1)}_{k,\vec{\kappa} - \vec{\kappa} '}}{\vec{\kappa}' \cdot \vec{\Upsilon}^{(0)}} \left( \PD{F^{(1)}_{j,\vec{\kappa}'}}{\partialnit{P}_k} - \frac{F^{(1)}_{j,\vec{\kappa} '}}{\vec{\kappa}' \cdot \vec{\Upsilon}^{(0)}} \PD{(\vec{\kappa}' \cdot \vec{\Upsilon}^{(0)}) } {\partialnit{P}_k}  \right) - \frac{\vec{\kappa}' \cdot \vec{f}^{(1)}_{\vec{\kappa} - \vec{\kappa}'}}{\vec{\kappa}' \cdot \vec{\Upsilon}^{(0)}} F^{(1)}_{j,\vec{\kappa}'}
		\biggr) \Biggr) e^{i \vec{\kappa} \cdot \vec{\partialnit{q}}},
	\end{split}
\end{align}
where $\{\cdot\}$ is used to denote the non-resonant oscillatory part of a product of functions.
Thus we can remove the oscillatory part of $\partialnit{F}_j^{(2)}$ by choosing
\begin{align}
	\begin{split}
		\partialnit{Y}^{(2)}_{j,\vec{\kappa}} = & \frac{i}{\vec{\kappa} \cdot \vec{\Upsilon}^{(0)}} \Biggl( 
		F_{j,\vec{\kappa}}^{(2)} + \PD{\avg{\partialnit{Y}_j^{(1)}}}{\partialnit{P}_k} F^{(1)}_{k,\vec{\kappa}} - i \PD{\avg{F_j^{(1)}}}{\partialnit{P}_k} \frac{F^{(1)}_{k,\vec{\kappa}}}{\vec{\kappa} \cdot \vec{\Upsilon}^{(0)}}
		\\ & +   \sum_{\vec{\kappa}' \in R} \biggl(   i \frac{F^{(1)}_{k,\vec{\kappa} - \vec{\kappa} '}}{\vec{\kappa}' \cdot \vec{\Upsilon}^{(0)}} \left( \PD{F^{(1)}_{j,\vec{\kappa}'}}{\partialnit{P}_k} - \frac{F^{(1)}_{j,\vec{\kappa} '}}{\vec{\kappa}' \cdot \vec{\Upsilon}^{(0)}} \PD{(\vec{\kappa}' \cdot \vec{\Upsilon}^{(0)}) } {\partialnit{P}_k}  \right) - \frac{\vec{\kappa}' \cdot \vec{f}^{(1)}_{\vec{\kappa} - \vec{\kappa}'}}{\vec{\kappa}' \cdot \vec{\Upsilon}^{(0)}} F^{(1)}_{j,\vec{\kappa}'}
		\biggr) \Biggr).
	\end{split}
\end{align}

\subsection{Freedom in the averaged pieces}
With the non-resonant oscillatory pieces of the NIT equations of motion removed, terms in the equations of motion become
\begin{subequations}
	\begin{gather}
		\hat{F}_j^{(1)} = \avg{F^{(1)}_j} + \sum_{N \neq 0} F^{(1)}_{j,N \vec{\kappa}_{\text{res}}} e^{i N \partialnit{q}_\perp} , \quad 
		\partialnit{f}_i^{(1)} = \avg{f^{(1)}_i} + \sum_{N \neq 0} f^{(1)}_{i,N \vec{\kappa}_{\text{res}}} e^{i N \partialnit{q}_\perp} - \PD{\Upsilon^{(0)}_i}{\nit{P}_j} \avg{Y_j^{(1)}}, \tag{\theequation a-b}
	\end{gather}
\end{subequations}
and 
\begin{align}
	\begin{split}
		\partialnit{F}_j^{(2)} = &  \avg{F^{(2)}_j} + \sum_{N \neq 0} F^{(2)}_{j,N \vec{\kappa}_{\text{res}}} e^{i N \partialnit{q}_\perp} +  \avg{\PD{\osc{Y}_j^{(1)}}{\partialnit{q}_i} \osc{f}_i^{(1)}} + \avg{\PD{\osc{\partialnit{Y}}_j^{(1)}}{\partialnit{P}_k} \osc{F}_k^{(1)}}\\ 
		& + \PD{\avg{\partialnit{Y}_j^{(1)}}}{\partialnit{P}_k} \avg{F_k^{(1)}} -\PD{\avg{F_j^{(1)}}}{\partialnit{P}_k} \avg{\partialnit{Y}_k^{(1)}}
	\end{split}
\end{align}

Note that we still have freedom to set the averaged pieces of the transformation functions $\avg{\partialnit{Y}_j^{(1)}}$, $\avg{\partialnit{Y}_j^{(2)}}$, $\avg{\partialnit{X}_i^{(1)}}$, and $\avg{\partialnit{W}^{(1)}}$ to be anything we choose.
As before, we make the simplest choice: $\avg{\partialnit{Y}_j^{(1)}}=\avg{\partialnit{Y}_j^{(2)}}=\avg{\partialnit{X}_i^{(1)}} = \avg{\partialnit{W}^{(1)}} = 0$, as this makes it easy to compare between OG and NIT inspirals. 
It also has the benefit of drastically reducing the terms in our equations of motion to
\begin{subequations}
	\begin{gather}
		\nit{F}_j^{(1)} = \avg{F^{(1)}_j} + \sum_{N \neq 0} F^{(1)}_{j,N \vec{\kappa}_{\text{res}}} e^{i N q_\perp} , \quad 
		\Upsilon_i^{(1)} = \avg{f^{(1)}_i} + \sum_{N \neq 0} f^{(1)}_{i,N \vec{\kappa}_{\text{res}}} e^{i N q_\perp}, \tag{\theequation a-b}
	\end{gather}
\end{subequations}
and 
\begin{align}
	\nit{F}_j^{(2)} & = \avg{F^{(2)}_j} + \sum_{N \neq 0} F^{(2)}_{j,N \vec{\kappa}_{\text{res}}} e^{i N q_\perp}+ \avg{\PD{\osc{Y}_j^{(1)}}{\nit{q}_i} \osc{f}_i^{(1)}} + \avg{\PD{\osc{Y}_j^{(1)}}{\nit{P}_k} F_k^{(1)}}
\end{align}

\subsection{Evolution of extrinsic quantities}
The last thing to add this formulation is the evolution of the extrinsic quantities. 
Thankfully, both the $t$ and $\phi$ geodesic equations are separable with respect to $r$ and $z$ and so
\begin{equation}
	s_k^{(0)} = \sum_{\vec{\kappa} \in \mathbb{Z}^N} s^{(0)}_{k,\vec{\kappa}} e^{i \vec{\kappa} \cdot \vec{q}}
	= \sum_{N} \left(s^{(0)}_{k,(N,0)} e^{i N qr} + s^{(0)}_{k,(0,N)} e^{i N qz} \right).
\end{equation}
This means that the geodesic rates of change of the extrinsic quantities $s_k^{(0)}$ have no dependence on $q_\perp$. Thus any term in the NIT transformations or equations of motion proportional to $1/\left( \kappa_{\text{res}} \cdot \Upsilon^{(0)}\right)$ will be multiplied by 0 and so all of our terms remain finite.
As such, we can continue using the Full NIT expressions for these terms.

\section{Switch NIT transition condition} \label{section:transition_condition}
In this section we give the derivation of the switching criterion between the Partial NIT and the  Full NIT equations of motion. The criterion is chosen so that no more accuracy can be gained by prolonging numerical integration of the Partial NIT. 

We only give a brief description of the derivation as applicable strictly to the algorithm and system of equations considered here. It is important to stress that the scalings discussed here change when one considers a different scheme which, for instance, has access to different orders of the GSF (or a different approximation scheme for the equations of motion altogether), executes NITs to different orders for the variables involved, or optimizes different quantities such as the accuracy of the variables when leaving the resonance rather than the global inspiral phase. A more general and detailed discussion allowing for other algorithm choices will appear in a forthcoming paper. 

\subsection{Singularities in the inverse NIT}

We begin by examining the structure of the inverse NIT given in Eqs.\eqref{eq:invNITdef}. 
The most important feature is that the denominators $\sim \vec{\kappa}_{\rm res}\cdot \vec{\Upsilon}^{(0)} = \Upsilon_\perp$ become small near resonance and the series starts to diverge for some constant integer vector $\vec{\kappa}_{\rm res}$. 

One needs to decide where to make the switch to the Partial NIT equations where such divergences do not appear. We define a power index $\beta>0$ so that the switch is executed when $ \vec{\kappa}_{\rm res}\cdot \vec{\Upsilon}^{(0)} \propto \epsilon^\beta$. This is equivalent to saying that we switch to the partial NIT at a (Mino) time $\sim \epsilon^{\beta-1} T$ before hitting the exact point $\vec{\kappa}_{\rm res}\cdot \vec{\Upsilon}^{(0)} = 0$, where $T$ is a dimensionful factor with the dimension of Mino time (1/length in $G=c=1$ units). This is because at that point we have 
\begin{align}
	\vec{\kappa}_{\rm res}\cdot \vec{\Upsilon}^{(0)} \sim -\epsilon^{\beta} \vec{\kappa}_{\rm res}\cdot  \frac{\partial \vec{\Upsilon}^{(0)}}{\partial P_j} \avg{F^{(1)}_j} T \sim  -\epsilon^\beta \Upsilon_\perp' T \,.
\end{align}
where all functions are evaluated on the resonance where $\vec{\kappa}_{\rm res}\cdot \vec{\Upsilon}^{(0)} =0$.

By examining the divergences in Eqs. \eqref{eq:NIT_Y}, \eqref{eq:NIT_X}, and \eqref{eq:NIT_Y2}, we see that the highest-order divergences are $1/(\vec{\kappa}_{\rm res}\cdot \Upsilon_{(0)})$ for $\tilde{Y}^{(1)}$, $1/(\vec{\kappa}_{\rm res}\cdot \vec{\Upsilon}^{(0)})^2$ for $\tilde{X}^{(1)}$, and $1/(\vec{\kappa}_{\rm res}\cdot \vec{\Upsilon}^{(0)})^3$ for $\tilde{Y}^{(2)}$.  One can show that the largest divergence in $\tilde{Y}^{(n)} \sim 1/(\vec{\kappa}_{\rm res}\cdot \vec{\Upsilon}^{(0)})^{2n-1} $ and the largest divergence in $\tilde{X}^{(n)} \sim 1/(\vec{\kappa}_{\rm res}\cdot \vec{\Upsilon}^{(0)})^{2n} $, where $n$ is the order of the inverse NIT. As such, we see that the NIT series carried out to infinite order necessarily diverges unless $\beta<1/2$. Even though the optimal value of $\beta$ will be determined later, it is important to remember that $\beta \in (0,1/2)$ in any case to understand the weighing of terms appearing in later expansions. 

Note also that even though the leading-order terms can be in principle computed, already at this order there will appear sub-leading singular terms corresponding to unknown orders of the self-force. Instead of introducing convoluted constructions, we simply truncate the NIT at second order in the orbital elements and at first order in the phases here. 

\subsection{Handover error} 
The optimal value of $\beta$ depends on the finite order to which we carry out the NIT. By neglecting $\tilde{Y}^{(3)}, \tilde{X}^{(2)}$ terms in the NIT, we are neglecting singular terms of the form
\begin{align}\label{eq:Handover_Errors}
	& \tilde{Y}^{(3)} \sim  \left(\frac{\partial \Upsilon^{(0)}}{\partial P}\right)^2 \left(\frac{\partial F^{(1)}}{\partial q}\right)^2  \frac{F^{(1)}_{\vec{\kappa}_{\rm res.}}}{(\vec{\kappa}_{\rm res}\cdot \Upsilon_{(0)})^5} + \mathcal{O}((\vec{\kappa}_{\rm res}\cdot \Upsilon_{(0)})^{-4})\,,
	\\
	& \tilde{X}^{(2)} \sim  \left(\frac{\partial \Upsilon^{(0)}}{\partial P}\right)   \left(\frac{\partial \Upsilon_\perp}{\partial P}\right)  \left(\frac{\partial F^{(1)}}{\partial q}\right) \frac{F^{(1)}_{\vec{\kappa}_{\rm res.}}}{(\vec{\kappa}_{\rm res}\cdot \Upsilon_{(0)})^4}+ \mathcal{O}((\vec{\kappa}_{\rm res}\cdot \Upsilon_{(0)})^{-3})\,,
\end{align}
where from this point onward we suppress summation indices and factors of order one for simplicity. Another simplification that we make is the definition of the dimensionless factor $\delta$, which can be understood as the ratio of the fluctuating or resonant part and the $q$-averaged part of any function appearing in the expressions. In particular, we assume that $A_{\vec{\kappa}_{\rm res.}} \sim \delta \langle A \rangle$ and $\left( \partial A/\partial q \right) \sim \delta \avg{A} $ and so on. This simplified ``universal'' scaling allows for a more tractable computation. 

We now assume that we transform from the Full NIT variables $\tilde{P},\tilde{q}$ back to OG variables $P,q$ using only $Y^{(2)},X^{(1)}$ and hand them over to the partial NIT integration. The leading hand-over error for $P,q$ respectively then scales as 
\begin{align}
	& \Delta P_{\rm h\text{-}o} \sim \epsilon^{3}  Y^{(3)} \sim \epsilon^{3-5\beta} \delta^3 \left(\Upsilon^{(0)}{}'\right)^2 \avg{F^{(1)} } (\Upsilon_\perp')^{-5} T^{-5} \,,\\
	& \Delta q_{\rm h\text{-}o} \sim \epsilon^{2}  X^{(2)} \sim \epsilon^{2-4\beta} \delta^2 \left(\Upsilon^{(0)}{}'\right) (\Upsilon_\perp')^{-3} T^{-4}\,.
\end{align} 

\subsection{Error of partial NIT evolution}

  The evolution equations are known only to some finite order in powers of $\epsilon$ and, as such, inevitably accumulate error when integrated over the time $\sim \epsilon^{\beta-1} T$. Here we estimate this secular error.
  
   We start by Taylor-expanding the functions $\hat{P}(\lambda),\hat{q}_{\perp}(\lambda)$ around the exact Mino time $\lambda_{\rm res}$ when $\Upsilon_\perp = 0$ as
\begin{subequations} \label{eq:Pqresexp}
  \begin{align}
  	& \hat{P}(\lambda_{\rm res} + \epsilon^{\beta - 1} T) = \sum_{k=0}^\infty \frac{1}{k!} \frac{\mathrm{d}^k\! \hat{P}}{\mathrm{d} \lambda^k}\Big|_{\lambda=\lambda_{\rm res}} \epsilon^{k(\beta - 1)} T^k\,,
  	\\
  	& \hat{q}_\perp(\lambda_{\rm res} + \epsilon^{\beta - 1} T) = \sum_{k=0}^\infty \frac{1}{k!} \frac{\mathrm{d}^k\! \hat{q}_\perp}{\mathrm{d} \lambda^k}\Big|_{\lambda=\lambda_{\rm res}} \epsilon^{k(\beta - 1)} T^k\,,
  \end{align}
\end{subequations}
where the terms in the Taylor expansion can be evaluated by iterating the partial NIT equations of motion \eqref{eq:res_transformed_EoM} at $\lambda_{\rm res}$. It can be shown that this Taylor series is divergent for $\beta<1/2$ since the shortest time-scale of the partial NIT equations is $\sim \sqrt{\epsilon}$. As such, this expansion is only useful for asymptotic analysis.

The key assumption that we make here is that we are able to evaluate $\hat{F}^{(2)}(\hat{P},q_\perp)$ as a function of $q_\perp$ accurately only exactly at resonance. This is because this corresponds to an average second-order flux averaged over a single resonant orbit at fixed $q_\perp$. Away from exact resonance, we assume to be able to evaluate only the full average $\langle\hat{F}^{(2)}\rangle_{q_\perp}$. As a consequence, we assume that we are unable to evaluate derivatives of oscillating parts of the second-order forcing term $\partial \osc{\hat{F}}^{(2)}/\partial \hat{P}$ and similar quantities. Furthermore, we assume that we have no information on the second order forcing terms on the phases such as the second order resonant phase term $f^{(2)}_\perp$. The leading-order unknown terms in the Taylor series \eqref{eq:Pqresexp} are then respectively
\begin{subequations}
	\label{eq:errpNIT}
\begin{align}
	&  \Delta \hat{P} = \frac{1}{2} T^2 \epsilon^{1+2\beta}\left(\osc{\hat{F}}^{(2)'} + \frac{\partial \hat{F}^{(1)}}{\partial q_\perp} \hat{f}^{(2)}_\perp \right)
	\,,
	\\
	& \Delta \hat{q}  = T \epsilon^{1+\beta} \hat{f}^{(2)}_\perp 
	\,.
\end{align}
\end{subequations}

\subsection{Balancing the errors}
The errors in \eqref{eq:errpNIT} cannot be removed by a longer integration of the partial NIT; they will keep accumulating the longer we integrate the equations. As such, it makes no sense to keep using the partial NIT equations when this error becomes larger than the hand-over error. However, we have two types of estimates, the error in phase $\Delta q$, and the error in the orbital elements $\Delta P$, and balancing the error each of these gives different optimal values for $\beta$. At this point we assume that when the inspiral leaves the resonance, it is still a $1/\epsilon$ time before plunge and that our primary goal is to obtain accurate phase at plunge. In that case, one can show that balancing the error in the orbital elements yields better accuracy in the phase at plunge. That is, we solve for $T$ and $\beta$ from the balance
\begin{subequations}
	\label{eq:errpBalance}
	\begin{align}
		 \Delta P_{\rm h\text{-}o}   & \sim  \Delta \hat{P} 
		\,,
		\\
		  \epsilon^{3-5\beta} \delta^3 \left(\Upsilon^{(0)}{}'\right)^2 \avg{F^{(1)} } (\Upsilon_\perp')^{-5} T^{-5}  &\sim 
		T^2 \epsilon^{1+2\beta}\left(\osc{\hat{F}}^{(2)'} + \frac{\partial \hat{F}^{(1)}}{\partial q_\perp} \hat{f}^{(2)}_\perp \right)
		\,.
	\end{align}
\end{subequations}
From this we get that the optimal choice of the power index:  $ \mr^{3-5 \beta} \sim \mr^{ 1 + 2 \beta} \rightarrow \beta = 2/7$. We also obtain our relation for the switching timescale to be:
\begin{equation}
	T = \left[ \frac{ \delta^2} {\left(\Upsilon'_\perp \right)^5} \frac{\left(\Upsilon'^{(0)}\right)^2 \avg{F^{(1)}} }{ \avg{F'^{(2)}}  + \avg{F^{(1)}} \avg{f^{(2)}_\perp} }\right]^{1/7}.
\end{equation}
We then also make the estimate that $f_\perp^{(2)} \sim f_\perp^{(1)}$, resulting in Eq.~\eqref{eq:res_timescale} which is used in our practical implementation (note that both $f_\perp^{(2)} \sim f_\perp^{(1)}$ are $\mathcal{O}(1)$ quantities when the mass ratio prefactors are removed).

Using this value of $\beta$, we can estimate that the total error in the orbital elements incurred from this switching procedure is given by:
\begin{equation}
	 \Delta P \sim \Delta \hat{P} + \Delta \hat{P}_{\rm h\text{-}o} \sim  \epsilon^{11/7} \delta^{11/7}  \left( \avg{F'^{(2)}}  + \avg{F^{(1)}} \avg{f^{(1)}_\perp}\right)^{5/7} \left(\Upsilon^{(0)}{}'\right)^{4/7} \left(\Upsilon'_\perp \right)^{-10/7},
\end{equation}
which corresponds to the $\mr^{11/7}$ we see in Fig.~\ref{fig:TransitionVsResNITConvergenceOrbitalElements}.

Likewise, the resulting error in the phases after evolving for a time of $\mr^{-1}$ is dominated by the error in the orbital elements and is given by:
\begin{subequations}
\begin{align}
	\Delta q_{\text{final}} & \sim  \frac{\Upsilon^{(0)}}{\epsilon \Upsilon'^{(0)}} \Delta \hat{P} + \Delta \hat{q} 
	\\
	& \sim  \epsilon^{4/7}  \delta^{11/7}  \left( \avg{F'^{(2)}}  + \avg{F^{(1)}} \avg{f^{(1)}_\perp}\right)^{5/7} \left(\Upsilon^{(0)}{}'\right)^{-3/7} \left(\Upsilon'_\perp \right)^{-10/7}\Upsilon^{(0)}  + \mathcal{O}(\mr^{9/7})
\end{align} 
\end{subequations}
corresponding to the $\mr^{4/7}$ we see in Fig.~\ref{fig:TransitionVsResNITConvergencePhases}. This is a dramatic improvement over the $\mr^{-1/2}$ error one would incur from incorrectly modelling the orbital resonance seen in Fig.~\ref{fig:NITvsOGONRResonanceConvergencePhases}, while minimising the time spend evaluating the Partial NIT equations of motion. In contrast, if we chose $\beta>2/7$ or $\beta = 2/7+\gamma$ with $\gamma>0$, the term $\Delta P_{\rm h-o}$ would dominate the error and we would obtain $\Delta q_{\rm final} \sim \epsilon^{4/7 - 5 \gamma}$. On the other hand, choosing a $\beta<2/7$, or $\gamma<0$ would lead to the dominance of the $\Delta \hat{P}$ term and $\Delta q_{\rm final} \sim \epsilon^{4/7 +2 \gamma}$.

\end{document}